\documentclass[prx,showpacs,twocolumn,floats,10pt,aps,citeautoscript,longbibliography,superscriptaddress]{revtex4-2}
\usepackage{dcolumn}
\usepackage{microtype} % improves pdf generation
\usepackage{graphicx} % figures
\usepackage[usenames,dvipsnames]{xcolor} % text colors

\usepackage{amsmath}
\usepackage{amssymb}
\usepackage{bbm} % double 1 for unit matrix \bbm{1}
\usepackage{empheq} % boxed align, rcases
\renewcommand{\vec}{\boldsymbol} % vector as bold symbol
\allowdisplaybreaks % page break in Equations

\usepackage{scalerel,stackengine} 
\newcommand\pig[1]{\scalerel*[5pt]{\big#1}{%
\ensurestackMath{\addstackgap[1.5pt]{\big#1}}}}

\newcommand\pigg[1]{\scalerel*[5pt]{\bigg#1}{%
\ensurestackMath{\addstackgap[1.5pt]{\bigg#1}}}}

\newcommand{\TK}{T_{\mathrm{K}}} % Kondo temperature
\newcommand{\mi}{\mathrm{i}} % imaginary unit
\newcommand{\md}{\mathrm{d}} % differential 
\newcommand{\TO}{\mathcal{T}} % time-ordering operator
\newcommand{\nint}{\!\int\!} % narrow integral without boundaries
\newcommand{\nbint}[2]{\!\int_{#1}^{#2}\!} % narrow integral with boundaries
\newcommand{\sseta}{^{\raisebox{-0.2pt}{{\tiny[}}\eta\raisebox{-0.2pt}{{\tiny]}}}} % superscript eta with reduced vertical space
\newcommand{\ub}[1]{{\ensuremath{\underline{#1}}}} % under bar
\newcommand{\ovb}[1]{{\ensuremath{\overline{#1}}}} % over bar
\newcommand{\ovh}[1]{{\ensuremath{\hat{#1}}}} % over hat
\newcommand{\ovwt}[1]{{\ensuremath{\widetilde{#1}}}} % wide over tilde
\newcommand{\ovwh}[1]{{\ensuremath{\widehat{#1}}}} % wide over hat
\newcommand{\pdag}{{\vphantom{dagger}}} % ensures aligned subscripts on d^\dagger_\sigma d^\pdag_\sigma 
\newcommand*{\ndots}{\kern-0.15em.\kern-0.05em.\kern-0.05em.} % narrow lower dots
\newcommand*{\nidots}{.\kern-0.05em.\kern-0.05em.} % narrow lower dots in subscript
\newcommand*{\ncdots}{\kern-0.15em\cdot\kern-0.2em\cdot\kern-0.2em\cdot\kern-0.15em} % narrow centraldots

\newcommand{\Eq}[1]{Eq.~\eqref{#1}}
\newcommand{\Eqs}[1]{Eqs.~\eqref{#1}}
\newcommand{\Fig}[1]{Fig.~\ref{#1}}

\def\ZF/{\mbox{ZF}} % zero-temperature formalism
\def\MF/{\mbox{MF}} % Matsubara formalism
\def\KF/{\mbox{KF}} % Keldysh formalism

\def\Ac{\mathcal{A}}
\def\Bc{\mathcal{B}}
\def\Gc{\mathcal{G}}
\def\Hc{\mathcal{H}}
\def\Kc{\mathcal{K}}
\def\Oc{\mathcal{O}}
\def\Sc{\mathcal{S}}

\RequirePackage[
  hyperindex,colorlinks,bookmarksnumbered,
  plainpages=true,pdfstartview=FitH]{hyperref}
\hypersetup{urlcolor={blue!50!black}, linkcolor={red!50!black},citecolor={blue!50!black}} 
\usepackage{hyperref}
\usepackage[all]{hypcap}

\begin{document}

\title{Multipoint correlation functions: spectral representation and numerical evaluation}
\author{Fabian B.~Kugler}
\thanks{These authors contributed equally to this work.}
\affiliation{Arnold Sommerfeld Center for Theoretical Physics, 
Center for NanoScience,\looseness=-1\,  and 
Munich Center for \\ Quantum Science and Technology,
Ludwig-Maximilians-Universit\"at M\"unchen, 80333 Munich, Germany}
\affiliation{Department of Physics and Astronomy, Rutgers University, Piscataway, New Jersey 08854, USA}
\author{Seung-Sup B.~Lee}
\thanks{These authors contributed equally to this work.}
\affiliation{Arnold Sommerfeld Center for Theoretical Physics, 
Center for NanoScience,\looseness=-1\,  and 
Munich Center for \\ Quantum Science and Technology,
Ludwig-Maximilians-Universit\"at M\"unchen, 80333 Munich, Germany}
\author{Jan von Delft}
\affiliation{Arnold Sommerfeld Center for Theoretical Physics, 
Center for NanoScience,\looseness=-1\,  and 
Munich Center for \\ Quantum Science and Technology,
Ludwig-Maximilians-Universit\"at M\"unchen, 80333 Munich, Germany}

\date{\today}

\begin{abstract}
The many-body problem is usually approached from one of two perspectives:
the first originates from an action and is based on Feynman diagrams,
the second is centered around a Hamiltonian and deals with quantum states and operators.
The connection between results obtained in either way is made through spectral (or Lehmann) representations,
well known for two-point correlation functions.
Here, we complete this picture by deriving generalized spectral representations for multipoint correlation functions that apply in all of the commonly used many-body frameworks:
the imaginary-frequency Matsubara and the real-frequency zero-temperature and Keldysh formalisms.
Our approach separates spectral from time-ordering properties and thereby elucidates the relation between the three formalisms.
The spectral representations of multipoint correlation functions consist of partial spectral functions and convolution kernels.
The former are formalism independent but system specific;
the latter are system independent but formalism specific.
Using a numerical renormalization group (NRG) method described in the accompanying paper,
we present numerical results for selected quantum impurity models.
We focus on the four-point vertex (effective interaction) obtained for the single-impurity Anderson model and for the dynamical mean-field theory (DMFT) solution of the one-band Hubbard model.
In the Matsubara formalism, we analyze the evolution of the vertex down to very low temperatures and describe the crossover from strongly interacting particles to weakly interacting quasiparticles.
In the Keldysh formalism, we first benchmark our results at weak and infinitely strong interaction and then reveal the rich real-frequency structure of the DMFT vertex in the coexistence regime of a metallic and insulating solution.
\end{abstract}

\maketitle

\section{Introduction}

\subsection{Multipoint correlation functions}

A major element in the ongoing challenge of the quantum many-body problem
is to extend our understanding, our analytical and numerical control, 
from the single- to the many-particle level.
One-particle correlation functions, describing the propagation of a particle 
in a potentially highly complex, interacting environment, 
are clearly important. However, almost since the beginning of interest in the many-body problem, two-particle correlation functions have played an equally important role. They describe the effective interaction between two particles in the many-body environment, response functions to optical or magnetic probes, collective modes, bound states, and pairing instabilities, to name but a few,
and are essential ingredients in Landau's Fermi-liquid theory \cite{Nozieres1997}.

For a long time, two-particle or four-point ($4$p) functions could only be computed by perturbative means, 
later including resummation and renormalization group schemes as well.
Recently, the advance of numerical techniques to compute such functions fully nonperturbatively, albeit only locally,
has opened a new chapter.
For various impurity models (general ones as well as those arising in 
dynamical mean-field theory (DMFT) \cite{Georges1996}),
it was even found that $4$p functions can exhibit divergences in strongly correlated regimes
\cite{Schaefer2013,Janis2014,Ribic2016,Schaefer2016,Gunnarsson2016,Vucicevic2018,Thunstroem2018,Chalupa2018,Melnick2020,Springer2020,Reitner2020}.
These divergences occur for a special class of $4$p functions, namely two-particle irreducible vertices,
and have renewed the interest in thoroughly understanding the properties of $4$p functions.

The properties of multi- or $\ell$-point ($\ell$p) functions depend on the theoretical framework employed.
Technically, $\ell$p functions are correlation functions of $\ell$ operators with $\ell$ arguments.
If the operators are taken at different times, the Fourier-transformed function depends on frequencies.
Clearly, these times and frequencies are real numbers, and the early works directly dealt with these \textit{real} frequencies.
Later, the many-body theory was revolutionized by the Matsubara technique, based on a Wick rotation from real to \textit{imaginary} times \cite{Abrikosov1975}.
This significantly facilitates numerical approaches, and 
the vast majority of numerical work on $\ell$p functions is based on the Matsubara formalism (\MF/).
The periodicity in imaginary times yields functions of discrete imaginary or Matsubara frequencies.
Physical results are obtained in the end by performing an analytic continuation, back from imaginary to real frequencies.
Yet, crucially, the \textit{numerical} analytic continuation from imaginary to real frequencies is
an ill-conditioned problem \cite{Gubernatis1991};
though it may work fairly well for docile $2$p functions, it is unfeasible for higher-point objects.

Formally, real-frequency frameworks are as well established as the \MF/:
the zero-temperature formalism (\ZF/) works with
ground-state expectation values of time-ordered operators \cite{Abrikosov1975}.
Finite-temperature real-frequency dynamics, both in and out of equilibrium,
are treated in the Keldysh formalism (KF), 
based on a doubled time contour \cite{Schwinger1961,Keldysh1964}.
Due to the problematic analytic continuation and 
growing interest in nonequilibrium dynamics,
the \KF/ has gained much popularity in recent years.
However, there are hardly any numerical real-frequency results of $\ell$p functions depending on more than one frequency.
Conceptually, it is therefore of great interest to extend the understanding of $\ell$p functions 
that has been reached for imaginary frequencies to the real(-frequency) world.

\subsection{Our approach}

In this work, we achieve a deepened understanding of $\ell$p functions, 
applying naturally in \textit{any} of the real- or imaginary-time frameworks.
The central principle of our approach
is to separate properties due to time ordering from spectral properties of the system.
The former govern the analytic structure of $\ell$p functions;
the latter are the key objects of the numerical evaluation.
While exploiting this separation is standard practice for $2$p functions, to our knowledge
it has not yet been pursued systematically for $\ell$p functions.
Here, we derive spectral representations of general $\ell$p functions for arbitrary $\ell$, 
in all three frameworks (\ZF/, \MF/, \KF/).
We illustrate the power of our approach by presenting numerical
results for various local $4$p examples.

The spectral representations are based on 
a density matrix, which defines the quantum averages,
and an expansion in eigenstates of the Hamiltonian, which determines the time evolution. The spectral representations have similar forms in all three frameworks. 
Indeed, the spectral or Lehmann representation of $2$p functions is arguably the most transparent 
way of demonstrating the analytic relation between Matsubara and retarded propagators \cite{Mahan2000}.
Here, we achieve a similar level of transparency:
all $\ell$p functions are expressed by a sum over \textit{formalism-independent} 
``partial'' spectral functions,
convolved with \textit{formalism-specific} kernels.
The partial spectral functions (PSFs) 
serve as unique and compact porters of
the \textit{system-specific} information.
Convolving them with the rather 
simple and \textit{system-independent} kernels yields correlation functions in the desired form.
Through this convolution, the correlation functions acquire their shape and characteristic long tails.
In fact, the numerical storage of $\ell$p functions, depending on $\ell \!-\! 1$ frequencies, can be problematic
\cite{Boehnke2011,Li2016,Wentzell2020,Shinaoka2017,Shinaoka2018,Shinaoka2020}.
In this regard, the PSFs are advantageous since they have compact support, 
bounded by the largest energy scale in the system.

For a set of external time arguments of a correlation function $\Gc(t_1, \ndots, t_\ell)$, 
the time-ordering prescription determines the corresponding order of operators in the expectation value.
In our approach, we first consider \textit{all} possible operator orderings, i.e.,
$\langle \Oc_\ovb{1}(t_\ovb{1}) \cdots \Oc_\ovb{\ell}(t_\ovb{\ell}) \rangle$
for all permutations 
$p(1, \, \ndots \, ,\ell) \!=\! (\ovb{1}, \, \ndots \, , \ovb{\ell})$.
The Fourier transform of each of these $\ell!$ expectation values yields a PSF;
the collection of all PSFs describes the spectrum of the system (on the $\ell$p level).
Second, for given external times, some operator orderings are specified, depending on the formalism.
In the \ZF/ and \MF/, this is always one ordering; in the \KF/ using the Keldysh basis, it may be a linear combination of multiple ones.
The ordering is specified by kernel functions which are multiplicative in the time domain.
In the frequency domain, they become convolution kernels,
mapping PSFs to contributions to $\Gc(\omega_1, \ndots, \omega_\ell)$.

The spectral representations clearly reveal the similarities and differences of the various $\ell$p functions.
In the most complex setup of the \KF/ with Keldysh basis, where each argument has an
extra Keldysh index $1$ or $2$,
we find that those components with a single Keldysh index equal $2$ at position $\eta$, dubbed $\mathcal{G}\sseta$, 
have the simplest structure.
Indeed, they are the \textit{(fully) retarded} objects which can be obtained from Matsubara $\ell$p functions via a suitable analytic continuation,
$\mi\omega_i \to \omega_i \pm \mi 0^+$ \cite{Weldon2005a,*Weldon2005b}. 
This does not apply to the remaining Keldysh components.
However, we find that their convolution kernels $K$ are linear combinations of the retarded kernels $K\sseta$.
Thereby, our spectral representations offer a direct way of discussing the relation 
between the Matsubara and \textit{all} Keldysh $\ell$p functions in explicit detail---%
this will be the topic of a forthcoming publication \cite{Ge2020b}.

To illustrate our approach,
we numerically evaluate the spectral representation of local $4$p functions 
for selected quantum impurity models via the numerical renormalization group (NRG) \cite{Bulla2008}.
Since its invention by Wilson in 1975 \cite{Wilson1975}, NRG has become the gold standard for solving impurity models.
It has the unique advantage of allowing one to 
(i) directly compute real- and imaginary-frequency results without the need for analytic continuation,
(ii) reach arbitrarily low temperatures with marginal increase in the numerical costs,
(iii) access vastly different energy scales, zooming in on the lowest excitation energies.
Our NRG scheme is based on the full density matrix (fdm) NRG \cite{Weichselbaum2007,Peters2006}
in an efficient tensor-network formulation \cite{Weichselbaum2012a,Weichselbaum2012b,Weichselbaum2020}, 
with an additional, iterative structure to finely resolve regimes of 
frequencies $|\omega_i| \!\ll\! |\omega_j|$, $i \!\neq\! j$.
The rather intricate prescription for how to do this for $3$p and $4$p functions 
is explained in the accompanying paper \cite{Lee2021}. 

\subsection{Structure of this paper}

Our spectral representations of $\ell$p functions and the NRG scheme to evaluate them have numerous potential 
applications,
such as finite-temperature formulations of Fermi-liquid theory and
nonlocal extensions of DMFT in the \MF/ at low temperatures
or in real-frequency \KF/ implementations.
We elaborate on the applications in the concluding Sec.~\ref{sec:concl:outlook}.

The rest of the paper is organized as follows.
In Sec.~\ref{sec:spectral_representation}, we derive the spectral representations in a general setting,
for arbitrary $\ell$, a given density matrix, and any time-independent Hamiltonian. 
After motivating our approach and introducing notation for $\ell \!=\! 2$, 
we subsequently derive our results in the \ZF/, \MF/, and \KF/, the latter both in the contour and Keldysh bases. 
The main results of this section are the spectral representations,
involving the PSFs \eqref{eq:S_w} and summarized 
in \Eqs{eq:G_w_RF} and \eqref{eq:K_w_RF}
for the \ZF/, 
in \Eqs{eq:G_iw_KS_IF} 
and \eqref{eq:K_Omega_IF_compact} or \eqref{eq:K_Omega_IF} for the \MF/, 
and in 
\Eqs{eq:Kc_KF_Keldysh_via_Keta}, \eqref{eq:G_w_KF_Keldysh} or \eqref{eq:G_w_KF_eta_j_total}
for the \KF/ in the Keldysh basis.
In Sec.~\ref{sec:quantities}, we briefly describe the quantities of interest for the numerical evaluation, i.e.,
local $4$p correlation and vertex functions.
Our numerical results are contained in Sec.~\ref{sec:results}.
We start with a simple model for x-ray absorption in metals
treated in the \ZF/ and analyze power laws in the $4$p vertex.
Proceeding with the Anderson impurity model (AIM) in the \MF/,
we first benchmark our results at intermediate temperatures against Monte Carlo data and then extend these results to lower temperatures to enter the Fermi-liquid regime and deduce the quasiparticle interaction.
Thereafter, we treat the AIM in the \KF/, first testing our method at weak and infinitely strong interaction,
before moving to the intermediate, strongly interacting regime.
Finally, we present results for the DMFT solution of the one-band Hubbard model and compare the \MF/ and \KF/ vertex for both the metallic and insulating solution.
In Sec.~\ref{sec:conclusion}, we summarize our results and give an outlook on applications.
Appendices~\ref{app:RF_G2} to \ref{app:KF} are devoted to exemplary calculations in each of the \ZF/, \MF/, and \KF/. 
Appendix~\ref{app:explicit_amputation} describes details needed for amputating external legs 
when computing the \ZF/ or \KF/ $4$p vertex, 
and App.~\ref{sec:app:HA_2ndOrder} discusses an example of anomalous parts 
in both the \MF/ and \KF/.

\section{Spectral representation}
\label{sec:spectral_representation}

\subsection{Motivation of partial spectral functions}
\label{sec:motivation}

To set the stage and introduce notation,
we review the standard derivation of spectral representations for
$2$p correlators. 
We denote complete sets of energy eigenstates
by underlined integers, e.g.\ $\{ | \ub{1} \rangle \}$,
with eigenvalues $E_{\ub{1}}$. 
We use calligraphic or roman symbols for operators
or their matrix elements, 
$A_{\ub{1}\ub{2}} = \langle \ub{1}| \Ac | \ub{2}\rangle $.
Expectation values are obtained through the density matrix $\varrho$,
which, in thermal equilibrium at temperature $T=1/\beta$, 
directly follows from the Hamiltonian $\Hc$:
\begin{align}
\langle \Ac \rangle
=
\mathrm{Tr} [\varrho \Ac ]
,
\quad
\varrho = e^{-\beta \mathcal{H}}/Z
,
\quad
Z = \mathrm{Tr} [ e^{-\beta \mathcal{H}} ]
.
\end{align}
For instance, we have 
$\langle \Ac \Bc \rangle = \sum_{\ub{1}\ub{2}} 
\rho_\ub{1} A_{\ub{1}\ub{2}} B_{\ub{2} \ub{1}} $, with 
$\rho_\ub{1} = \langle \ub{1}| \varrho | \ub{1} \rangle = e^{-\beta E_{\ub{1}}}/Z$. 
The \ZF/ assumes $\rho_{\ub{1}} = \delta_{\ub{1} g}$ at $T=0$
with a nondegenerate ground state $|g\rangle$ \cite{Brouder2008};
the \MF/ and \KF/ work at any $T$.
Further, the \KF/ can also be used with a nonequilibrium density matrix.
Yet, for simplicity, we do not consider such cases explicitly. 

In the \ZF/ or \MF/, operators obey Hamiltonian evolution
in real time $t$ or imaginary time $\tau$, respectively:
\begin{align}
\Ac(t)
= 
e^{\mi \mathcal{H}t} \Ac e^{-\mi \mathcal{H}t}
, \qquad 
\Ac(\tau) 
& = 
e^{\mathcal{H}\tau} \Ac e^{-\mathcal{H}\tau}.
\end{align}
The corresponding time-ordered ($\TO$) correlators and 
their Fourier transforms in the \ZF/ or \MF/ are defined as 
\begin{subequations}
\begin{alignat}{2}
G(t) & =-\mi \langle \TO \Ac(t)\Bc \rangle , \quad
& 
G(\omega) & =\nbint{-\infty}{\infty} \md t \, e^{\mi \omega t} G(t)
, 
\\
G(\tau) & = - \langle \TO \Ac(\tau) \Bc \rangle , \quad 
\; & 
G(\mi \omega) & = \nbint{0}{\beta} \md \tau \, e^{\mi \omega \tau} G(\tau) ,
\end{alignat}
\end{subequations}
where, depending on context, $\omega$ denotes a continuous real frequency or a discrete Matsubara frequency.
(For brevity, we distinguish $G$ in the \ZF/ and \MF/ solely through its arguments, $t$ vs.\ $\tau$ or $\omega$ vs.\ $\mi \omega$.)
Textbook calculations, based on judicious insertions of the identity in the form 
$\mathbbm{1} = \sum_{\ub{1}} |\ub{1}\rangle \langle \ub{1}| 
= \sum_{\ub{2}} |\ub{2}\rangle \langle \ub{2}|$, 
yield the following Lehmann representation of the Fourier-transformed \ZF/ correlator:
\begin{align}
\nonumber
G(\omega)
& =
-\mi
\nbint{0}{\infty} 
\md t \, e^{\mi \omega t} \langle \Ac(t) \Bc \rangle
-\mi \zeta
\nbint{-\infty}{0} 
\md t \, e^{\mi \omega t} \langle \Bc \Ac(t) \rangle
\\
& =
\sum_{\ub{1}\,\ub{2}}
A_{\ub{1}\ub{2}} B_{\ub{2}\ub{1}}
\Big(
\frac{\rho_{\ub{1}}}{\omega^+ - E_{\ub{2}\ub{1}}}
-\zeta
\frac{\rho_{\ub{2}}}{\omega^- - E_{\ub{2}\ub{1}}}
\Big) 
.
\label{eq:G2_RF}
\end{align}
Here, we used 
$\zeta \!=\! 1$ ($\zeta \!=\! -1$) for bosonic (fermionic) operators,
$\omega^\pm = \omega \pm \mi 0^+$ for convergence of real-time integrals,
and $E_{\ub{2}\ub{1}} \!=\! E_{\ub{2}} \!-\! E_{\ub{1}}$.
The \MF/ correlator is obtained as
\begin{subequations}
\begin{align}
G(\mi\omega)
& =
-
\nbint{0}{\beta}
\md \tau \, e^{\mi \omega \tau} \langle \Ac(\tau) \Bc \rangle
\label{eq:G2_IF-integral}
\\ 
& = 
\sum_{\ub{1}\,\ub{2}}
\rho_{\ub{1}} A_{\ub{1}\ub{2}} B_{\ub{2}\ub{1}}
 \frac{1 - e^{\beta(\mi \omega -E_{\ub{2}\ub{1}})}}
 {\mi\omega - E_{\ub{2}\ub{1}}}
\label{eq:G2_IF-intermediate}
\\
& =
\sum_{\ub{1}\,\ub{2}}
A_{\ub{1}\ub{2}} B_{\ub{2}\ub{1}}
\frac{\rho_{\ub{1}} - \zeta \rho_{\ub{2}}}{\mi\omega - E_{\ub{2}\ub{1}}}
.
\label{eq:G2_IF}
\end{align}
\end{subequations}
In the bosonic case, where $\mi \omega$ can equal zero, 
terms with both $\mi \omega = 0$ and $E_{\ub{2}\ub{1}}=0$ (if present) can be (analytically) dealt with by taking the limit $E_{\ub{2}\ub{1}} \to 0$
in \Eq{eq:G2_IF-intermediate}, yielding $(1 \!-\! e^{-\beta E_{\ub{2}\ub{1}}})/E_{\ub{2}\ub{1}} \to \beta$.
It will sometimes be convenient
to make this ``anomalous'' case explicit, writing
\begin{align}
G (\mi\omega)
& =
\sum_{\ub{1}\,\ub{2}}
A_{\ub{1}\ub{2}} B_{\ub{2}\ub{1}}
\begin{cases}
{\displaystyle \frac{\rho_{\ub{1}} - \zeta \rho_{\ub{2}}}{\mi\omega - E_{\ub{2}\ub{1}}}}
, & 
\mi \omega - E_{\ub{2}\ub{1}} \neq 0
, 
\\
-\beta \rho_{\ub{1}}
, & \quad 
\textrm{else}
.
\end{cases}
\label{eq:G2_incl_anomalous}
\end{align}

We can also consider $2$p correlators arising in the \KF/,
e.g.\ $G^{+-}(t) = -\mi \langle \Ac(t) \Bc \rangle$, or the retarded propagator
$G^{21}(t) = -\mi \theta(t) \langle [\Ac(t),\Bc]_{-\zeta} \rangle$, where $\theta$ denotes 
the step function, $[\cdot,\cdot]_-$ a commutator, and $[\cdot,\cdot]_+$
an anticommutator. 
Their Fourier transforms read
\begin{subequations}
\label{eq:G2_KF}
\begin{align}
G^{+-}(\omega)
& 
=
\sum_{\ub{1}\,\ub{2}}
\rho_{\ub{1}} A_{\ub{1}\ub{2}} B_{\ub{2}\ub{1}}
\Big(
\frac{1}{\omega^+ \!-\! E_{\ub{2}\ub{1}}}
-
\frac{1}{\omega^- \!-\! E_{\ub{2}\ub{1}}}
\Big)
,
\\
G^{21}(\omega)
& 
=
\sum_{\ub{1}\,\ub{2}}
A_{\ub{1}\ub{2}} B_{\ub{2}\ub{1}}
\frac{\rho_{\ub{1}} - \zeta \rho_{\ub{2}}}{\omega^+ - E_{\ub{2}\ub{1}}}
.
\end{align}
\end{subequations}
Evidently, the imaginary-time and the retarded correlator
are connected by the well-known analytic continuation,
$G^{21}(\omega) = G(\mi \omega \to \omega^+)$ \cite{Baym1961}.
They are often expressed through the ``standard'' spectral function $S_{\mathrm{std}}$:
\begin{subequations}
\begin{align}
S_{\mathrm{std}}(\omega)
& =
\sum_{\ub{1}\,\ub{2}}
A_{\ub{1}\ub{2}} B_{\ub{2}\ub{1}}
(\rho_{\ub{1}} - \zeta \rho_{\ub{2}})\delta(\omega - E_{\ub{2}\ub{1}})
,
\label{eq:Sstd}
\\
G(\mi \omega) 
& = 
\int \md \omega' \, \frac{S_{\mathrm{std}}(\omega')}{\mi \omega - \omega'}
, 
\label{eq:G2_IF-Sstd}
\\
G^{21}(\omega) 
& = 
\int \md \omega' \, \frac{S_{\mathrm{std}}(\omega')}{\omega^+ - \omega'}
.
\end{align}
\end{subequations}
It would be convenient to have spectral representations
capable of describing $G(\omega)$ and $G^{+-}(\omega)$, too.
To this end, we define the partial spectral functions (PSFs)
\begin{align}
S[\Ac,\Bc](\omega)
& = 
\sum_{\ub{1}\,\ub{2}}
\rho_{\ub{1}} A_{\ub{1}\ub{2}} B_{\ub{2}\ub{1}}
\delta(\omega - E_{\ub{2}\ub{1}})
.
\end{align}
Clearly, we can reconstruct $S_{\mathrm{std}}$ from $S$ according to
\begin{align}
S_{\mathrm{std}}(\omega)
& = 
S[\Ac,\Bc](\omega)
- \zeta S[\Bc,\Ac](-\omega)
.
\end{align}
Furthermore, $S$ can be used to express all correlators encountered so far:
\begin{subequations}
\begin{flalign}
G(\omega) 
& = 
\nint \md \omega' 
\left[ \frac{S[\Ac,\Bc](\omega')}{\omega^+ \!-\! \omega'}
- \zeta
\frac{S[\Bc,\Ac](-\omega')}{\omega^- \!-\! \omega'} \right]
,
\hspace{-0.5cm}
&
\\
G(\mi\omega) 
& =
\nint \md \omega' 
\left[\frac{S[\Ac,\Bc](\omega')}{\mi\omega \!-\! \omega'}
- \zeta
\frac{S[\Bc,\Ac](-\omega')}{\mi\omega \!-\! \omega'} \right]
,
\hspace{-0.5cm}
&
\label{eq:G2_IF-S}
\\
G^{21}(\omega) 
& =
\nint \md \omega' 
\left[\frac{S[\Ac,\Bc](\omega')}{\omega^+ \!-\! \omega'}
- \zeta
\frac{S[\Bc,\Ac](-\omega')}{\omega^+ \!-\! \omega'} \right]
,
\hspace{-0.5cm}
\label{eq:G2_w_KF_retarded}
&
\\
G^{+-}(\omega) 
& =
\nint \md \omega' S[\Ac,\Bc](\omega')
\left[
\frac{1}{\omega^+ \!-\! \omega'}
- 
\frac{1}{\omega^- \!-\! \omega'}
\right] 
&
\nonumber
\\
& =
-2\pi \mi S[\Ac,\Bc](\omega)
.
&
\label{eq:G2_w_KF_+-}
\end{flalign}
\end{subequations}

For the bosonic case, the representations
\eqref{eq:G2_IF-Sstd} and \eqref{eq:G2_IF-S} 
involve a subtlety: 
To correctly reproduce the anomalous term of \Eq{eq:G2_incl_anomalous},
the integral over $\omega'$ must be performed first
and the limit $E_{\ub{2}\ub{1}} \to 0$ taken only thereafter.
If, instead, $S_{\mathrm{std}}(\omega)$ is simplified first
by using $E_{\ub{2}\ub{1}} =0$ to conclude that 
$\rho_{\ub{1}} - \rho_{\ub{2}} = 0$, the anomalous terms are missed.
To ensure that \Eq{eq:G2_incl_anomalous} is always correctly reproduced,
including its anomalous terms, we refine
the representation \eqref{eq:G2_IF-S} 
by using a kernel in which the anomalous case 
$\mi \omega - \omega' =0$ is specified separately:
\begin{subequations}
\label{eq:G2_IF_total}
\begin{flalign}
G & (\mi \omega)
=
\sum_p
\zeta^p
\nint \md \omega_{\ovb{1}}' \,
K(\mi \omega_{\ovb{1}} - \omega_{\ovb{1}}' )
S[\Oc^{\ovb{1}},\Oc^{\ovb{2}}](\omega_{\ovb{1}}') 
,
\hspace{-0.3cm}
&
\label{eq:G2_IF-kernel}
\\
K & (\mi \omega - \omega')
= 
\begin{cases}
1/(\mi \omega - \omega')
, \qquad\quad 
\mi \omega - \omega' \neq 0
, 
\\
- \beta/2
, \qquad\qquad\quad\ \,
\textrm{else}
.
\end{cases}
\hspace{-0.3cm}
&
\label{eq:K2_IF}
\end{flalign}
\end{subequations}
The sum in \Eq{eq:G2_IF-kernel} is over 
the two permutations of two indices, 
$p = (\ovb{1} \, \ovb{2}) \in \{(1\,2), (2\,1)\}$, 
with corresponding signs
$\zeta^{(12)}=1$, $\zeta^{(21)}=\zeta$.
Furthermore, $\Oc^{\ovb{1}}$, $\omega_{\ovb{1}}$,
and $\omega'_{\ovb{1}}$ 
are components of operator and frequency tuples
defined as $(\Oc^1,\Oc^2)=(\Ac,\Bc)$, 
$(\omega_1,\omega_2) = (\omega,-\omega)$
and $(\omega'_1,\omega'_2)$, respectively, 
the latter being (dummy) integration variables. 
Spelled out, the representation \eqref{eq:G2_IF_total} 
yields
\begin{align}
\nonumber
G (\mi \omega)
& = \sum_{\ub{1}\,\ub{2}} 
\pig[
\rho_{\ub{1}} A_{\ub{1}\ub{2}} B_{\ub{2}\ub{1}}
K(\mi \omega - E_{\ub{2}\ub{1}} )
\\[-3mm]
& \qquad \qquad 
+ \zeta \rho_{\ub{1}} B_{\ub{1}\ub{2}} A_{\ub{2}\ub{1}}
K(-\mi \omega - E_{\ub{2}\ub{1}} ) 
\pig]
,
\end{align}
which matches \Eq{eq:G2_incl_anomalous} after relabeling $\ub{1} \!\leftrightarrow\! \ub{2}$ in the second term. 
The anomalous part, relevant only for $\zeta \!=\! 1$ and $\omega \!=\! 0$, 
receives equal contributions from both terms, 
since $E_{\ub{2}\ub{1}} \!=\! 0$ implies $\rho_{\ub{1}} \!=\! \rho_{\ub{2}}$.
For $\ell$p functions, the treatment
of anomalous terms is more involved, as bosonic frequencies
can arise through several combinations of fermionic ones.
The formalism to be developed below will enable us to
deal with such subtleties in a systematic fashion.

In the following sections, we derive spectral representations for $\ell$p functions. We start with the \ZF/, where the derivation
is most compact. Subsequently, we show that the same PSFs also arise in the \MF/. 
Finally, we extend our analysis to the \KF/ in the contour and Keldysh bases.

\subsection{Zero-temperature formalism}
\label{sec:RF}

To obtain concise expressions for $\ell$p functions, we introduce further compact notation. 
\textit{A priori}, such functions 
of $\ell$ operators $\vec{\Oc} \!=\! (\Oc^1, \ndots, \Oc^\ell)$ depend
on $\ell$ time or frequency arguments, 
$\vec{t} \!=\! (t_1, \ndots, t_\ell)$ or 
$\vec{\omega} \!=\! (\omega_1, \ndots, \omega_\ell)$.
Permutations of such ordered tuples are denoted 
$\vec{\Oc}_p \!=\! (\Oc^{\ovb{1}}, \ndots, \Oc^{\ovb{\ell}})$, 
$\vec{t}_p \!=\! (t_{\ovb{1}}, \ndots, t_{\ovb{\ell}})$,
and $\vec{\omega}_p \!=\! (\omega_{\ovb{1}}, \ndots, \omega_{\ovb{\ell}})$. 
Here, by definition, the permutation 
$p(1 2 \, \ndots \ell) \!=\! (\ovb{1} \ovb{2} \, \ndots \ovb{\ell} ) $
[or $p \!=\! (\ovb{1} \ovb{2} \, \ndots \ovb{\ell} )$ for short]
acts on the index tuple $(12 \, \ndots \ell)$ by replacing 
$i$ by $p(i) \!=\! \ovb{i}$ in slot $i$.
Note that $p$ moves $i$ to slot $j \!=\! p^{-1}(i)$, 
replacing $j$ there by $p(j) \!=\! i$.
If $p \!=\! (312)$, e.g.,
then $(t_1,t_2,t_3)_p \!=\! (t_3,t_1,t_2)$.
We also use the shorthands
\begin{align*}
\textstyle
\vec{\omega} \! \cdot \! \vec{t}
=
\sum_{i=1}^\ell \omega_i t_i
, \quad 
\md^\ell t
= 
\prod_{i=1}^\ell \md t_i 
, \quad 
\md^\ell \omega
= 
\prod_{i=1}^\ell \md \omega_i
.
\end{align*}
Because of time-translation invariance, $\ell$p functions actually 
depend on only $\ell \!-\! 1$ independent time or frequency arguments. 
Nevertheless, as anticipated above, it will be helpful
to use all $\ell$ frequencies, 
it being understood that their sum equals zero. 
With the shorthand 
$\omega_{i \cdots j} \!=\! \sum_{n=i}^{j} \omega_n$,
we can express the corresponding energy-conservation relations as
\begin{align}
\omega_{1 \cdots \ell} = 0 
, \qquad 
\omega_{1 \cdots i} = - \omega_{i+1 \cdots \ell} 
. 
\end{align}

We use calligraphic symbols, $\Gc$, $\Kc$, $\Sc$, for functions of all $\ell$ arguments and roman symbols, 
$G$, $K$, $S$, for functions of the independent $\ell \!-\! 1$ arguments. 
In the time domain, we define the time-ordered \ZF/ correlator as 
\begin{align}
\Gc(\vec{t})
& =
(-\mi)^{\ell-1}
\langle \TO \prod_{i=1}^\ell \Oc^i(t_i) \rangle
.
\label{eq:RF_timedomain} 
\end{align}
It is invariant under a uniform shift of all times, 
e.g.\ $t_i \to t_i \!- t_\ell$,
and thus depends on only $\ell \!-\! 1$ time differences, 
$(t_1 \!- t_\ell, \ndots, t_{\ell-1} \!- t_\ell, 0)$.
Accordingly, in the frequency domain,
\begin{align}
\Gc(\vec{\omega})
& =
\nint \md^\ell t \, e^{\mi \vec{\omega} \cdot \vec{t}} 
\Gc(\vec{t})
=
2\pi \delta(\omega_{1\cdots \ell}) G(\vec{\omega})
.
\label{eq:RF_frequencydomain}
\end{align}
We write $G(\vec{\omega})$, the part remaining after
factoring out $2\pi \delta(\omega_{1\cdots \ell})$,
with $\ell$ frequency arguments,
it being understood that they satisfy 
energy conservation, $\omega_{1 \cdots \ell} = 0$.

\begin{figure}
\includegraphics[width=\linewidth]{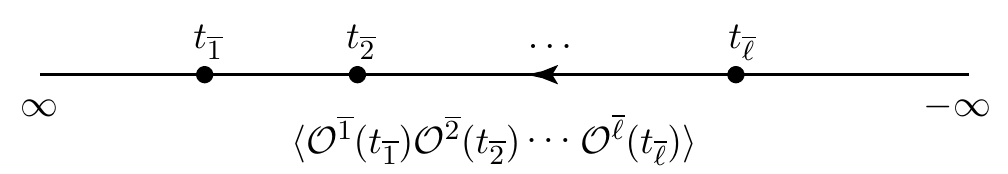}
\caption{\ZF/ time ordering. For a given
$\ell$-tuple of times $\vec{t} = (t_1, \ndots, t_\ell)$, only
that permutation $p$ in \Eq{eq:Gc_t_KS_RF_total} yields a nonzero
contribution, $\Kc(\vec{t}_p) \neq 0$, for which the permuted
times, $\vec{t}_p = (t_{\ovb{1}}, \ndots, t_{\ovb{\ell}})$,
satisfy $t_{\ovb{1}} > t_{\ovb{2}} > \ncdots > t_{\ovb{\ell}}$.
In the corresponding operator product, $\Sc [\vec{\Oc}_p](\vec{t}_p)$, larger times appear to the left of smaller ones.
Thus, the right-to-left order of operators in the product matches the order in which their times appear on the time axis, drawn with times increasing toward the left.}
\label{fig:TimeOrderingRF} 
\end{figure}

The effect of the time-ordering procedure
in the definition of $\Gc(\vec{t})$ can be expressed as a sum over permutations 
involving products of step functions $\theta$:
\begin{subequations}
\label{eq:Gc_t_KS_RF_total}
\begin{align}
 \Gc(\vec{t})
& =
\sum_p
\vec{\zeta}^p
\Kc(\vec{t}_p) \, 
\Sc [\vec{\Oc}_p] (\vec{t}_p) 
, 
\label{eq:Gc_t_KS_RF}
\\
\Kc(\vec{t}_p)
& = \prod_{i=1}^{\ell-1} \Big[ 
- \mi \theta(t_{\ovb{i}}-t_{\ovb{i+1}}) 
\Big]
,
\label{eq:Kc_t_RF}
\\
\Sc [\vec{\Oc}_p](\vec{t}_p)
& = 
\langle \prod_{i=1}^\ell \Oc^{\ovb{i}}(t_{\ovb{i}}) \rangle
.
\label{eq:Sc_t}
\end{align}
\end{subequations}
In \Eq{eq:Gc_t_KS_RF}, the sum is over all
permutations $p$ of the $\ell$ indices labeling operators and times. For a given choice of times, only
one permutation yields a nonzero result, namely the one
which arranges the operators in time-ordered fashion (larger
times left of smaller times, cf.\ \Fig{fig:TimeOrderingRF}). 
The sign $\vec{\zeta}^p$ is $+1$ ($-1$) if 
$\vec{\Oc}_p$ differs from $\vec{\Oc}$ by an 
even (odd) number of transpositions of fermionic operators. 
Our Eqs.~\eqref{eq:Gc_t_KS_RF_total} conveniently 
separate properties due to time ordering, contained
in the kernel $\Kc$, from spectral properties 
involving eigenenergies and matrix elements, contained in 
$\Sc$. 
This separation is well known for $\ell \!=\! 2$; for larger
$\ell$, it was pointed out in 1962 by Kobe \cite{Kobe1962},
but without elaborating its consequences in great detail.
The guiding principle of this paper is to
systematically exploit this separation---on the one hand,
to unravel the analytic structure of $\ell$p correlators,
arising from $\Kc$, on the other hand, to facilitate
their numerical computation, involving mainly 
$\Sc$.

For notational simplicity, 
we start by considering the identity permutation, $p=\mathrm{id}$, 
in \Eq{eq:Gc_t_KS_RF_total};
generalizing to arbitrary $p$ afterward will be straightforward.
The multiplicative structure of
Eq.~\eqref{eq:Gc_t_KS_RF} translates into an $\ell$-fold convolution in frequency space:
\begin{align}
\Gc_{\mathrm{id}} & (\vec{\omega})
=
\nint \md^\ell \omega' \,
\Kc( \vec{\omega} - \vec{\omega}')
\Sc[\vec{\Oc}](\vec{\omega}')
.
\label{eq:Gc_w_RF_p=id}
\end{align}
We may thus Fourier transform $\Kc$ and $\Sc$ separately. 
We begin with the Fourier transform of $\Kc$:
\begin{align}
\Kc (\vec{\omega})
& =
\nint \md^\ell t \, e^{\mi \vec{\omega} \cdot \vec{t} }
\prod_{i=1}^{\ell-1} \Big[ -\mi \theta(t_i-t_{i+1})\Big] 
.
\end{align}
It is convenient to use $t'_i = t_i - t_{i+1}$ for
$i < \ell$ and $t'_\ell = t_\ell$ 
as independent integration variables. Then, 
$t_i = t'_{i\cdots \ell}$ (short for $\sum_{j=i}^\ell t'_j$)
and 
$\vec{\omega} \cdot \vec{t} = 
\sum_{i=1}^{\ell} \omega_{1 \cdots i} t_i'$, 
such that
\begin{subequations}
\label{subeq:K_w_RF}
\begin{align}
\Kc (\vec{\omega})
& = 
\nint 
\md t'_\ell \, e^{\mi \omega_{1 \cdots \ell} \, t'_\ell}
\prod_{i=1}^{\ell-1} \Big[ 
-\mi \nint \md t_i' \, e^{\mi \omega_{1 \cdots i} t_i'} \theta(t_i') 
\Big]
\label{eq:Kc_w_RF_p=id}
\\
& =
2\pi \delta(\omega_{1 \cdots \ell}) K(\vec{\omega}) , 
\label{eq:Kc_w_RF_p=id_delta}
\\ 
K(\vec{\omega}) & = 
\prod_{i=1}^{\ell-1} (\omega^+_{1 \cdots i} )^{-1}
.
\label{eq:K_w_RF_p=id}
\end{align}
\end{subequations}
The frequencies $\omega_{1 \cdots i}^+ = \omega_{1 \cdots i} + \mi 0^+$ 
again contain infinitesimal imaginary parts to ensure convergence of real-time integrals.
As for $G(\vec{\omega})$, the arguments of $K(\vec{\omega})$
are understood to satisfy energy conservation, 
$\omega_{1 \cdots \ell} = 0$.
A similar result, with imaginary parts sent to zero, was reached by Kobe \cite{Kobe1962}.
Importantly, however, for numerical calculations, the imaginary parts are necessarily finite
and their mutual relations must be treated carefully---%
this is discussed in Sec.~\ref{sec:KF:retarded_kernel}.

We next turn to the Fourier transform of $\Sc$, 
which will yield an $\ell$p PSF.
Including in its definition a factor $(2\pi)^\ell$ omitted in 
\Eq{eq:Gc_w_RF_p=id}, we have 
\begin{align}
\Sc[\vec{\Oc}](\vec{\omega})
& =
\nint 
\frac{\md^\ell t}{(2\pi)^\ell}
e^{\mi \vec{\omega} \cdot \vec{t} } 
\langle \prod_{i=1}^\ell \Oc^i(t_i) \rangle
\nonumber \\
& =
\sum_{\ub{1}, \nidots, \ub{\ell}}
\rho_{\ub{1}} \prod_{i=1}^\ell \Big[ 
O^i_{\ub{i} \, \ub{i+1}} 
\,
\delta(\omega_i - E_{ \ub{i+1} \, \ub{i}} )
\Big]
.
\label{eq:Sc_w_p=id}
\end{align}
Here and henceforth, the index $\ell \!+\! 1$ is identified with $1$. 
Since the dependence of $\Kc$ on frequencies enters
through the variables $\omega_{1 \cdots i}$,
it is convenient to express $\Sc$
through these, too. 
The conditions $\omega_i \!=\! E_{\ub{i+1} \, \ub{i}}$
implied by the $\delta$ functions are equivalent to 
$\omega_{1 \cdots i} \!=\! \sum_{j=1}^i E_{\ub{j+1} \, \ub{j}} \!=\! E_{\ub{i} \ub{1}}$, and particularly
$\omega_{1 \cdots \ell} \!=\! E_{ \ub{\ell+1} \, \ub{1} } \!=\! E_{\ub{1}\ub{1}} \!=\! 0$. We thus obtain
\begin{subequations}
\label{eq:Sc_w_p=id_delta_and_nodelta}
\begin{flalign}
\Sc[\vec{\Oc}](\vec{\omega})
& = 
 \delta(\omega_{1\cdots \ell})
S [\vec{\Oc}] (\vec{\omega}) 
, 
&
\label{eq:Sc_w_p=id_delta}
\\
S [\vec{\Oc}] (\vec{\omega})
& = \!
\sum_{\ub{1}, \nidots, \ub{\ell}} 
\rho_{\ub{1}} \prod_{i=1}^{\ell-1} \! \Big[ 
O^i_{\ub{i} \, \ub{i+1}} 
\,
\delta(\omega_{1 \cdots i} \!-\! E_{ \ub{i+1}\, \ub{1} } )
\Big] 
O^\ell_{\ub{\ell}\ub{1}} 
.
\hspace{-1cm} & 
\label{eq:S_w_p=id}
\end{flalign}
\end{subequations}
Our \Eqs{eq:Sc_w_p=id} and \eqref{eq:Sc_w_p=id_delta_and_nodelta} are compact
Lehmann representations for the PSFs, with $\omega_i$ serving as placeholder
for $E_{\ub{i+1} \ub{i}}$, and
$\omega_{1 \cdots i}$ for $E_{\ub{i+1} \ub{1}}$, respectively.

We now insert \Eqs{eq:Kc_w_RF_p=id_delta} and 
\eqref{eq:Sc_w_p=id_delta}
into
\Eq{eq:Gc_w_RF_p=id} for $\Gc (\vec{\omega})$.
The corresponding $G (\vec{\omega})$,
extracted as in \Eq{eq:RF_frequencydomain},
has the form of an $(\ell \!-\! 1)$-fold convolution:
\begin{align}
G_{\mathrm{id}} (\vec{\omega})
= \nint \md^{\ell-1} \omega' \,
K( \vec{\omega} - \vec{\omega}' )
S[\vec{\Oc}](\vec{\omega}') .
\label{eq:G_w_RF_p=id}
\end{align}
As already mentioned, for each of 
$G$, $K$, and $S$, the $\ell$ frequency 
arguments are understood to sum to zero.

Next, we consider the case of an arbitrary permutation $p$
in \Eq{eq:Gc_t_KS_RF_total}.
The Fourier transforms of 
functions with permuted arguments, $\Kc(\vec{\omega}_p)$
and $\Sc[\vec{\Oc}_p](\vec{\omega}_p)$, readily 
follow from those given above for $\Kc(\vec{\omega})$
and $\Sc[\vec{\Oc}](\vec{\omega})$.
For a given $p$: $i \to \ovb{i}$, 
the integration measure and 
exponent in the Fourier integral are invariant under relabeling times and frequencies 
as $t_i \to t_{\ovb{i}}$ and
$\omega_i \to \omega_{\ovb{i}}$, 
i.e.\ 
$\int \md^\ell t \, e^{\mi \vec{\omega} \cdot \vec{t}} =
\int \md^\ell t_p \, e^{\mi \vec{\omega}_p \cdot 
\vec{t}_p}$. 
Hence, the above discussion applies 
unchanged, except for the index relabeling 
$i \to \ovb{i}$ on times, frequencies, and
operator superscripts. Depending on context, 
it may or may not be useful to additionally
rename the dummy summation indices as 
$(\ub{\ovb{1}} \, \ub{\ovb{2}} \, \ndots \ub{\ovb{\ell}})$.

For instance, the permuted version of \Eq{eq:Sc_w_p=id}
can be written in either of the following forms:
\begin{subequations}
\begin{align}
\Sc[\vec{\Oc}_p](\vec{\omega}_p)
& =
\sum_{\ub{1}, \nidots, \ub{\ell}}
\rho_{\ub{1}} \prod_{i=1}^\ell \Big[ 
O^{\ovb{i}}_{\ub{i} \, \ub{i+1}} 
\,
\delta(\omega_{\ovb{i}} - E_{ \ub{i+1} \, \ub{i}} )
\Big]
\\
& =
\sum_{\ub{1}, \nidots, \ub{\ell}}
\rho_{\ub{\ovb{1}}} \prod_{i=1}^\ell \Big[ 
O^{\ovb{i}}_{\ub{\ovb{i}} \, \ub{\ovb{i+1}}} 
\,
\delta(\omega_{\ovb{i}} - E_{ \ub{\ovb{i+1}} \, \ub{\ovb{i}}} )
\Big]
. 
\label{eq:Sc_w_explicit}
\end{align}
\end{subequations}
The first choice ensures that the density matrix carries the same index, $\rho_\ub{1}$, irrespective of $p$. 
This is helpful for the NRG implementation 
\cite{Lee2021},
used to obtain the numerical results in Sec.~\ref{sec:results}.
The second choice yields matrix elements 
$O^{\ovb{i}}_{\ub{\ovb{i}}\ub{\ovb{i +1}}}$ 
whose subscript indices are linked to the superscripts.
This is often convenient for obtaining analytical results.
It shows, e.g.,
that PSFs whose arguments are cyclically related are proportional
to each other. To be explicit, let 
$p \!=\! (\ovb{1} \, \ndots \ovb{\ell})$ and 
$p_\lambda \!=\! 
(\ovb{\lambda} \, \ndots \ovb{\ell}
\, \ovb{1} \, \ndots \ovb{\lambda \!-\! 1})$
be cyclically related permutations, 
e.g.\ $(\ovb{1} \ovb{2} \ovb{3} \ovb{4})$ and 
$(\ovb{3}\ovb{4} \ovb{1} \ovb{2})$.
According to \Eq{eq:Sc_w_explicit},
the corresponding PSFs, 
$\Sc[\vec{\Oc}_{p}](\vec{\omega}_{p}) $ and 
$\Sc[\vec{\Oc}_{p_\lambda}](\vec{\omega}_{p_\lambda})$,
differ only in the indices on the density matrix, which
appears as $\rho_{\ub{\ovb{1}}}$ or $\rho_{\ub{\ovb{\lambda}}}$,
respectively; they otherwise 
contain the \textit{same} product $\prod_i$, written in two
different, cyclically related orders. 
Since $\rho_{\ub{\ovb{\lambda}}} \!=\! 
\rho_{\ub{\ovb{1}}} e^{-\beta E_{\ub{\ovb{\lambda}}\ub{\ovb{1}}}}$
in thermal equilibrium
and the $\delta$ functions enforce 
$E_{\ub{\ovb{\lambda}}\ub{\ovb{1}}} \!=\! \omega_{\ovb{1} \cdots \ovb{\lambda-1}}$, 
we obtain the cyclicity relation
\begin{align}
\Sc[\vec{\Oc}_{p_\lambda}](\vec{\omega}_{p_\lambda})
= 
\Sc[\vec{\Oc}_{p}](\vec{\omega}_{p}) 
e^{- \beta \omega_{\ovb{1} \cdots \ovb{\lambda-1}}} 
. 
\label{eq:Sc_w_cyclicity}
\end{align}
This relation is useful for analytical arguments and, in particular,
allows one to reduce the number of PSFs in the spectral representation from $\ell!$ to $(\ell \!-\! 1)!$. 
However, we here refrain from doing so, since it would increase the complexity of the kernels and modify their role in the spectral representation by introducing Boltzmann factors.

We are now ready to present our final results for
the Fourier transform of \Eqs{eq:Gc_t_KS_RF_total}.
It involves a sum over permuted versions of 
\Eq{eq:G_w_RF_p=id}:
\begin{align}
G (\vec{\omega})
= \sum_p \vec{\zeta}^p
\nint \md^{\ell-1} \omega'_p \,
K( \vec{\omega}_p - \vec{\omega}_p' )
S[\vec{\Oc}_p](\vec{\omega}_p')
.
\label{eq:G_w_RF}
\end{align}
Here,
$\vec{\omega}'_p \!=\! (\omega'_\ovb{1}, \ndots, \omega'_\ovb{\ell})$ is a permuted version of 
$\vec{\omega}' \!=\! (\omega'_1, \ndots, \omega'_\ell)$, with 
$\omega'_{1 \cdots \ell} \!=\! \omega'_{\ovb{1} \cdots \ovb{\ell}} \!=\! 0$ understood, 
and the integral is over the first $\ell \!-\! 1$ independent components of $\vec{\omega}'_p$. 
Alternatively, the integration variables can also be chosen independent of $p$ 
(by using the same set of $\ell \!-\! 1$ independent components of $\omega'$ for all $p$), 
in which case the measure will be written as $\md^{\ell-1} \omega'$.
The permuted \ZF/ convolution kernels [\Eq{eq:K_w_RF_p=id}] are given by
\begin{align}
K(\vec{\omega}_p - \vec{\omega}_p')
& =
\prod_{i=1}^{\ell-1} (\omega^+_{\ovb{1} \cdots \ovb{i}} - \omega_{\ovb{1} \cdots \ovb{i}} ')^{-1}
,
\label{eq:K_w_RF}
\end{align}
and the permuted PSFs [\Eq{eq:S_w_p=id}] read
\begin{flalign}
S [\vec{\Oc}_p] (\vec{\omega}'_p)
& = 
\! \sum_{\ub{1}, \nidots, \ub{\ell}} \!
\rho_{\ub{1}} \prod_{i=1}^{\ell-1} \! 
\Big[ O^{\ovb{i}}_{\ub{i} \, \ub{i+1}} 
\,
\delta(\omega'_{\ovb{1}\cdots \ovb{i}} 
\! - \! E_{ \ub{i+1} \, \ub{1} } )
\Big] O^{\ovb{\ell}}_{\ub{\ell}\ub{1}} . 
\hspace{-1cm} & 
\label{eq:S_w}
\end{flalign}
Equations \eqref{eq:G_w_RF} to \eqref{eq:S_w} give the spectral representation for \ZF/ $\ell$p correlators. 
Combining them, we obtain
\begin{align}
G(\vec{\omega})
& =
\sum_p \vec{\zeta}^p
 \sum_{\ub{1}, \nidots, \ub{\ell}}
\frac{
\rho_{\ub{1}}
\prod_{i=1}^{\ell}
O^{\ovb{i}}_{\ub{i} \ub{i+1}}
}
{
\prod_{i=1}^{\ell-1}
(\omega^+_{\ovb{1} \cdots \ovb{i}} - E_{\ub{i+1} \, 
\ub{1}} )
}.
\label{eq:RFSpectralRepresentationExplicit}
\end{align}
For $\ell \!=\! 2$, this
reproduces \Eq{eq:G2_RF}; see App.~\ref{app:RF_G2}.

We conclude this section with a remark on connected correlators.
These are relevant in many contexts, 
particularly for the $4$p vertices 
in Sec.~\ref{sec:results}.
The connected correlator $\Gc^{\mathrm{con}}$ follows from the full correlator $\Gc$ by subtracting 
the disconnected part $\Gc^{\mathrm{dis}}$;
the latter is a sum over products of lower-point correlators.
Through the spectral representation,
we can transfer the notion of a connected and disconnected part from $\Gc$ onto the PSFs,
$\Sc^{\mathrm{con}} \!=\! \Sc \!-\! \Sc^{\mathrm{dis}}$.
By linearity, $\Gc^{\mathrm{con}}$ will follow by combining $\Sc^{\mathrm{con}}$ with the same kernels $\Kc$
as for the full correlators.

Let us consider explicitly the case of four fermionic 
creation or annihilation operators (cf.\ Sec.~\ref{sec:quantities}).
Then,
\begin{align}
\Sc^{\mathrm{dis}} [\vec{\Oc}_p](\vec{t}_p)
& =
\langle \Oc^{\ovb{1}}(t_{\ovb{1}}) \Oc^{\ovb{2}}(t_{\ovb{2}}) \rangle 
\langle \Oc^{\ovb{3}}(t_{\ovb{3}}) \Oc^{\ovb{4}}(t_{\ovb{4}}) \rangle
\nonumber
\\ 
& \ +
\zeta
\langle \Oc^{\ovb{1}}(t_{\ovb{1}}) \Oc^{\ovb{3}}(t_{\ovb{3}}) \rangle 
\langle \Oc^{\ovb{2}}(t_{\ovb{2}}) \Oc^{\ovb{4}}(t_{\ovb{4}}) \rangle
\nonumber
\\ 
& \ +
\langle \Oc^{\ovb{1}}(t_{\ovb{1}}) \Oc^{\ovb{4}}(t_{\ovb{4}}) \rangle 
\langle \Oc^{\ovb{2}}(t_{\ovb{2}}) \Oc^{\ovb{3}}(t_{\ovb{3}}) \rangle
,
\label{eq:Sc_t-dis}
\end{align}
since expectation values of an odd number of operators vanish.
Each factor on the right describes a time-dependent $2$p PSF,
e.g.\
$\Sc [\Oc^{\ovb{1}},\Oc^{\ovb{2}}](t_\ovb{1}, t_\ovb{2}) 
\Sc [\Oc^{\ovb{3}},\Oc^{\ovb{4}}](t_\ovb{3}, t_\ovb{4})$
for the first summand.
The expansion of $\Sc^{\mathrm{dis}}$ perfectly matches the
corresponding expansion of $\Gc^{\mathrm{dis}}$,
since the kernel $\Kc$ in 
$\Gc^{\mathrm{dis}}(\vec{t})
\!=\! \sum_p \vec{\zeta}^p
\Kc(\vec{t}_p)
\Sc^{\mathrm{dis}} [\vec{\Oc}_p] (\vec{t}_p)$
[cf.\ \Eqs{eq:Gc_t_KS_RF_total}]
ensures that each factor in \Eq{eq:Sc_t-dis} is already time ordered.
Evidently, \Eq{eq:Sc_t-dis} can be easily Fourier transformed.
The first summand, e.g., yields
$\Sc [\Oc^{\ovb{1}},\Oc^{\ovb{2}}](\omega_\ovb{1}, \omega_\ovb{2}) 
\Sc [\Oc^{\ovb{3}},\Oc^{\ovb{4}}](\omega_\ovb{3}, \omega_\ovb{4})$,
which contains 
$\delta(\omega_{\ovb{12}})\delta(\omega_{\ovb{34}})
\!=\!
\delta(\omega_{\ovb{12}})\delta(\omega_{\ovb{1234}})$.
Upon factoring out the $\delta$ function ensuring energy conservation,
we have in total:
\begin{flalign}
& S^{\mathrm{dis}} [\vec{\Oc}_p](\vec{\omega}_p)
\!=\! 
S [\Oc^{\ovb{1}},\Oc^{\ovb{2}}](\omega_\ovb{1}, \omega_\ovb{2}) 
S [\Oc^{\ovb{3}},\Oc^{\ovb{4}}](\omega_\ovb{3}, \omega_\ovb{4}) 
\delta(\omega_{\ovb{12}})
&
\nonumber
\\ 
& \qquad + 
\zeta
S [\Oc^{\ovb{1}},\Oc^{\ovb{3}}](\omega_\ovb{1}, \omega_\ovb{3}) 
S [\Oc^{\ovb{2}},\Oc^{\ovb{4}}](\omega_\ovb{2}, \omega_\ovb{4}) 
\delta(\omega_{\ovb{13}})
& 
\nonumber
\\ 
& \qquad + 
S [\Oc^{\ovb{1}},\Oc^{\ovb{4}}](\omega_\ovb{1}, \omega_\ovb{4}) 
S [\Oc^{\ovb{2}},\Oc^{\ovb{3}}](\omega_\ovb{2}, \omega_\ovb{3}) 
\delta(\omega_{\ovb{14}})
.
\hspace{-1cm}
&
\label{eq:Sc_w-dis} 
\end{flalign}
This resembles the form of $G^{\mathrm{dis}}(\vec{\omega})$ [cf.~\Eq{eq:disc_part}], which will be needed in Sec.~\ref{sec:quantities} when defining the $4$p vertex.

\subsection{Matsubara formalism}

The derivation of spectral representations 
in the \MF/ is analogous to that in the \ZF/,
and it utilizes the same real-frequency PSFs 
from Eq.~\eqref{eq:S_w}. 
All arguments regarding time-translational invariance and energy conservation from above still hold.
Hence, we also use the same notation in terms of $\Gc$ and $G$.
Nevertheless, there are subtleties arising in the \MF/.
The first concerns (composite) bosonic frequencies that may be zero 
and lead to anomalous terms as in Eq.~\eqref{eq:G2_incl_anomalous} \cite{Shvaika2016,Shvaika2006}.
The second stems from the fact that imaginary times can be chosen positive and thus always larger than the time set to zero.
This leads to only $(\ell \!-\! 1)!$ permutations of operators.
Yet, the nontrivial boundary conditions of the imaginary-time integral provide $\ell$ terms,
thus making up for the total of $\ell!$ terms.
For us, it will be convenient to directly work with $\ell!$ summands.
In fact, it turns out that only the terms arising from the trivial lower boundary of the imaginary-time integral contribute; 
all others cancel out and need not be computed explicitly.
Without this trick, the calculations are more tedious, but still straightforward;
see App.~\ref{app:IF_direct} for $\ell \!=\! 3$ and $4$.

\begin{figure}
\includegraphics[width=0.8\linewidth]{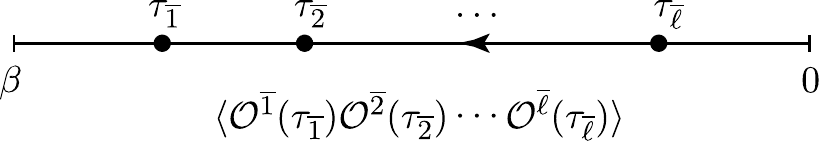}
\caption{\MF/ time ordering is similar to \ZF/ ordering (Fig.~\ref{fig:TimeOrderingRF}),
except that the times $\tau_\ovb{i}$ are constrained to the interval $(0,\beta)$.}
\label{fig:TimeOrderingIF}
\end{figure}

We start from the \MF/ $\ell$p function,
\begin{align}
\Gc(\vec{\tau})
& =
(-1)^{\ell-1}
\langle \TO \prod_{i=1}^\ell \Oc^i(\tau_i) \rangle
, 
\label{eq:Gc_tau_IF}
\end{align}
with operators time ordered on the interval $\tau_i \in (0, \beta)$
(cf.\ \Fig{fig:TimeOrderingIF}) and periodic or antiperiodic boundary conditions for
$\Gc$, depending on the bosonic or fermionic nature of the corresponding operators.
We wish to compute 
\begin{align}
\Gc(\mi \vec{\omega}) 
= 
\nbint{0}{\beta} \md^\ell \tau \, 
e^{\mi \vec{\omega} \cdot \vec{\tau}} \Gc(\vec{\tau})
= 
\beta \delta_{\omega_{1 \cdots \ell},0}
\, G(\mi \vec{\omega}) 
, 
\label{eq:Gc_iw_IF}
\end{align}
where the Kronecker $\delta$ follows from translational invariance in $\tau$.
As for the \ZF/, we express $\Gc(\vec{\tau})$ 
as a sum over permutations involving products of step functions $\theta$:
\begin{subequations}
\label{eq:Gc_tau_KS_IF_total}
\begin{align}
\Gc(\vec{\tau})
& =
\sum_p
\vec{\zeta}^p
\Kc(\vec{\tau}_p) \, 
\Sc [\vec{\Oc}_p] (-\mi \vec{\tau}_p) 
, 
\\
\Kc(\vec{\tau}_p)
& = \prod_{i=1}^{\ell-1} \Big[ 
- \theta(\tau_{\ovb{i}}-\tau_{\ovb{i+1}}) 
\Big]
,
\\
\Sc [\vec{\Oc}_p](- \mi \vec{\tau}_p)
& = 
\langle \prod_{i=1}^\ell \Oc^{\ovb{i}}(\tau_{\ovb{i}}) \rangle
\label{eq:Sc_tau}
\\
& = 
\nint \md^\ell \omega'_p \, e^{- \vec{\omega}'_p \cdot 
\vec{\tau}_p} \Sc[\vec{\Oc}_p] (\vec{\omega}'_p) 
.
\label{eq:Sc_tau_FT}
\end{align}
\end{subequations}
Here, \Eq{eq:Sc_tau}
is the analytic continuation
$t_{\ovb{i}} \!\to\! -\mi \tau_{\ovb{i}}$
of \Eq{eq:Sc_t} for $\Sc [\vec{\Oc}_p](\vec{t}_p)$.
Performing this continuation in the real-frequency Fourier representation of the
latter yields \Eq{eq:Sc_tau_FT},
where $\Sc[\vec{\Oc}_p] (\vec{\omega}'_p)$
is the PSF \eqref{eq:S_w}
from the \ZF/ discussion. 
Inserting \Eqs{eq:Gc_tau_KS_IF_total} 
into \eqref{eq:Gc_iw_IF} yields
\begin{subequations}
\begin{flalign}
\Gc(\mi\vec{\omega})
& =
\sum_p \vec{\zeta}^p
\nint \md^{\ell} \omega_p' \,
\Kc(\mi \vec{\omega}_p - \vec{\omega}_p')
\Sc[\vec{\Oc}_p](\vec{\omega}_p')
\hspace{-0.8cm} & 
\\
& =
\sum_p \vec{\zeta}^p
\nint \md^{\ell-1} \omega_p' \,
\Kc(\mi \vec{\omega}_p - \vec{\omega}_p')
S[\vec{\Oc}_p](\vec{\omega}_p') ,
\hspace{-0.8cm} &
\label{eq:Gc_iw_KS_IF}
\\
\Kc(\vec{\Omega}_p) 
& = \int \md^\ell \tau_p \,
e^{\vec{\Omega}_p \cdot \vec{\tau}_p} 
\Kc (\vec{\tau}_p) 
,
\hspace{-0.8cm} &
\end{flalign}
\end{subequations}
where $\vec{\Omega}_p \!=\! \mi \vec{\omega}_p \!-\! \vec{\omega}_p'$,
and $\omega'_{\ovb{1} \cdots \ovb{\ell}} \!=\! 0$, 
set by the $\delta$ functions in $\Sc$,
is understood implicitly from \Eq{eq:Gc_iw_KS_IF} onward.

The $\theta$ functions in $\Kc(\vec{\tau}_p)$
enforce $\tau_{\ovb{i}} > \tau_{\ovb{i+1}}$, such that
\begin{align}
\Kc (\vec{\Omega}_p)
& = 
\nbint{0}{\beta} \md \tau_{\ovb{\ell}}
\prod_{i=\ell-1}^1 \Big[ - \nbint{\tau_{\ovb{i+1}}}{\beta} \md \tau_\ovb{i} \Big] 
e^{ \vec{\Omega}_p \cdot \vec{\tau}_p} ,
\end{align}
where the integrals are arranged in ``descending'' order, 
$\int \md \tau_{\ovb{\ell}} \, \md \tau_{\ovb{\ell-1}} \, \ndots \, \md \tau_\ovb{1}$. 
Using $\tau'_\ovb{i} \!=\! \tau_\ovb{i} - \tau_\ovb{i+1}$ for 
$i \!<\! \ell$ and $\tau'_{\ovb{\ell}} \!=\! \tau_{\ovb{\ell}}$ 
as integration variables,
with $\tau_\ovb{i} \!=\! \tau'_{\ovb{i} \dots \ovb{\ell}}$
and $\vec{\Omega}_p \! \cdot \vec{\tau}_p \! 
\!=\! \sum_{i=1}^\ell \Omega_{\ovb{1} \cdots \ovb{i}} \tau'_\ovb{i}$,
we have
\begin{align}
\Kc & (\vec{\Omega}_p)
= 
\nbint{0}{\beta} \! \md \tau'_{\ovb{\ell}} \,
e^{\Omega_{\ovb{1} \cdots \ovb{\ell}} \tau'_{\ovb{\ell}}} 
\! \prod_{i=\ell-1}^1 \!
\Big[
- \nbint{0}{\beta - \tau'_{\ovb{i+1} \cdots \ovb{\ell}}} 
\! \md \tau'_\ovb{i} \,
e^{\Omega_{\ovb{1} \cdots \ovb{i}} \tau'_\ovb{i}} 
\Big] 
.
\label{eq:Kc_Omega_integral}
\end{align}
Since \Eq{eq:Gc_iw_IF} for $\Gc(\mi \vec{\omega})$
contains a factor $\beta \delta_{\omega_{1 \cdots \ell},0}$, 
the integrals in \Eq{eq:Kc_Omega_integral}
must yield a result of the form
\begin{align}
\Kc (\vec{\Omega}_p)
& = 
\beta \delta_{\omega_{\ovb{1} \cdots \ovb{\ell},0}}
K (\vec{\Omega}_p) + \Kc'(\vec{\Omega}_p) 
. 
\label{eq:Kc_Omega_K_and_Kcprime}
\end{align}
Inserted into \Eq{eq:Gc_iw_KS_IF}
for $\Gc(\mi \vec{\omega})$, the $K$ term reproduces the discrete
$\delta$ function in \Eq{eq:Gc_iw_IF},
yielding 
\begin{align}
G(\mi\vec{\omega})
& =
\sum_p \vec{\zeta}^p
\nint \md^{\ell-1} \omega_p' \,
K(\mi \vec{\omega}_p - \vec{\omega}_p') \, 
S[\vec{\Oc}_p](\vec{\omega}_p')
,
\label{eq:G_iw_KS_IF}
\end{align}
while the $\Kc'$ term must yield zero
when summed over $p$, 
\begin{align}
\sum_p \vec{\zeta}^p
\nint \md^{\ell-1} \omega_p' \,
\Kc'(\mi \vec{\omega}_p - \vec{\omega}_p')
S[\vec{\Oc}_p](\vec{\omega}_p')
= 0 
. 
\label{eq:upperboundaries-vanish}
\end{align}
The $K$ term in \Eq{eq:Kc_Omega_K_and_Kcprime} 
comes from the outermost 
integral over $\tau'_\ovb{\ell}$ in \Eq{eq:Kc_Omega_integral}, 
involving $e^{\Omega_{\ovb{1}\cdots \ovb{\ell}}\tau'_\ovb{\ell}}
= e^{\mi \omega_{\ovb{1}\cdots \ovb{\ell}}\tau'_\ovb{\ell}}$ 
(recall $\omega'_{\ovb{1} \cdots \ovb{\ell}} = 0$). 
However, this integral generates the prefactor 
$\beta \delta_{\omega_{\ovb{1} \cdots \ovb{\ell}},0}$
\textit{only} for terms with no 
further dependence on $\tau'_\ovb{\ell}$.
These terms, which arise solely from the lower integration boundaries of all subsequent integrals, 
give $K$:
\begin{align}
K(\vec{\Omega}_p) 
& = 
\prod_{i=1}^{\ell-1} 
\Big[ - 
\nbint{0}{}
\md \tau'_\ovb{i} \, 
e^{\Omega_{\ovb{1} \cdots \ovb{i}} \tau'_{\ovb{i}}}
\Big]
. 
\label{eq:K_iw_integral}
\end{align}
Conversely, all terms arising from one or more upper integration boundaries, 
which depend on $\tau'_\ovb{\ell}$ via
$\tau'_{\ovb{i+1} \cdots \ovb{\ell}}$, contribute
to $\Kc'$. We need not compute them explicitly, since,
by \Eq{eq:upperboundaries-vanish}, their contributions cancel. 

We now evaluate the integrals \eqref{eq:K_iw_integral} for $K$. We temporarily exclude the anomalous term arising if some exponents vanish and indicate this by putting a tilde on $\ovwt{K}$ (and 
$\ovwt{G}$):
\begin{align}
\ovwt{K}(\vec{\Omega}) 
& = 
\prod_{i=1}^{\ell-1}
\big( \Omega_{\ovb{1} \cdots \ovb{i}} \big)^{-1} 
= 
\prod_{i=1}^{\ell-1}
\big( \mi \omega_{\ovb{1} \cdots \ovb{i}} 
- \omega'_{\ovb{1} \cdots \ovb{i}} \big)^{-1}
.
\label{eq:Ktilde_Omega_IF}
\end{align}
Inserting \Eqs{eq:Ktilde_Omega_IF} and \eqref{eq:S_w} for $S$
into \Eq{eq:G_iw_KS_IF}, we obtain
\begin{align}
\ovwt{G}(\mi\vec{\omega})
& =
\sum_p \vec{\zeta}^p
 \sum_{\ub{1}, \nidots, \ub{\ell}}
\frac{
\rho_{\ub{1}}
\prod_{i=1}^{\ell}
O^{\ovb{i}}_{\ub{i} \ub{i+1}}
}
{
\prod_{i=1}^{\ell-1}
[ \mi \omega_{\ovb{1} \cdots \ovb{i}} - E_{\ub{i+1} \, 
\ub{1}} ]
}
.
\label{eq:Gtilde_iw_IF}
\end{align}
This compact expression 
is our first main result
for \MF/ $\ell$p functions. For $\ell = 2$, 
it yields [upon relabeling summation indices as in \Eq{eq:Sc_w_explicit}]
\begin{align}
\ovwt{G}(\mi \omega) 
= 
\sum_{\ub{1} \, \ub{2}} 
O^1_{\ub{1} \ub{2}} O^2_{\ub{2} \ub{1}} 
\bigg[
\frac{\rho_{\ub{1}}}{\mi \omega_1 - E_{\ub{2}\ub{1}}}
+ 
\frac{\zeta \rho_{\ub{2}}}{\mi \omega_2 - E_{\ub{1}\ub{2}}} 
\bigg]
. 
\end{align}
Since $-\omega_2 = \omega_1 = \omega$, this 
reproduces \Eq{eq:G2_IF}. 
The cases $\ell=3$ and $\ell =4$ are verified by explicit computation in App.~\ref{app:IF_direct}.
They also agree with published results (which use less compact notation):
Eq.~(A.2) of Ref.~\onlinecite{Oguri2001} and
Eq.~(3.11) of Ref.~\onlinecite{Shvaika2006} for $\ell = 3$;
Eq.~(A.2) of Ref.~\onlinecite{Eliashberg1961},
Eq.~(A1) of Ref.~\onlinecite{Toschi2007},
Eq.~(A.2) of Ref.~\onlinecite{Hafermann2009},
and Eq.~(3.3) of Ref.~\onlinecite{Shvaika2016} for $\ell = 4$.
(Refs.~\onlinecite{Shvaika2006}, \onlinecite{Hafermann2009}, and \onlinecite{Shvaika2016}
also discuss anomalous contributions, 
reproduced by \Eq{eq:K_Omega_IF_compact} below.)

For applications, we will be mostly interested in fermionic systems and $\ell$p functions with $\ell \!\leq\! 4$.
These can still contain bosonic operators, e.g., as bilinears of fermionic operators.
However, for arbitrary $2$p functions, $3$p functions with only one bosonic operator, and fermionic $4$p functions, 
$\omega_{\ovb{1} \cdots \ovb{i}}$, with $i \!<\! \ell$, 
produces at most one bosonic frequency.
In this case, at most one frequency 
in the denominators of \Eq{eq:Gtilde_iw_IF} can vanish, 
say $\omega_{\ovb{1} \cdots \ovb{j}}$, 
for some $j \!<\! \ell$.
One can show (see App.~\ref{app:IF_direct}) that the corresponding anomalous 
terms are fully included using the following kernel in \Eq{eq:G_iw_KS_IF}:
\begin{align}
K(\vec{\Omega}_p) =
\prod_{i=1}^{\ell-1}
\frac{1}{\Omega_{\ovb{1}\cdots\ovb{i}}}
\; - \; 
\frac{\beta}{2}
 \sum_{j=1}^{\ell-1} \delta_{\Omega_{\ovb{1}\cdots\ovb{j}},0}
\prod^{\ell-1}\limits_{\substack{i=1 \\ i \neq j}}
\frac{1}{\Omega_{\ovb{1}\cdots\ovb{i}}}
.
\label{eq:K_Omega_IF_compact}
\end{align}
Here, $\delta_{\Omega_{\ovb{1}\cdots\ovb{j}},0}$ is symbolic notation indicating that the anomalous second term contributes 
only if $\Omega_{\ovb{1}\cdots\ovb{j}} = 0$
(i.e.\ if both $\omega_{\ovb{1}\cdots\ovb{j}} = 0$
and $\omega'_{\ovb{1}\cdots\ovb{j}} = 0$, the latter
due to a degeneracy, $E_{\ovb{\ub{j+1}}}= E_{\ovb{\ub{1}}}$, 
in the spectrum).
In this case, $1/\Omega_{\ovb{1}\cdots\ovb{j}}$ diverges,
but the product $\prod_{i \neq j}$ duly excludes 
such factors. The regular first term needs no such exclusion,
since, if $\Omega_{\ovb{1}\cdots\ovb{j}} \to 0$, 
then also $\Omega_{\ovb{i+1} \cdots \ovb{\ell}} \to 0$, 
and the $1/\Omega_{\ovb{1}\cdots\ovb{i}}$ 
divergence is canceled by a
$- 1/\Omega_{\ovb{i+1} \cdots \ovb{\ell}}$ divergence 
stemming from a cyclically related permutation. We
confirm this in App.~\ref{app:IF_direct}
by treating nominally vanishing denominators as infinitesimal
and explicitly tracking the cancellation of divergent terms. This procedure yields the following expression for the full kernel:
\begin{align}
K(\vec{\Omega}_p) =
\begin{cases}
\displaystyle
\prod_{i=1}^{\ell-1}
\Omega_{\ovb{1}\cdots\ovb{i}}^{-1}
, \hspace{2.6cm} \text{if} \quad 
\prod_{i=1}^{\ell-1}
\Omega_{\ovb{1}\cdots\ovb{i}}
\neq 0 
, 
\\[-3mm]
\\
\displaystyle
- \frac{1}{2} 
\Big[
\beta + 
\sum^{\ell-1}\limits_{\substack{i=1 \\ i \neq j}}
\Omega_{\ovb{1}\cdots\ovb{i}}^{-1}
\Big]
\prod^{\ell-1}\limits_{\substack{i=1 \\ i \neq j}}
\Omega_{\ovb{1}\cdots\ovb{i}}^{-1} \, 
, \! \quad
\textrm{if} \! \quad \Omega_{\ovb{1}\cdots\ovb{j}} = 0
. \vspace{-4mm} \phantom{.}
\end{cases}
\label{eq:K_Omega_IF}
\end{align}
The kernels~\eqref{eq:K_Omega_IF_compact} and \eqref{eq:K_Omega_IF} 
yield equivalent results. The former is well suited for analytical work;
the latter is more convenient for numerical computations,
as it is manifestly free from divergences.
The spectral representation given by
\Eqs{eq:G_iw_KS_IF}, 
\eqref{eq:S_w}, and
\eqref{eq:K_Omega_IF_compact} or
\eqref{eq:K_Omega_IF} 
is our final result for \MF/ $\ell$p functions.

\subsection{Keldysh formalism}

The Keldysh formalism (\KF/) \cite{Schwinger1961,Keldysh1964} is based on ordering operators on a doubled time contour 
involving a forward and a backward branch.
In the contour basis, each time argument carries an extra index specifying 
which branch the corresponding operator resides on.
The Keldysh basis employs linear combinations of such contour-ordered correlators.
Before discussing these two options in turn, we introduce 
a \textit{fully retarded} kernel, 
a very useful object through which all other \KF/ kernels can be expressed.
While doing so, we carefully discuss the imaginary parts of complex frequencies, 
needed for convergence of real-time integrals. 
We present a choice of imaginary parts which is consistent both 
if they are infinitesimal, as assumed in analytical work, or finite, as needed for numerical computations.

\subsubsection{Fully retarded kernel}
\label{sec:KF:retarded_kernel}

Operators on the forward branch of the Keldysh contour are time ordered while those on the backward branch are anti-time-ordered.
In \Eq{eq:Kc_t_RF} of the \ZF/, we already encountered the time-ordered kernel.
In the \KF/, it will be useful to have a kernel that combines 
$\eta \!-\! 1$
anti-time-ordering and $\ell \!-\! \eta$ time-ordering
factors in the form
\begin{align}
\Kc^{[\eta]}(\vec{t}_p)
& =
\prod_{i=1}^{\eta-1} \Big[ \mi \theta(t_{\ovb{i+1}}-t_\ovb{i}) \Big]
\prod_{i=\eta}^{\ell-1} \Big[ -\mi \theta(t_{\ovb{i}}-t_{\ovb{i+1}}) \Big]
,
\label{eq:fully_retarded_kernel}
\end{align}
for $1 \!\leq\! \eta \!\leq\! \ell$.
As usual, a product over an empty set, with lower limit larger than upper limit, 
is defined to equal unity.
Note how the doubled time contour of the \KF/ 
(cf.\ Fig.~\ref{fig:TimeOrderingKF} below)
is reflected in \Eq{eq:fully_retarded_kernel}:
the successive (from right to left) time-ordering and
anti-time-ordering factors 
single out 
$t_{\ovb{\eta}}$, the $\eta$th component of $\vec{t}_p$,
as largest time.
The kernel $\Kc\sseta(\vec{t}_p)$ is (fully) retarded w.r.t.\ $t_{\ovb{\eta}}$;
i.e., it is nonzero only for $t_{\ovb{i}} \!<\! t_{\ovb{\eta}}$, $i \!\neq\! \eta$.

\begin{figure}
\includegraphics[width=\linewidth]{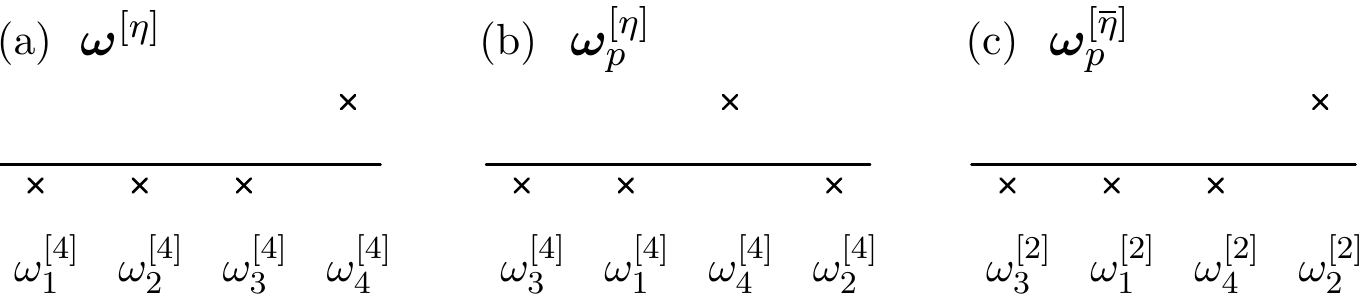}
\caption{Complex frequency tuples, 
for $\ell=4$, $\eta = 4$, $p=(3142)$, in which case 
$p^{-1}=(2413)$, $\mu = p^{-1}(\eta) = 3$, 
$\ovb{\eta} = p(\eta) = 2$.} 
\label{fig:ComplexFrequencyTuples}
\end{figure}

To Fourier transform $K\sseta(\vec{t}_p)$,
we first consider the identity permutation, $\ovb{i} \!=\! i$,
requiring no subscripts $p$ or overbars. 
(The general case will follow by suitably reinstating overbars at the end.)
In the Fourier integral, we can take the perspective that 
the largest time $t_\eta$ runs over the entire real axis,
while all other $t_i$ are constrained by $t_i \!<\! t_\eta$.
It is thus natural to use the integral over $t_\eta$ for energy conservation and,
exploiting time-translational invariance, consider all other time dependencies $t_i \!-\! t_\eta$ as advanced
(i.e.\ contributing only for $t_i \!-\! t_\eta < 0$):
\begin{flalign}
& \Kc^{[\eta]}(\vec{\omega}) 
=
\nint \md^\ell t \, e^{\mi \vec{\omega} \cdot \vec{t}} \Kc^{[\eta]}(\vec{t})
=
\nint \md t_\eta \, e^{\mi \omega_{1 \cdots \ell} t_\eta}
\\
& \ \times \!
\prod_{i \neq \eta} \! \bigg[
\nbint{-\infty}{0} \!\!\!\!\!\: \md (t_i \!-\! t_\eta) \, 
e^{\mi \omega_i (t_i - t_\eta) }
\bigg]
\Kc^{[\eta]}(t_1 \!-\! t_\eta, \ndots, 0, \ndots, t_\ell \!-\! t_\eta)
.
\nonumber
\end{flalign}
The $t_{i \neq \eta}$ integrals can be regularized, 
without affecting the $t_\eta$ integral,
by replacing the real tuple $\vec{\omega}$
by a complex tuple $\vec{\omega}\sseta$
with components $\omega\sseta_i$, having
finite imaginary parts. 
We thus shift
$\omega_i \!\to\! \omega\sseta_i$
[see Fig.~\ref{fig:ComplexFrequencyTuples}(a)],
where
\begin{align}
\omega^{[\eta]}_i = \omega_i + \mi \gamma^{[\eta]}_i 
, \quad 
\gamma^{[\eta]}_{i \neq \eta} < 0
, \quad 
\gamma^{[\eta]}_\eta = - {\textstyle \sum_{i \neq \eta}} 
\gamma^{[\eta]}_i
. 
\label{eq:choice_imaginary_parts}
\end{align}
This assigns a negative imaginary part
to each frequency $\omega\sseta_{i \neq \eta}$
multiplying $t_i \!-\! t_\eta$, 
as in the $2$p case, whereas $\omega\sseta_{\eta}$ has a positive imaginary part.
The superscript indicates that this choice depends on $\eta$.
By construction $\gamma\sseta_{1 \cdots \ell} \!=\! 0$ holds, 
ensuring $\omega\sseta_{1 \cdots \ell} \!=\! \omega_{1 \cdots \ell}$.
Then, the $t_\eta$
integral yields $2\pi\delta(\omega_{1 \cdots \ell})$,
ensuring $\omega_{1 \cdots \ell} \!=\! 0$ 
and thus $\omega\sseta_{1 \cdots \ell} \!=\! 0$, too.
We need not distinguish individual $\gamma_{i\neq \eta}$
and hence choose $\gamma\sseta_{i\neq \eta} \!=\! - \gamma_0$, 
$\gamma\sseta_\eta \!=\! (\ell \!-\! 1)\gamma_0$.
Nevertheless, we keep the index $i$ for a compact notation of sums like 
$\gamma\sseta_{i \cdots j} \!=\! \sum_{n=i}^j \gamma\sseta_n$.
For numerics, $\gamma_0 \!>\! 0$ should be kept small but finite. For analytical
arguments, it can be taken infinitesimal. 

Since $\Kc\sseta$ depends only on time differences, with $t_\eta$ singled out as the largest time, 
we take $t_\eta$ and $t_i' \!=\! t_i \!-\! t_{i+1}$ for $i \!<\! \ell$ as our $\ell$ 
integration variables. 
Then,
$t_{i<\eta} \!=\! t_\eta \!+\! \sum_{j=i}^{\eta-1} t_j'$ 
and 
$t_{i > \eta} \!=\! t_\eta \!-\! \sum_{j=\eta}^{i-1} t_i'$, 
and the Fourier exponent is
\begin{align}
\vec{\omega}^{[\eta]} \cdot \vec{t} 
= 
\omega_{1 \cdots \ell} \, t_\eta 
+ 
\sum_{i=1}^{\eta-1} 
\omega^{[\eta]}_{1 \cdots i} \, t_i'
- 
\sum_{i=\eta}^{\ell-1} 
\omega^{[\eta]}_{i+1 \cdots \ell} \, t_i'
.
\end{align}
Consequently, we find
$\Kc\sseta(\vec{\omega}) = 2\pi\delta(\omega_{1 \cdots \ell}) K\sseta(\vec{\omega})$, 
with
\begin{align}
K^{[\eta]}(\vec{\omega}) 
& =
\prod_{i=1}^{\eta-1} \bigg[ \mi
\nbint{-\infty}{0} \md t_i' \, e^{\mi \omega^{[\eta]}_{1 \cdots i} t_i' }
\bigg]
\nonumber
\\
& \ \ \times
\prod_{i=\eta}^{\ell-1} \bigg[
- \mi
\nbint{0}{\infty} \md t_i' \,
e^{-\mi \omega^{[\eta]}_{i+1 \cdots \ell} t_i' }
\bigg]
\nonumber
\\ 
& = 
\prod_{i=1}^{\eta-1} \pigg[ \frac{1}{\omega^{[\eta]}_{1 \cdots i}} \pigg]
\prod_{i=\eta}^{\ell-1} \pigg[ \frac{-1}{\omega^{[\eta]}_{i+1 \cdots \ell}} \pigg]
= 
\prod_{i=1}^{\ell-1} \frac{1}{\omega^{[\eta]}_{1 \cdots i}}
,
\label{eq:Keta-unpermuted-final}
\end{align}
having used $\omega\sseta_{1 \cdots \ell}=0$ in the last step.

Now, consider a general permutation $p$. 
To compute $K\sseta(\vec{\omega}_p)$, the Fourier
transform of $K\sseta(\vec{t}_p)$, we need permuted versions of the complex 
frequencies in \Eq{eq:choice_imaginary_parts}.
We define 
$\vec{\omega}\sseta_p \!=\! (\omega\sseta_{\ovb{1}}, \ndots , \omega\sseta_{\ovb{\ell}})$
as the tuple $(\vec{\omega}\sseta)_p$ obtained by permuting the components
of $\vec{\omega}\sseta$, including their imaginary parts, according to $p$. 
This moves $\omega_\eta\sseta$,
the component with positive imaginary part,
to slot $\mu \!=\! p^{-1}(\eta)$
[see Fig.~\ref{fig:ComplexFrequencyTuples}(b)].
Hence, the components of $\vec{\omega}\sseta_p$ have a positive
imaginary part in the slot $i \!=\! \mu$, 
with the entry $\ovb{i} \!=\! \ovb{\mu} \!=\! \eta$, 
and negative imaginary parts in all other slots:
$\omega\sseta_{\ovb{i}} \!=\! \omega_{\ovb{i}} \!+\! \mi (\ell -1)\gamma_0$ for $\ovb{i} \!=\! \eta$,
$\omega\sseta_{\ovb{i}} \!=\! \omega_{\ovb{i}} \!-\! \mi \gamma_0$ for $\ovb{i} \!\neq\! \eta$.
Moreover, $\omega\sseta_{\ovb{1} \cdots \ovb{\ell}} \!=\! 0$,
and the imaginary part of $\omega\sseta_{\ovb{1} \cdots \ovb{i}}$
is negative for $1 \! \leq \! i \! < \! \mu$ and positive for
$\mu \! \leq \! i \! < \! \ell$ 
(yielding $\omega\sseta_{\ovb{1} \cdots \ovb{i}} \!=\! \omega^\mp_{\ovb{1} \cdots \ovb{i}}$, 
respectively, if $\gamma_0$ is infinitesimal).

By \Eq{eq:fully_retarded_kernel}, $K\sseta(\vec{t}_p)$ is nonzero only if its largest time argument is $t_{\ovb{\eta}}$, 
the $\eta$th component of $\vec{t}_p$. To achieve convergent
Fourier integrals for $K\sseta(\vec{\omega}_p)$, we should 
thus add negative imaginary parts to all frequencies
except the Fourier partner of $t_{\ovb{\eta}}$, i.e.\ $\omega_{\ovb{\eta}}$, 
sitting in slot $\eta$ of 
$\vec{\omega}_p$.
This is achieved by using the complex tuple $\vec{\omega}^{[\ovb{\eta}]}_p$.
Indeed, with superscript $\ovb{\eta}$, the positive imaginary part 
ends up in slot $\mu \!=\! p^{-1}(\ovb{\eta}) \!=\! \eta$ 
[see Fig.~\ref{fig:ComplexFrequencyTuples}(c)]. 
Substituting $\vec{\omega}^{[\ovb{\eta}]}_p$ in place of $\vec{\omega}\sseta$ on the right of \Eq{eq:Keta-unpermuted-final}, we obtain
\begin{align}
K^{[\eta]}(\vec{\omega}_p) 
= 
\prod_{i=1}^{\ell-1} 
\frac{1}{\omega^{[\ovb{\eta}]}_{\ovb{1} \cdots {\ovb{i}}}} .
\label{eq:Keta_finite_gamma}
\end{align}
This is our main result
for general retarded kernels, applicable for both finite or infinitesimal imaginary parts.
Among the denominators, $\eta \!-\! 1$ ($\ell \!-\! \eta$) of them
have a negative (positive) imaginary part,
for all permutations $p$.
Indeed, for infinitesimal imaginary parts, \Eq{eq:Keta_finite_gamma} reads 
\begin{align}
K^{[\eta]}(\vec{\omega}_p) 
= 
\prod_{i=1}^{\eta-1} \bigg[
\frac{1}{\omega^-_{\ovb{1} \cdots {\ovb{i}}}} 
\bigg] 
\,
\prod_{i=\eta}^{\ell-1} 
\frac{1}{\omega^+_{\ovb{1} \cdots {\ovb{i}}}} 
.
\label{eq:Keta_infinitesimal_gamma}
\end{align}
However, the selection of the components of $\vec{\omega}$ 
that receive a positive or negative imaginary shift in \Eq{eq:Keta_infinitesimal_gamma}
depends on $p$.
A specific component $\omega_i$ may receive a positive shift for one permutation and a negative shift for another.
If $\eta \!=\! 1$, e.g., only denominators of the form $\omega^+_{\ovb{1} \cdots {\ovb{i}}}$,
with positive imaginary parts, occur in \Eq{eq:Keta_infinitesimal_gamma}.
For a permutation with $\ovb{1} \!=\! 1$, one of them equals $\omega_1^+$;
with $\ovb{\ell} \!=\! 1$, another equals $\omega_{\ovb{1} \cdots \ovb{\ell-1}}^+ \!=\! - (\omega_1^-)$.
Hence, a simple sum $\sum_p K\sseta (\vec{\omega}_p)$
has no well-defined analytical structure w.r.t.\ $\omega_1$,
or more generally w.r.t.\ $\vec{\omega}$. 
However, we will later obtain suitable combinations of retarded kernels with different $\eta$ that do,
yielding a retarded correlator.

To conclude this section, we note that the above approach
to construct $K\sseta$ can also be used to find
the \ZF/ kernel with finite imaginary parts, as needed for numerics.
Indeed, for $\eta \!=\! 1$,
\Eq{eq:fully_retarded_kernel} exactly matches the \ZF/ time-ordered kernel of \Eq{eq:Kc_t_RF},
$\Kc^{[1]} \!=\! \Kc$.
This kernel has both a largest time $t_\ovb{1}$ \textit{and} a smallest time $t_\ovb{\ell}$, 
and the energy-conservation integral may be performed using
either (i) $t_\ovb{1}$ or (ii) $t_\ovb{\ell}$. 
Choice (i) corresponds to the above discussion; 
hence, the $\eta \! = \! 1$ version of 
\Eq{eq:Keta_finite_gamma} can be used for $K(\vec{\omega}_p)$ 
for an expression with finite imaginary parts instead of \Eq{eq:K_w_RF}.
For choice (ii),
the dependence on all $t_{\ovb{i}} \!-\! t_{\ovb{\ell}}$, $i \!<\! \ell$,
is retarded (contributing only for $t_{\ovb{i}} \!-\! t_{\ovb{\ell}} \!>\! 0$),
as in \Eq{eq:Kc_w_RF_p=id}.
To regularize the corresponding time integrals there, 
we replace the frequencies 
$\omega_{\ovb{i}}$ in \Eqs{subeq:K_w_RF} 
by $\bigl(\omega^{[\ovb{\ell}]}_{\ovb{i}}\bigr)^\ast$, 
such that those with $i \!<\! \ell$ all have positive imaginary parts.
The two choices can be summarized as 
\begin{subequations}
\label{eq:K_w_RF_finite_gamma}
\begin{align}
K(\vec{\omega}_p) 
& \,\overset{\textrm{(i)}}{=}\;
\prod_{i=1}^{\ell-1}
\frac{1}{ \omega^{[\ovb{1}]}_{\ovb{1} \cdots \ovb{i}} } 
=
\prod_{i=1}^{\ell-1}
\frac{1}{\omega_{\ovb{1} \cdots \ovb{i}} + (\ell - i) \, \mi \gamma_0 }
,
\label{eq:K_w_RF_finite_gamma_i}
\\
K(\vec{\omega}_p) 
& \overset{\textrm{(ii)}}{=}\,
\prod_{i=1}^{\ell-1}
\frac{1}{ \big( \omega^{[\ovb{\ell}]}_{\ovb{1} \cdots \ovb{i}} \big)^* } 
=
\prod_{i=1}^{\ell-1}
\frac{1}{\omega_{\ovb{1} \cdots \ovb{i}} + i \, \mi \gamma_0 }
,
\label{eq:K_w_RF_finite_gamma_ii}
\end{align}
\end{subequations}
since, for $i \!<\! \ell$, we have
$\gamma^{[\ovb{1}]}_{\ovb{1} \cdots \ovb{i}} = (\ell-i)\gamma_0$
and
$\gamma^{[\ovb{\ell}]}_{\ovb{1} \cdots \ovb{i}} = - i \gamma_0 $
for any permutation.
Thus, both choices give
positive imaginary shifts accompanying
all $\omega_{\ovb{1} \cdots \ovb{i}}$ ($i \!<\! \ell$),
consistent with $\omega_{\ovb{1} \cdots \ovb{i}} \!+\! \mi 0^+$ used in \Eq{eq:K_w_RF}.
They are both legitimate and yield equivalent results for $\gamma_0 \!\to\! 0$.
For $\ell \!=\! 2$, they are identical even at finite $\gamma_0$.
For $\ell \!>\! 2$ and finite $\gamma_0$, they lead to qualitatively similar results, with slight quantitatively differences that decrease with $\gamma_0$.
For the curves shown in
Fig.~\ref{fig:RF_XEM}, e.g., both choices yield indistinguishable results.

\subsubsection{Contour basis}

In the contour basis, time-dependent operators $\Oc^i(t_i^{c_i})$ are defined on the Keldysh double time contour.
The contour index $c_i$ on 
the time argument $t_i^{c_i}$ specifies the branch, 
with $c_i \!=\! -$ or $+$ for an operator residing on the 
forward or backward branch, respectively. Correspondingly, 
a contour-ordered $\ell$p function, defined as 
\begin{align}
\Gc^{\vec{c}}(\vec{t})
& =
(-\mi)^{\ell-1}
\langle \TO_{\rm c} \prod_{i=1}^\ell \Oc^i(t_i^{c_i}) \rangle
, 
\end{align}
carries $\ell$ contour indices, 
$\vec{c} \!=\! c_1 \ncdots \, c_\ell$, 
one for each operator.
The contour-ordering operator $\TO_{\rm c}$ rearranges the operators such 
that those on the forward branch are all time ordered, those on
the backward branch anti-time-ordered;
the former are applied first, i.e., they all sit to the right of the latter.
It also provides a sign change for each transposition of two fermionic operators incurred while reordering.

As for the \ZF/, we express $\Gc^{\vec{c}}(\vec{t})$ as 
\begin{align}
\Gc^{\vec{c}}(\vec{t})
& =
\sum_p
\vec{\zeta}^p
\Kc^{ \vec{c}_p }( \vec{t}_p )
\Sc[\vec{\Oc}_p]( \vec{t}_p ) 
.
\label{eq:Gc_t_KS_KF_contour}
\end{align}
In this sum over permutations of $\ell$ indices, each summand is a 
product of a kernel $\Kc^{\vec{c}}$, 
enforcing contour ordering, and the PSF
$\Sc$ of \Eq{eq:Sc_t}, containing the time-dependent operators.
For a given choice of times $\vec{t}$ and contour
indices $\vec{c}$, only
one permutation yields a nonzero result, namely the one
which arranges the operators in contour-ordered fashion.

\begin{figure}
\includegraphics[width=\linewidth]{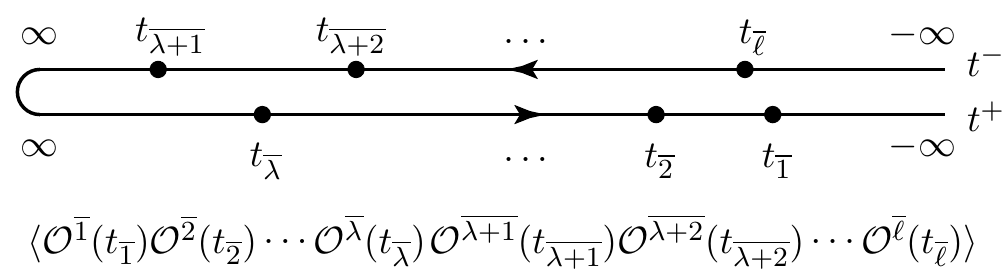}
\caption{\KF/ time ordering. The 
$\ell$-tuples of times $\vec{t} = (t_1, \ndots, t_\ell)$ and
contour indices $\vec{c} = (c_1, \ndots, c_\ell)$ specify
an $\ell$-tuple $(t_1^{c_1}, \ndots, t^{c_\ell}_\ell)$,
with $t_i^{\mp}$ on the forward or backward 
branch, respectively. Consider such 
an $\ell$-tuple, with $\lambda$ contour indices equal to $+$, the others $-$. 
Then, only that permutation $p$ in \Eq{eq:Gc_t_KS_KF_contour} yields a nonzero
contribution, $\Kc^{\vec{c}_p}(\vec{t}_p) \neq 0$, 
denoted $\Kc^{[\lambda,\ell-\lambda]}(\vec{t}_p)$, for which the permuted
$\ell$-tuple $(t_{\ovb{1}}^{c_{\ovb{1}}}, \ndots, 
t_{\ovb{\ell}}^{c_{\ovb{\ell}}})$ has the form 
$(t_{\ovb{1}}^+, \ndots, t_{\ovb{\lambda}}^+,
t_{\ovb{\lambda+1}}^-, \ndots , t_{\ovb{\ell}}^-)$, with 
$t_{\ovb{1}} < t_{\ovb{2}} < \ncdots < t_{\ovb{\lambda}}$
and 
$t_{\ovb{\lambda+1}} > t_{\ovb{\lambda+2}} > \ncdots > 
t_{\ovb{\ell}}$.
In the corresponding operator product $\Sc [\vec{\Oc}_p](\vec{t}_p)$,
all (forward-branch) $t^-$ times appear to the right of all (backward-branch) $t^+$ times, 
with larger $t^-$ to the left of smaller $t^-$, and smaller $t^+$ to the left of larger $t^+$. 
In other words, the right-to-left order of operators in the product matches
the order in which their times appear on the contour.
}
\label{fig:TimeOrderingKF}
\end{figure}

Given $\vec{c}$, let $\lambda$ denote
the number of its $+$ entries, i.e. the number
of operators on the backward branch. 
Contour ordering places all $\lambda$ operators on the backward branch 
to the left of all $\ell \!-\! \lambda$
operators on the forward branch. Hence,
only those components of $\Kc^{\vec{c}_p} (\vec{t}_p)$ are nonzero for which 
$\vec{c}_p \!=\! c_{\ovb{1}} \ncdots c_{\ovb{\ell}}$ 
has the form $+ \ncdots + \! - \ncdots \, -$, with
$\lambda$ entries of $+$ followed by $\ell \! - \! \lambda$ entries of $-$, 
$0 \!\leq\! \lambda \!\leq\! \ell$.
We use the shorthand $\vec{c}_p \!=\! [\lambda, \ell-\lambda]$
for this structure, e.g.\ 
$[0,3] \!=\! -\!-\!-$ or $[2,2] \!=\! +\!+\!- -$.
Then, the successive time-ordering and anti-time-ordering 
rules on the two branches
(from right to left) are implemented by the kernel 
\begin{align}
\Kc^{[\lambda,\ell-\lambda]} (\vec{t}_p)
& 
=
(-\mi)^{\ell-1}
\!
\prod_{i=1}^{\lambda-1}
\!
\Big[
\theta(t_{\ovb{i+1}} \!\!\;-\!\!\; t_{\ovb{i}})
\Big]
\!\!\:
\prod_{i=\lambda+1}^{\ell-1} 
\!\!
\theta(t_{\ovb{i}} \!\!\;-\!\!\; t_{\ovb{i+1}})
,
\label{eq:general-contour-thetas}
\end{align}
cf.\ \Fig{fig:TimeOrderingKF}.
Again, a product over an empty set, with lower limit larger than upper limit, equals unity. 
The superscript of $\Kc$, standing for $\vec{c}_p = [\lambda,\ell-\lambda]$, reflects the number of $+$ and $-$ entries
in $\vec{c}$ and hence is the same for all 
nonzero components $\Kc^{\vec{c}_p}$
associated with a given $\Kc^{\vec{c}}$.

For $\lambda \!=\!0$, \Eq{eq:general-contour-thetas} yields the time-ordered kernel $\Kc$ [\Eq{eq:Kc_t_RF}], 
which also matches the fully retarded kernel $\Kc^{[1]}$ from \Eq{eq:fully_retarded_kernel}, 
$\Kc^{[0,\ell]} \!=\! \Kc \!=\! \Kc^{[1]}$.
The other extremal case, $\lambda \!=\! \ell$, yields a retarded kernel, too:
$\Kc^{[\ell,0]} \!=\! (-1)^{\ell-1} \Kc^{[\ell]}$.
Yet, for intermediate $0 \!<\! \lambda \!<\! \ell$, 
\Eq{eq:general-contour-thetas} has only $\ell \!-\! 2 $ factors,
while \Eq{eq:fully_retarded_kernel} has $\ell \!-\! 1$.
We can nevertheless express $\Kc^{[\lambda, \ell-\lambda]}$ through
retarded kernels by
inserting in \Eq{eq:general-contour-thetas}
a factor of unity in the form
$1 = \theta( t_{\ovb{\lambda}} - t_{\ovb{\lambda+1}} ) 
+ \theta( t_{\ovb{\lambda+1}} - t_{\ovb{\lambda}} )$.
Including all prefactors, this yields
$(-1)^{\lambda-1} \Kc^{[\lambda]} \!+\! (-1)^{\lambda} \Kc^{[\lambda+1]}$.
In summary:
\begin{subequations}
\label{eq:K_KF_contour}
\begin{flalign}
\Kc^{[0,\ell]} 
& = 
\Kc^{[1]}
,
\qquad\qquad\quad\
\Kc^{[\ell,0]} = (-1)^{\ell-1} \Kc^{[\ell]}
,
\hspace{-0.5cm} & 
\\
\Kc^{[\lambda,\ell-\lambda]} 
& = 
(-1)^{\lambda-1} ( \Kc^{[\lambda]} - \Kc^{[\lambda+1]} )
,
\qquad
0 \!<\! \lambda \!<\! \ell
.
\hspace{-0.5cm} & 
\end{flalign}
\end{subequations}

To Fourier transform $\Kc^{[\lambda,\ell-\lambda]}$,
it suffices to consider the identity permutation, $\ovb{i} = i$.
Since the Fourier transform is linear, we can directly infer 
$K^{[\lambda,\ell-\lambda]}(\vec{\omega})$ from \Eqs{eq:K_KF_contour}
and \eqref{eq:Keta_finite_gamma}. For $\lambda \in \{0, \ell\}$, we have
\begin{subequations}
\label{subeq:K_w_KF_contour_p=id}
\begin{align}
K^{[0,\ell]}(\vec{\omega})
& =
\prod_{i=1}^{\ell-1} \frac{1}{\omega^{[1]}_{1 \cdots i}}
,
\quad
K^{[\ell,0]}(\vec{\omega})
=
\prod_{i=1}^{\ell-1} \frac{-1}{\omega^{[\ell]}_{1 \cdots i}}
.
\end{align}
The cases $0 \!<\! \lambda \!<\! \ell$ mix $K^{[\lambda]}$ and $K^{[\lambda+1]}$
and thus $\omega^{[\lambda]}$ and $\omega^{[\lambda+1]}$.
For all $i \notin \{\lambda,\lambda \!+\! 1\}$,
we have $\gamma_i^{[\lambda]} = \gamma_i^{[\lambda+1]}$,
implying $\omega_i^{[\lambda]} = \omega_i^{[\lambda+1]}$,
so that $\omega_{1\cdots i}^{[\lambda]} = \omega_{1\cdots i}^{[\lambda+1]}$ for $i \!<\! \lambda$
and $\omega_{i\cdots \ell}^{[\lambda]} = 
\omega_{i \cdots \ell}^{[\lambda+1]}$ for $i \!>\! \lambda \!+\! 1$. 
Moreover, $\omega_{1 \cdots i}^{[\lambda+1]} = 
- \omega_{i + 1 \cdots \ell}^{[\lambda+1]}$, by complex frequency
conservation.
We thus find 
\begin{flalign}
K^{[\lambda, \ell-\lambda]} (\vec{\omega})
& = 
\prod_{i=1}^{\lambda-1} \pigg[ 
\frac{-1}{\omega^{[\lambda]}_{1 \cdots i}} 
\pigg]
\,
\Delta(\vec{\omega})
\,
\prod_{i=\lambda+1}^{\ell-1} \pigg[
\frac{-1}{\omega^{[\lambda+1]}_{i+1 \cdots \ell}} 
\pigg] 
,
\hspace{-1cm} &
\nonumber
\\
\nonumber
\Delta(\vec{\omega}) 
& = 
\frac{-1}{\omega^{[\lambda]}_{\lambda+1\dots \ell}} -
\frac{1}{\omega^{[\lambda+1]}_{1 \cdots \lambda}} 
&
\\
& = 
\frac{1}{ \omega_{1 \cdots \lambda} + \mi \gamma^{[\lambda]}_{\lambda+1 \cdots \ell}}
-
\frac{1}{\omega_{1 \cdots \lambda} - \mi \gamma^{[\lambda+1]}_{1 \cdots \lambda}} 
.
\hspace{-1cm} & 
\end{flalign}
\end{subequations}
If all imaginary parts are sent to zero, $\gamma_0 \to 0$,
the first and second terms of $\Delta(\vec{\omega})$ yield 
$-\mi \pi \delta(\omega_{1 \cdots \lambda} )\pm \mathcal{P}(1/\omega_{1 \cdots \lambda})$, 
and their sum $\Delta(\vec{\omega}) \to -2 \pi \mi \delta( \omega_{1 \cdots \lambda} )$. 
One thereby obtains a second
$\delta$ function, next to the overall $\delta(\omega_{1 \cdots \ell})$ ensuring energy conservation.
The second $\delta$ function is indeed to be expected:
in the Fourier integral for $0 \!<\! \lambda \!<\! \ell$, 
time-translation invariance can be exploited on each branch separately. 
This generates two $\delta$ functions, one for the sum of frequencies on each branch,
$\omega_{1 \cdots \lambda}$ and $\omega_{\lambda+1 \cdots \ell}$.
Together, they give
$
\delta(\omega_{1 \cdots \lambda})\delta(\omega_{\lambda+1 \cdots \ell})
= 
\delta(\omega_{1 \cdots \lambda})\delta(\omega_{1 \cdots \ell})
$.

For a general permutation $p$, the permuted kernel $K^{\vec{c}_p}(\vec{\omega}_p)$ is 
nonzero only for contour indices of the form 
$ \vec{c}_p = [\lambda,\ell \!-\!\lambda]$ 
($c_{\ovb{i}}=+$ if $i \le \lambda$, 
$c_{\ovb{i}}=-$ if $i > \lambda$).
Then, $K^{[\lambda,\ell-\lambda]}(\vec{\omega}_p)$,
obtained via \Eq{eq:Keta_finite_gamma} for
$K^{[\lambda]}(\vec{\omega}_p)$, is given by
\Eqs{subeq:K_w_KF_contour_p=id}, 
with $\omega^{[\lambda]}_{1 \cdots i}$ 
replaced by $\omega^{[\ovb{\lambda}]}_{\ovb{1} \cdots \ovb{i}}$ there.
With this, we can proceed with the contour-ordered correlation functions.
Again, the multiplicative structure of
\Eq{eq:Gc_t_KS_KF_contour} for $\Gc^{\vec{c}} (\vec{t})$
yields a convolution in frequency space:
\begin{subequations}
\begin{flalign}
\Gc^{\vec{c}} (\vec{\omega})
& =
\sum_p
\vec{\zeta}^p \!
\nint \md^\ell \omega_p' \,
\Kc^{ \vec{c}_p } ( \vec{\omega}_p - \vec{\omega}_p')
\Sc[\vec{\Oc}_p](\vec{\omega}_p')
,
\hspace{-0.5cm}
&
\\
G^{\vec{c}} (\vec{\omega})
& =
\sum_p \vec{\zeta}^p \!
\nint \md^{\ell-1} \omega_p' \,
K^{ \vec{c}_p }( \vec{\omega}_p - \vec{\omega}_p' )
S[\vec{\Oc}_p](\vec{\omega}_p')
.
\hspace{-0.5cm}
&
\label{eq:G_w_KS_KF_contour}
\end{flalign}
\end{subequations}
We extracted $G^{\vec{c}}$ from $\Gc^{\vec{c}}$
using the $\delta$ functions
in $\Kc$ and $\Sc$. 
The PSFs \eqref{eq:S_w}
are the same as for the \ZF/ and \MF/.

As a simple example, consider the correlator $G^{+-} (t)= 
- \mi \langle \Ac(t) \Bc \rangle$ from Sec.~\ref{sec:motivation}:
$\Ac(t)$ sits to the left of $\Bc$, since 
the contour indices $\vec{c} = +-$ 
place $\Ac(t)$ on the backward and $\Bc$ on the forward branch.
The sum $\sum_p$ from 
\Eq{eq:G_w_KS_KF_contour} involves
two components 
of $K^{\vec{c}_p}$, namely 
$K^{+-} (\omega_1 - \omega'_1) =
- 2 \pi \mi \delta(\omega_1 - \omega'_1)$ and $K^{-+} = 0$. Thus,
\Eq{eq:G_w_KS_KF_contour} yields 
$G^{+-}(\omega) = - 2 \pi \mi S[A,B](\omega)$, reproducing
Eq.~\eqref{eq:G2_w_KF_+-}.

\subsubsection{Keldysh basis}
\label{sec:KeldyshBasis}

Correlators in the Keldysh basis $\Gc^{\vec{k}}$
carry Keldysh indices, 
$\vec{k} = k_1 \ncdots k_\ell$, with $k_i \in \{1,2\}$. 
They are obtained from correlators in the contour basis
by applying a linear transformation $D$ to
each contour index \cite{Keldysh1964},
\begin{subequations}
\begin{align}
\Gc^{\vec{k}} (\vec{t})
& = \sum_{c_1 \nidots c_\ell}
\prod_{i=1}^\ell \Big[ D^{k_i c_i} \Big]
\Gc^{\vec{c}} (\vec{t}) ,
\\
D^{kc} & = 
\frac{1}{\sqrt{2}}
\Big(
\begin{array}{cc}
1 & -1 \\
1 & \phantom{-}1
\end{array}
\Big)_{kc}
= 
{\textstyle \frac{1}{\sqrt{2}}} (-1)^{k\cdot \delta_{c,+}} 
, 
\label{eq:Keldysh_rotation_matrix}
\end{align}
\end{subequations}
with $c= -$ or $+$ giving the first or second column of $D$, respectively.
In Eq.~\eqref{eq:Gc_t_KS_KF_contour}, the dependence on
$\vec{c}$ resides solely in $\Kc^{\vec{c}}$; thus, the Keldysh rotation yields
(with a conventional prefactor)
\begin{subequations}
\begin{align}
\Gc^{\vec{k}}(\vec{t})
& =
\frac{2}{2^{\ell/2}}
\sum_p
\vec{\zeta}^p
\Kc^{ \vec{k}_p }( \vec{t}_p )
\Sc[\vec{\Oc}_p]( \vec{t}_p ) 
, 
\\
\Kc^{\vec{k}_p}( \vec{t}_p ) 
& =
\frac{2^{\ell/2}}{2}
\sum_{c_\ovb{1}, \nidots, c_\ovb{\ell}}
\prod_{i=1}^\ell \Big[ D^{k_\ovb{i} c_\ovb{i}} \Big]
\Kc^{\vec{c}_p}( \vec{t}_p ) 
\label{eq:Kc_t_KF_Keldysh_1}
\\
& =
\frac{1}{2} \sum_{\lambda = 0}^{\ell}
(-1)^{k_{\ovb{1} \cdots \ovb{\lambda}}} \, 
\Kc^{[\lambda, \ell-\lambda]}( \vec{t}_p ) 
.
\label{eq:Kc_t_KF_Keldysh_2}
\end{align}
\end{subequations}
To perform the sum over all $\vec{c}_p$ in \Eq{eq:Kc_t_KF_Keldysh_1},
we recalled that the kernels $\Kc^{\vec{c}_p}$ are nonzero only
if $\vec{c}_p$ has the form $\vec{c}_p = [\lambda, \ell \!-\! \lambda]$, with $\lambda \in [0,\ell]$. 
For these, 
$\sqrt{2} D^{k_{\ovb{i}} c_{\ovb{i}}}$ equals $(-1)^{k_{\ovb{i}}}$ 
for $i \le \lambda$ and 1 otherwise,
yielding the factor $(-1)^{k_{\ovb{1} \cdots \ovb{\lambda}}}$.
We used
$k_{\ovb{1} \cdots \ovb{\lambda}} = \sum_{i=1}^\lambda k_{\ovb{i}}$,
as usual defining the sum over an empty set as zero,
$k_{1 \cdots 0}=0$.

Using \Eq{eq:K_KF_contour}, we can express the permuted \KF/ kernel through fully retarded ones: 
\begin{align}
\Kc^{\vec{k}_p} 
\! & = \!
\frac{1}{2}
\Big\{
\Kc^{[1]} 
+ 
\sum_{\lambda = 1}^{\ell-1}
(-1)^{k_{\ovb{1} \cdots \ovb{\lambda}}} 
\pig[
(-1)^{\lambda-1} \Kc^{[\lambda]}
\nonumber
\\
& \ \ 
+
(-1)^{\lambda} \Kc^{[\lambda+1]}
\pig]
+ 
(-1)^{k_{\ovb{1}\cdots \ovb{\ell}}} (-1)^{\ell-1} \Kc^{[\ell]}
\Big\}
\nonumber
\\
& = 
\sum_{\lambda = 1}^{\ell}
(-1)^{\lambda-1} (-1)^{k_{\ovb{1}\cdots \ovb{\lambda-1}}} 
\frac{ 1 + (-1)^{k_\ovb{\lambda} }}{2}
\Kc^{[\lambda]}
.
\label{eq:Kc_KF_Keldysh_via_Keta}
\end{align}
This result directly yields 
the Fourier transform $\Kc^{\vec{k}_p}(\vec{\omega}_p)
= 2 \pi \delta(\omega_{\ovb{1} \cdots \ovb{\ell}}) K^{\vec{k}_p}(\vec{\omega}_p)$, with $K^{[\lambda]} (\vec{\omega}_p)$ given by 
\Eq{eq:Keta_finite_gamma}. 
Thus, we know all ingredients of the Fourier
transform of $G^{\vec{k}}$, written as a convolution of $K^{\vec{k}}$
and $S$: 
\begin{align}
G^{\vec{k}} (\vec{\omega})
=
\frac{2}{2^{\ell/2}}
\sum_p \vec{\zeta}^p
\nint \md^{\ell-1} \omega_p' \,
K^{ \vec{k}_p }( \vec{\omega}_p - \vec{\omega}_p' )
S[\vec{\Oc}_p]( \vec{\omega}_p' ) .
\label{eq:G_w_KF_Keldysh}
\end{align}
Together, \Eqs{eq:G_w_KF_Keldysh}, \eqref{eq:Kc_KF_Keldysh_via_Keta},
and \eqref{eq:Keta_finite_gamma} give the spectral representation for \KF/ $\ell$p functions in the Keldysh basis, 
in a form well suited for numerical computations \cite{Lee2021}.

To obtain more analytical insight, it is fruitful to further simplify
\Eq{eq:Kc_KF_Keldysh_via_Keta}.
The fraction in its last line 
vanishes whenever $k_\ovb{\lambda} = 1$.
Hence, there is a cancellation pattern that depends on the number of 2's, say $\alpha$, contained in the 
composite Keldysh index 
$\vec{k}_p \!=\! k_\ovb{1} \ncdots k_\ovb{\ell}$.
To elaborate on this, we first consider the identity permutation,
$\ovb{i} = i$, for which the Keldysh indices
of $K^{\vec{k}_p}$ match those of $G^{\vec{k}}$.
Let $\eta_j$ denote the slot of the $j$th $2$ in $\vec{k}$, 
with $\eta_{j} \!<\! \eta_{j+1}$.
Then, $\vec{k}$ is uniquely specified by the ordered list 
$\vec{k} \!=\! [\eta_1 \, \ndots \, \eta_\alpha]$, e.g.\
$1111 \!=\! [\,]$, $2121 \!=\! [13]$. 
From \Eq{eq:Kc_KF_Keldysh_via_Keta}, it follows immediately that
$\Kc^{[\,]} \!=\! \Kc^{1 \cdots 1} \!=\! 0$.
Next, consider Keldysh indices containing a solitary 2
in slot $\eta$, i.e.\ $k_{\eta} \!=\! 2$ and 
$\vec{k} \!=\! [\eta]$. 
As anticipated by this notation, 
$\Kc^{\vec{k}}$ is then equal to the retarded kernel $\Kc\sseta$.
Indeed, \Eq{eq:Kc_KF_Keldysh_via_Keta} with $\vec{k} \!=\! [\eta]$
has only one nonzero summand, having $\lambda \!=\! \eta$, and
\begin{align}
(-1)^{\eta-1} (-1)^{k_{1 \cdots \eta-1}} \Kc^{[\eta]} 
=
\Kc^{[\eta]} 
,
\label{eq:Keta_identity}
\end{align}
since $k_{i \neq \eta} \!=\! 1$ implies
$k_{1 \cdots \eta-1}\! =\! \eta \! -\! 1$.
Finally, Keldysh indices with 
$2 \!\leq\! \alpha \!\leq\! \ell$ many $2$'s 
in slots $[\eta_1 \, \ndots \, \eta_\alpha]$ yield
a term of the form \eqref{eq:Keta_identity}
for each $\eta_j$:
\begin{align}
\Kc^{[\eta_1 \, \ndots \, \eta_\alpha]}
& = 
\sum_{j=1}^\alpha 
\underbrace{
(-1)^{\eta_j-1}
(-1)^{k_{1 \cdots \eta_j-1}}
}_{
(-1)^{j-1}
}
\Kc^{[\eta_j]} 
\label{eq:K_eta_1_to_eta_alpha}
.
\end{align}
To find the sign,
we used $k_{i \neq \eta_j} \!=\!1$, $k_{\eta_j}\!=\!2$, 
and the fact that the $\eta_j$'s are ordered, implying
$k_{1 \cdots \eta_j - 1} \!=\! \eta_j \!+\! j \!-\! 2$.

Now, consider a general permutation $p$. The permuted 
Keldysh index $\vec{k}_p$ contains the same number
of $2$'s as $\vec{k}$, but in different slots.
We can analogously express it through an ordered list 
as $\vec{k}_p \!=\! [\ovh{\eta}_1 \, \ndots \, \ovh{\eta}_\alpha]$,
where $\ovh{\eta}_j$ denotes the slot of the $j$th $2$ in $\vec{k}_p$,
with $\ovh{\eta}_j \!<\! \ovh{\eta}_{j+1}$. 
To find the $\ovh{\eta}_j$'s given the $\eta_j$'s, note that 
a 2 from slot $\eta_j$ in $\vec{k}$ is moved by $p$ to slot $\mu_j =
p^{-1}(\eta_j)$ in $\vec{k}_p$. 
The sequence 
$[\mu_1 \, \ndots \, \mu_\alpha]$ lists the new slots
of the $2$'s; placing its elements in increasing order yields $[\ovh{\eta}_1 \, \ndots \, \ovh{\eta}_\alpha]$.
For example, if $\vec{k}=1212 = [24]$,
then $p=(4123)$ yields $[\mu_1\mu_2]=[31]$ and $\vec{k}_p=2121=[13]$.

Combining \Eqs{eq:K_eta_1_to_eta_alpha}, \eqref{eq:G_w_KF_Keldysh},
and \eqref{eq:Keta_finite_gamma}, 
we finally obtain 
\begin{subequations}
\label{eq:G_w_KF_eta_j_total}
\begin{flalign}
G^{[\eta_1 \nidots \eta_\alpha]} (\vec{\omega})
& =
\frac{2}{2^{\ell/2}}
\sum_p \vec{\zeta}^p 
\nint \md^{\ell-1} \omega_p' \, & 
\label{eq:G_w_KF_eta_j}
\\
\nonumber 
& \quad \times
K^{[\ovh{\eta}_1 \nidots \ovh{\eta}_\alpha]}( \vec{\omega}_p - \vec{\omega}_p' )
S[\vec{\Oc}_p](\vec{\omega}_p')
, & 
\\
K^{[\ovh{\eta}_1 \, \ndots \, \ovh{\eta}_\alpha]} 
( \vec{\omega}_p \!-\! \vec{\omega}_p' )
&=
\sum_{j=1}^\alpha (-1)^{j-1} \! \prod_{i=1}^{\ell-1} 
\frac{1}{\omega^{[\ovb{\ovh{\eta}}_j]}_{\ovb{1} \cdots \ovb{i}} - \omega'_{\ovb{1} \cdots \ovb{i}} }
.
\hspace{-0.5cm} &
\label{eq:K_w_KF_eta_j}
\end{flalign}
\end{subequations}
In App.~\ref{app:KF_G2}, we show how
\Eqs{eq:G_w_KF_eta_j_total} reproduce the well-known results for $2$p \KF/ correlators.
Since the set of $\ovh{\eta}_j$ is obtained by ordering the set of 
$\mu_j = p^{-1}(\eta_j)$,
it follows 
that each $\ovb{\ovh{\eta}}_j$ in
\Eq{eq:K_w_KF_eta_j} is equal to some $\eta_{j'}$
(with $j'$ and $j$ related in a manner depending on $p$).
Hence, the external frequencies 
$\vec{\omega}$
enter the kernel through sets of complex frequencies
$\omega^{[\eta_{j'}]}_{\ovb{1} \cdots \ovb{i}}$,
whose imaginary parts are determined by the 
\textit{external} Keldysh indices 
$\vec{k} \!=\! [\eta_1 \, \ndots \, \eta_\alpha]$.

For a fully retarded correlator $G\sseta$, where $\alpha \!=\! 1$,
we have $\ovh{\eta} \!=\! \mu \!=\! p^{-1}(\eta)$, hence $\ovb{\ovh{\eta}} \!=\! \eta$, so that the right side of 
\Eq{eq:K_w_KF_eta_j} depends on
$\omega\sseta_{\ovb{1} \cdots \ovb{i}}$.
In this case, the permuted Keldysh indices 
$\vec{k}_p$
enter only intermediately and are not needed explicitly. 
Hence, $G\sseta$ can be expressed through a single set of complex frequencies 
$\vec{\omega}\sseta$, entering via the product
$\prod_{i=1}^{\ell-1}(\omega\sseta_{\ovb{1} \cdots \ovb{i}} - \omega'_{\ovb{1} \cdots \ovb{i}})^{-1}$. 
It follows that $G\sseta (\vec{\omega})$ is an analytic function of the variable $\omega_\eta$ in the upper half complex plane. 
To see this, note that, for each denominator containing 
$\omega\sseta_\eta \!=\! \omega_\eta \!+\! \mi \gamma\sseta_\eta$ 
in the sum $\omega\sseta_{\ovb{1} \cdots \ovb{i}}$
(i.e.\ for which $\eta \!\in\! \{\ovb{1}, \, \ndots , \ovb{i}\}$), the
latter has a positive imaginary part, $\gamma\sseta_{\ovb{1} \cdots \ovb{i}} \!>\! 0$. 
Thus, for infinitesimal $\gamma_0$, the corresponding denominator 
has the form 
$\omega_\eta \!+\! \mi 0^+ \!+\!$ real frequencies,
such that $\omega_\eta$ can be analytically continued into the upper half plane without encountering any singularities.
Accordingly, in the time domain, $G\sseta(\vec{t})$ is fully retarded w.r.t.\ $t_\eta$ (i.e.\ nonzero only for $t_i \!<\! t_\eta$, $i \!\neq\! \eta$) \cite{Lehmann1957,*Glaser1957,Evans1992,Baier1994}. 

The spectral representation \eqref{eq:G_w_KF_eta_j_total} 
constitutes our main result for \KF/ $\ell$p functions.
It is very compact, with the number of terms increasing with $\alpha$, 
the number Keldysh indices equal to $2$,
and offers insight into the analytical structure of Keldysh correlators, 
as we explain next.

For the correlators $G\sseta$ with a solitary Keldysh index $2$ in slot $\eta$,
\Eq{eq:K_w_KF_eta_j} for the kernel $K\sseta$ involves only one summand,
similar to its analogs in the \ZF/ and \MF/ (without anomalous terms),
\Eqs{eq:K_w_RF_finite_gamma_i} and \eqref{eq:Ktilde_Omega_IF}, respectively.
The \ZF/, \MF/, and \KF/ results read
\begin{subequations}
\label{eq:analytic_structure_RF_IF_KF}
\begin{align}
G(\vec{\omega})
& =
\sum_p \vec{\zeta}^p 
\nint 
\frac{
\md^{\ell-1} \omega_p' \,\,\, S[\vec{\Oc}_p](\vec{\omega}_p')
}{
\prod_{i=1}^{\ell-1} \pig[
\omega^{[\ovb{1}]}_{\ovb{1} \cdots \ovb{i}} 
\!
-
\omega'_{\ovb{1} \cdots \ovb{i}} 
\pig]
}
,
\label{eq:analytic_structure_RF}
\\
\widetilde{G}(\mi\vec{\omega})
& =
\sum_p \vec{\zeta}^p 
\nint 
\frac{
\md^{\ell-1} \omega_p' \,\,\, S[\vec{\Oc}_p](\vec{\omega}_p')
}{
\prod_{i=1}^{\ell-1} \pig[
\mi\omega_{\ovb{1} \cdots \ovb{i}} - \omega'_{\ovb{1} \cdots \ovb{i}} 
\pig]
}
,
\label{eq:analytic_structure_IF}
\\
\frac{2^{\ell/2}}{2}
G^{[\eta]} (\vec{\omega})
& =
\sum_p \vec{\zeta}^p 
\nint 
\frac{
\md^{\ell-1} \omega_p' \,\,\, S[\vec{\Oc}_p](\vec{\omega}_p')
}{
\prod_{i=1}^{\ell-1} \pig[
\omega^{[\eta]}_{\ovb{1} \cdots \ovb{i}} - \omega'_{\ovb{1} \cdots \ovb{i}} 
\pig]
}
,
\label{eq:analytic_structure_KF}
\end{align}
\end{subequations}
respectively, in terms of the PSFs $S$ from \Eq{eq:S_w}.
There is an important distinction between the two real-frequency correlators
of \Eqs{eq:analytic_structure_RF} and \eqref{eq:analytic_structure_KF}:
for the former, the imaginary part of 
each component of $\vec{\omega}^{[\ovb{1}]}$ 
depends on the permutation $p$;
for the latter,
the imaginary parts of $\vec{\omega}\sseta$ 
are independent of $p$, being
determined by the external index $\eta$ for \textit{all} permutations.
Hence, $G\sseta$ can be expressed through a single set of complex frequencies 
$\vec{\omega}\sseta$
(as stated before), while $G$ cannot.

Comparing $G\sseta$ to the \MF/ correlator \eqref{eq:analytic_structure_IF},
we observe that they exactly agree up to a replacement of frequencies:
\begin{align}
2^{\ell/2-1} G^{[\eta]}(\vec{\omega}) 
= 
\ovwt{G}(\mi\vec{\omega}) \big|_{\mi\vec{\omega} \to \vec{\omega}^{[\eta]}}
.
\label{eq:analytic_continuation_IF_to_fully_retarded}
\end{align}
This is the analytic continuation between 
\MF/ $\ell$p functions and (fully) retarded ones in the \KF/
\cite{Evans1992,Baier1994,Weldon2005a,*Weldon2005b}.
It generalizes the well-known $2$p relation
$G^{21/12} (\omega) = \ovwt{G}(\mi \omega \to \omega^\pm)$. 

By contrast, Keldysh correlators
with multiple 2's cannot be obtained from \MF/ ones by
\textit{direct} analytic continuation
because their kernels in \Eq{eq:K_w_KF_eta_j} 
involve two or more sets of frequencies, 
$\vec{\omega}^{[\eta_1]}$, $\vec{\omega}^{[\eta_2]}$, etc.,
having \textit{different} imaginary parts.
Therefore, they do not have any well-defined regions of analyticity in the space of complex frequencies.
One may also realize that the summation in \Eq{eq:K_w_KF_eta_j} 
for $\alpha \!>\! 1$ combines denominators of frequencies differing only by their infinitesimal imaginary parts
and thus leads to $\delta$ functions.
These are clearly at odds with any simple analytic structure.
Nonetheless, the spectral representations derived above offer a convenient
starting point for a systematic analysis of relations between different Keldysh correlators,
similar to the famous fluctuation-dissipation theorem for $\ell \!=\! 2$ [cf.\ \Eq{eq:G22_FDT}] 
\cite{Kubo1957,Martin1959,Wang2002}.
Such an analysis exceeds the scope of this work and will appear in a separate publication \cite{Ge2020b}.

\section{Quantities to compute}
\label{sec:quantities}

The next two sections are devoted to illustrating the potential
of our approach for computing $\ell$p functions with several exemplary applications. 
In the present section, we recall the definition of the $4$p vertex,
give the Hamiltonians of the relevant models, 
and summarize various analytical results available for them. 
In Sec.~\ref{sec:results}, we present numerical results for the $4$p vertex,
comparing them against benchmarks where available.

\subsection{Definition of the 4p vertex}

For the numerical evaluation of our spectral representations, 
we focus on local $4$p functions of fermionic creation or annihilation operators, $d_\sigma^\dag$ or $d_\sigma^\pdag$,
where $\sigma \in \{ \uparrow, \downarrow \}$ is the electronic spin.
We also need the $2$p correlator, which follows from \Eqs{eq:RF_timedomain}, \eqref{eq:Gc_tau_IF}
in the \ZF/, \MF/ using $\Oc^1 = d_\sigma^\pdag$, $\Oc^2 = d_\sigma^\dag$.
In this section (in contrast to previous ones), 
we display only the first $\ell-1$ time or frequency arguments of $\ell$p functions,
evoking time-translational invariance to set the $\ell$th time argument to zero
and $\omega_\ell = - \omega_{1\cdots \ell-1}$.

\begin{figure}
\includegraphics[width=\linewidth]{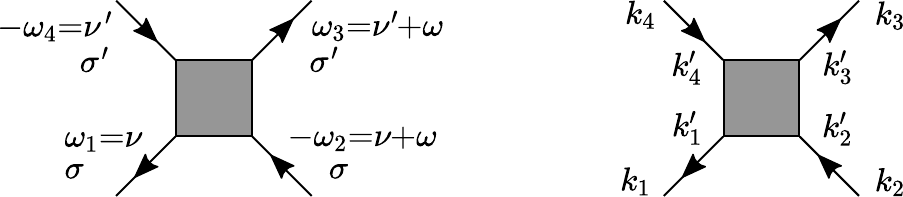}
\caption{%
Diagrammatic representation labeling the external legs of a $4$p function.
The left panel gives frequency and spin labels; 
one label per line suffices as the propagator is diagonal in spin and has only one independent frequency.
The right panel additionally shows two Keldysh indices per propagator.}
\label{fig:four-leg_diagram} 
\end{figure}

In systems with spin symmetry, the $2$p function, or \textit{propagator}, is diagonal in spin and simply reads
\begin{flalign}
G(\tau) 
& = 
- \langle \TO d_{\sigma}^\pdag(\tau) d_{\sigma}^\dag \rangle 
,
&
G(t) 
& = 
-\mi \langle \TO d_{\sigma}^\pdag(t) d_{\sigma}^\dag \rangle 
,
\hspace{-0.4cm}
&
\\
G(\mi\omega) 
& =
\nbint{0}{\beta} \md \tau \, e^{\mi\omega\tau} G(\tau)
,
&
G(\omega) 
& =
\nbint{-\infty}{\infty} \md t \, e^{\mi\omega t} G(t)
.
\hspace{-0.4cm}
&
\nonumber 
\end{flalign}
For the $4$p function, we use
$\vec{\Oc} = (d_\sigma^\pdag, d_\sigma^\dag, d_{\sigma'}^\pdag, d_{\sigma'}^\dag)$, i.e.,
\begin{flalign}
& G_{\sigma\sigma'}(\tau_1,\tau_2,\tau_3)
=
(-1)^3
\langle \TO d_{\sigma}^\pdag(\tau_1) d_{\sigma}^\dag(\tau_2) d_{\sigma'}^\pdag(\tau_3) d_{\sigma'}^\dag \rangle 
,
\hspace{-0.5cm}
&
\\
& G_{\sigma\sigma'}(\mi\omega_1,\mi\omega_2,\mi\omega_3) 
=
\nbint{0}{\beta} \md^3 \tau \, e^{\mi\sum_{i=1}^3 \omega_i\tau_i} G_{\sigma\sigma'}(\tau_1,\tau_2,\tau_3)
,
&
\nonumber
\end{flalign}
in the \MF/ and analogously in the \ZF/ [replacing $(-1)^3$ by $(-\mi)^3$].
We here use the Fourier exponent $\mi\sum_i \omega_i \tau_i$ with the same sign for all frequencies,
whereas one typically attributes alternating signs to annihilation and creation operators.
We can directly switch to the standard convention 
by expressing our final results in the particle-hole representation of frequencies
\cite{Rohringer2018}
(cf.\ \Fig{fig:four-leg_diagram}),
\begin{align}
\vec{\omega} = (\nu, -\nu\!-\! \omega, \nu'\! + \! \omega, -\nu') ,
\label{eq:freq_ph_representation}
\end{align}
involving minus signs for the frequencies 
$\omega_2$ and $\omega_4$, related to creation operators. 
We will use these new variables in the discussion of our results.
(An exception is the model for x-ray absorption
for which different operators $\vec{\Oc}$ are used; see Sec.~\ref{sec:Models} for details.)
However, for brevity, we retain the $\omega_i$ notation throughout this section.

The $4$p correlators can be decomposed into a connected (con) and disconnected (dis) part,
$G = G^{\mathrm{con}} + G^{\mathrm{dis}}$.
The latter corresponds to the independent
propagation of two particles and reads in the \MF/ and \ZF/, respectively, 
\begin{align}
G^{\mathrm{dis}}_{\sigma\sigma'}(\mi\omega_1,\mi\omega_2,\mi\omega_3)
& \!=\!
\beta
G(\mi\omega_1) G(\mi\omega_3)
(
\delta_{\sigma,\sigma'}
\delta_{\omega_{23},0} 
\!-\!
\delta_{\omega_{12},0}
) 
,
\nonumber
\\
G^{\mathrm{dis}}_{\sigma\sigma'}(\omega_1,\omega_2,\omega_3)
& \!=\!
2\pi\mi 
G(\omega_1) G(\omega_3)
[
\delta_{\sigma,\sigma'}
\delta(\omega_{23})
\!-\!
\delta(\omega_{12})
] 
.
\label{eq:disc_part}
\end{align}
By contrast, the connected part focuses on the mutual interaction. 
Yet, one can still factor out the propagation of each particle to and from the ``scattering event''
(``external legs'' in a diagrammatic language, see \Fig{fig:four-leg_diagram}).
This yields the effective interaction, i.e.\ the (full) 
$4$p vertex,
\begin{flalign}
F_{\sigma\sigma'}(\mi\omega_1,\mi\omega_2,\mi\omega_3)
& \!=\!
\frac{
G^{\mathrm{con}}_{\sigma\sigma'}(\mi\omega_1,\mi\omega_2,\mi\omega_3)
}{
G(\mi\omega_1) G(-\mi\omega_2) G(\mi\omega_3) G(-\mi\omega_4)
}
,
\hspace{-0.5cm} & 
\label{eq:vertex_part}
\end{flalign}
where $\omega_4=-\omega_{123}$,
and analogously in the \ZF/. 
Hence, once the disconnected part is removed---%
which can also be done on the level of the PSFs, see \Eq{eq:Sc_t-dis}---%
the vertex follows from a simple division of numbers.

In the \KF/, each operator is placed on either the forward or backward branch, according to its contour index $c_i \in \{ \pm \}$.
Thereby, an $\ell$p function acquires $2^\ell$ components. 
The Keldysh rotation exploits the fact that not all of these components are independent. 
For $\ell=2$, the resulting matrix structure in Keldysh indices $k_i \in \{1, 2 \}$ is
\begin{align}
G(\omega) =
\bigg( 
\begin{array}{cc}
0 & G^A \\
G^R & G^K
\end{array}
\bigg)
(\omega)
\label{eq:G2_Keldysh_matrix}
\end{align} 
in terms of the retarded ($R$), advanced ($A$), and Keldysh ($K$) component.

The matrix structure naturally carries over to \Eq{eq:disc_part}.
Because of $G^{11}=0$, the right hand side vanishes if the $4$p function has only one Keldysh index equal $2$;
i.e., the fully retarded components $G\sseta$ have no disconnected part.
The translation between connected correlator and vertex 
from \Eq{eq:vertex_part} now involves matrix multiplications, i.e.,
\begin{align}
& G_{\sigma\sigma'}^{\mathrm{con};k_1 k_2 k_3 k_4}(\omega_1,\omega_2,\omega_3)
=
G^{k_1 k_1'}(\omega_1)
G^{k_3 k_3'}(\omega_3)
\nonumber
\\
& \ \times
F_{\sigma\sigma'}^{k_1' k_2' k_3' k_4'}(\omega_1,\omega_2,\omega_3)
G^{k_2' k_2}(-\omega_2)
G^{k_4' k_4}(-\omega_4)
\label{eq:Keldysh_vertex_part}
\end{align}
with summation over $k_i'$.
One thus gets $F_{\sigma\sigma'}^{\vec{k}}$ from 
$G_{\sigma\sigma'}^{\mathrm{con};\vec{k}}$
by multiplying matrix inverses of the propagator~\eqref{eq:G2_Keldysh_matrix}.
With reference to \Fig{fig:four-leg_diagram},
the right Keldysh index $k_1'$ of $G^{k_1 k_1'}(\omega_1)$ 
corresponds to a creation operator and marks the beginning of the propagator line;
the left one, $k_1$, corresponds to an annihilation operator and marks the end of the propagator line.
Using $G^{1111}=0$, $G^{11}=0$ [cf.\ \Eq{eq:Kc_KF_Keldysh_via_Keta}] 
in \Eq{eq:Keldysh_vertex_part} directly implies $F^{2222}=0$.
One further finds that a vertex with only one Keldysh index equal $1$
in slot $\eta$, dubbed $F\sseta$,
is directly proportional to $G\sseta$.
Hence, we call them (fully) retarded as well.
Indeed, in these cases, only retarded or advanced propagators with $k_i \neq k_i'$ contribute to \Eq{eq:Keldysh_vertex_part}.

When using \Eqs{eq:Keldysh_vertex_part} to numerically extract $F$ from $G^{\mathrm{con}}$ by dividing out the external legs, 
the \textit{same} 
imaginary frequency shifts must be used for the external-leg $2$p correlators on the right as for the $4$p correlator 
$G^{\mathrm{con}}$ on the left. 
In App.~\ref{app:explicit_amputation}, we explain how this can be achieved.

\subsection{Models}
\label{sec:Models}

We compute local $4$p vertices for three impurity models.
The first describes x-ray absorption in metals, the second 
is the symmetric Anderson impurity model (AIM), and the third 
is a self-consistent AIM for the one-band Hubbard model (HM) in dynamical mean-field theory (DMFT). 
 
For x-ray absorption, we consider the following Hamiltonian,
to be called Mahan impurity model (MIM) \cite{Mahan1967} 
(mimicking the nomenclature customary for the AIM):
\begin{align}
\textstyle
\Hc_{\textrm{MIM}}
= 
\sum_{\epsilon} \epsilon \, c_{\epsilon}^\dag c_\epsilon 
+ |\epsilon_p | \, p p^\dagger
- U c^\dag c \, p p^\dag 
,
\end{align}
where $c \!=\! \sum_{\epsilon} c_\epsilon$. 
The first term describes a conduction band of spinless electrons with flat density of states $(1/2D) \theta (D - |\epsilon|)$ and half-bandwidth $D$, the second a localized core level with $\epsilon_p \!\ll\! -D$, filled in thermal equilibrium. 
An x-ray absorption process, described by $c^\dagger p$, 
transfers an electron from the core level into the conduction band, 
thereby turning on a local attractive scattering potential $-U < 0$ with $U \ll |\epsilon_p|$, described by the third term.
We define the absorption threshold $\omega_{\mathrm{th}}$
as the difference between the ground-state energies
of the subspaces with or without a core hole
($\omega_{\mathrm{th}}$ is of order $|\epsilon_p|$, but slightly smaller, since the hole-bath interaction is attractive). 
The x-ray absorption rate 
at an energy $\omega$ relative to the threshold
is given by the imaginary part of the particle-hole susceptibility,
the $2$p correlator
$\chi(\omega) \!=\! G[p^\dag c, c^\dag p ] (\omega + \omega_{\mathrm{th}})$.
The corresponding $4$p correlator is $G[p^\dag\!,c,c^\dag\! ,p](\nu, \nu',\omega)$, 
where, in the present context,
the definition~\eqref{eq:freq_ph_representation} of these frequencies
is replaced by 
\begin{align}
&
\vec{\omega} = (\nu + \omega_{\mathrm{th}}, 
-\nu\! + \! \omega, \nu'\!- \! \omega, -\nu' - \omega_{\mathrm{th}}) .
\label{eq:freq_ph_representation-XES}
\end{align}
For the arguments equated to $\omega_1$ and $\omega_4 = - \omega_{123}$,
associated with the operators $p^\dagger$ and $p$ switching from the no-hole to 
the one-hole subspace and back, we split off $\pm \omega_{\mathrm{th}}$, 
the energy differences
$E_{\ub{2}\ub{1}}$ and $E_{\ub{1}\ub{4}} $ associated with the
 transitions $\langle \ub{1} |p^\dagger| \ub{2} \rangle$ and $\langle \ub{4} |p| \ub{1} \rangle$ [cf.~\Eq{eq:S_w_p=id}]
between subspace ground states. 
Furthermore, the bosonic frequency $\omega$ is chosen to have
opposite sign compared to \Eq{eq:freq_ph_representation}, ensuring that 
$\omega_{12} \!=\! -\omega_{34} \!=\! \omega \!+\! \omega_{\mathrm{th}}$ 
matches the argument of the susceptibility 
$G[p^\dag c, c^\dag p ] (\omega + \omega_{\mathrm{th}})$.
 
The AIM is described by the Hamiltonian
\begin{align}
\nonumber
\Hc_\textrm{AIM} 
& = 
\sum_{\epsilon \sigma} \epsilon \, 
c_{\epsilon \sigma}^\dag c_{\epsilon \sigma}^\pdag 
+ 
\sum_\sigma \epsilon_d \, d^\dagger_\sigma d_\sigma^\pdag
+
U d_{\uparrow}^\dag d_{\uparrow}^\pdag d_{\downarrow}^\dag d_{\downarrow}^\pdag 
\\
\label{eq:AIM-define}
& 
\quad + 
\sum_{\epsilon} (V_\epsilon d^\dagger_\sigma c_{\epsilon \sigma}^\pdag + \textrm{H.c.}) 
.
\end{align}
It contains a band of spinful electrons,
a local level with energy $\epsilon_d$ and Coulomb repulsion $U$,
and a hybridization term, fully characterized by the hybridization function 
$\Delta (\nu) \!=\! \sum_{\epsilon} \pi |V_\epsilon|^2 \delta(\nu - \epsilon)$.
We take $\epsilon_d \!=\! -U/2$ 
and choose 
$\Delta (\nu) = \Delta \, \theta(D \!-\! |\nu|)$
box shaped.
The local density of states is given by the standard spectral function 
$S_\textrm{std}(\nu)$ associated with the $2$p correlator
$G[d_\sigma^\pdag,d_\sigma^\dag]$; 
the vertex $F_{\sigma\sigma'}$ 
follows from the $4$p correlator 
$G[d_\sigma^\pdag,d_\sigma^\dag, d_{\sigma'}^\pdag,d_{\sigma'}^\dag]$.

The one-band HM is a lattice model with
\begin{align}
\Hc_\textrm{HM} 
& = 
- t
\sum_{\langle i j \rangle \sigma} 
c_{i \sigma}^\dag c_{j \sigma}^\pdag 
+
U \sum_i 
c_{i \uparrow}^\dag c_{i \uparrow}^\pdag c_{i \downarrow}^\dag c_{i \downarrow}^\pdag 
,
\end{align}
where $\langle i j \rangle$ enumerate nearest-neighbor lattice sites,
$t$ is the hopping amplitude and $U$ the interaction strength.
In DMFT, the HM is mapped onto a self-consistently determined AIM \cite{Georges1996}.
The associated impurity degrees of freedom $d_{\sigma}^{(\dag)}$ experience 
the same local interaction $U$, while their coupling to the rest of the lattice is encoded in the hybridization function $\Delta(\nu)$. 
In this paper, we consider the Bethe lattice with infinite coordination number,
which yields a semicircular lattice density of states of half-bandwidth $D$ ($\propto\! t$),
and paramagnetic, spatially uniform phases.
Then, the DMFT self-consistency condition simply reads
$\Delta(\nu) \!=\! (D/2)^2 \pi S_{\mathrm{std}}(\nu)$.
Upon self-consistency, local correlators of the HM can be found 
from the corresponding AIM and thus analyzed in direct analogy.

\subsection{Analytic benchmarks}
\label{sec:benchmarks}

The MIM was thoroughly investigated by Nozi\`{e}res and collaborators in the \ZF/
\cite{Roulet1969,Nozieres1969a,Nozieres1969b}.
The particle-hole susceptibility has a power-law divergence
for frequencies above the threshold,
the celebrated x-ray--edge singularity \cite{Mahan1967,Ohtaka1990},
solved analytically \cite{Nozieres1969b} as
\begin{align}
\mathrm{Im} \, \chi(\omega) \propto \omega^{-\alpha}
, \qquad
\alpha = 2 \, \delta/\pi - (\delta/\pi)^2
.
\label{eq:XES_analytic}
\end{align}
Here, $\omega$ is the absorption frequency above the threshold $\omega_{\mathrm{th}}$,
and $\delta$ is the conduction-electron phase shift 
induced by the core-hole scattering potential.
It has the analytic expression \cite{Nozieres1969b}
$\delta \!=\! \arctan(\pi g)$,
with $g \!=\! U/(2D)$ here.
It can also be computed with NRG through $\delta \!=\! \pi \Delta_\mathrm{h}$,
where $\Delta_\mathrm{h}$ is the charge drawn in toward the scattering site by the core hole \cite{Weichselbaum2011,Muender2012}.

In the first paper of the series \cite{Roulet1969}, 
the ($2$p) susceptibility $\chi$ was deduced from the (full) $4$p vertex.
The latter actually contains a variety of power laws, 
as summarized by Eq.~(35) of Ref.~\onlinecite{Roulet1969}.
Plus, one can extract $\chi$, and thus the same power law~\eqref{eq:XES_analytic}, 
from the vertex by sending suitable frequencies to infinity.
To this end, we
consider the vertex $F$, 
related to the $4$p correlator $G[p^\dag,c,c^\dag,p]$,
whose bare part is $F_0 \!=\! - U$.
In the particle-hole representation of frequencies~\eqref{eq:freq_ph_representation},
$\chi$ then follows as \cite{Kunes2011,Wentzell2020,Kugler2018a} 
\footnote{
This relation can be seen in the following results Ref.~\onlinecite{Roulet1969}
(although not explicitly mentioned there):
In their Eq.~(35), $t_1$, $t_2$, $\alpha$, $\beta$ are logarithmic variables corresponding to
our $\nu$, $\nu'$, $\nu \!+\! \nu' \!-\! \omega$, $\omega$,
respectively.
One has, e.g., $t_1 \!=\! \ln \xi_0/\nu$ for $|\nu| \!\ll\! \xi_0$
and $t_1 \!=\! 0$ for $|\nu| \!\gg\! \xi_0$,
where $\xi_0$ is a cutoff.
Now, for $\nu^{(\prime)} \!\to\! \infty$, $t_{1,2} \!\to\! 0$
as well as $\alpha \!\to\! 0$, while
$\beta$ is large for small $\omega$.
Hence, in Eq.~(35) of Ref.~\onlinecite{Roulet1969},
$\gamma_1$ (particle-particle reducible) vanishes,
while $\gamma_2$ (particle-hole reducible)
reproduces $V^2\chi$ according to Eq.~(41).
Here, $V$ is the bare vertex, and the same relation
holds for the full vertex after subtracting $V$.}
\begin{align}
\textstyle
\chi(\omega) = \lim_{|\nu|,|\nu'|\to\infty} \, [ F(\nu,\nu',\omega) + U ] / U^2
.
\label{eq:XES_vertex_limit}
\end{align}
The limit of fermionic frequencies must be taken such that $|\nu|$, $|\nu'|$, 
\textit{and} $|\nu \pm \nu'|$ become arbitrarily large \cite{Wentzell2020}. 
In this limit, 
$ \mathrm{Im} \, F / U^2 = \mathrm{Im} \, \chi(\omega) \propto \omega^{-\alpha}$.

Analytic results are also available for the half filled AIM in the \MF/ 
in the limits of either weak or infinitely strong interaction.
In the former, one often simplifies 
$D \!\to\! \infty$ and $T \!=\! 0$
to find the bare particle-hole susceptibility \cite{Yosida1970}
\begin{align}
\chi_0(\mi\omega)
& =
\frac{2\Delta}{\pi|\omega|(|\omega|+2\Delta)}
\ln \frac{|\omega|+\Delta}{\Delta} 
.
\label{eq:bubble_IF}
\end{align}
Its particle-particle counterpart yields $-\chi_0$.
Thus, combining all three two-particle channels, the vertex in second-order perturbation theory follows as
(using $\bar{\uparrow} \! = \, \downarrow$, $\bar{\downarrow} \! = \, \uparrow$)
\begin{align}
F_{\sigma\sigma'}
& =
U
\delta_{\sigma,\bar{\sigma}'}
-
U^2
\nonumber
\\
& \ \ \times
[
\delta_{\sigma,\sigma'}
\chi_0(\mi\omega_{12})
+
\delta_{\sigma,\bar{\sigma}'}
\chi_0(\mi\omega_{13})
-
\chi_0(\mi\omega_{14})
]
.
\label{eq:vertex_SOPT}
\end{align}
Here, the first term $U \delta_{\sigma,\bar{\sigma}'}$ on the right is the \MF/ bare vertex $F_{0;\sigma\sigma'}$.
An expression analogous to Eq.~\eqref{eq:vertex_SOPT} holds for the local vertex of the weakly interacting HM.
In that case, Eq.~\eqref{eq:bubble_IF} must be computed for the appropriate
hybridization function $\Delta(\nu)$,
with a nontrivial frequency dependence resulting from the self-consistency condition.

In the opposite limit 
where $U/\Delta \!\to\! \infty$
in the AIM
(corresponding to $U/t \!\to\! \infty$ in the HM)
realizing the Anderson or Hubbard atom (HA),
the \MF/ vertex is known, too
\cite{Hafermann2009,Pairault2000,Rohringer2012,Wentzell2020}. 
In compact notation, we have
\begin{subequations}
\label{eq:vertex_AL}
\begin{align}
& F_{\uparrow\downarrow}
=
2u
+
\frac{u^3\sum_i(\mi\omega_i)^2}{\prod_i (\mi\omega_i)}
-
\frac{6u^5}{\prod_i (\mi\omega_i)}
\\
& \ +
\beta u^2
[
\delta_{\omega_{12}} \mathrm{th}
\!+\!
\delta_{\omega_{13}} (\mathrm{th}\!-\!1)
\!+\!
\delta_{\omega_{14}} (\mathrm{th}\!+\!1)
]
\frac{\prod_i (\mi\omega_i+u)}{\prod_i (\mi\omega_i)}
,
\nonumber
\\
& F_{\uparrow\uparrow}
=
\beta u^2
(
\delta_{\omega_{14}}
-
\delta_{\omega_{12}}
)
\frac{\prod_i (\mi\omega_i+u)}{\prod_i (\mi\omega_i)}
,
\end{align}
\end{subequations}
with
$u \!=\! U/2$,
$\delta_\omega \!=\! \delta_{\omega,0}$,
$\mathrm{th} \!=\! \tanh \beta u /2$,
and
$i \!\in\! \{1,2,3,4\}$.
Generally, $F_{\uparrow\uparrow}$ follows from $F_{\uparrow\downarrow}$ by SU(2) spin and crossing symmetry \cite{Rohringer2012}:
$
F_{\uparrow\uparrow}(\mi\vec{\omega}) 
\!=\! 
F_{\uparrow\downarrow}(\mi\vec{\omega}) 
\!-\! 
F_{\uparrow\downarrow}(\mi\vec{\omega}')
$,
where $\vec{\omega}'$ relates to $\vec{\omega}$ by exchanging either $\omega_1 \!\leftrightarrow\! \omega_3$ or $\omega_2 \!\leftrightarrow\! \omega_4$.

In the \KF/, the retarded vertex can be deduced from the analytic continuation
\eqref{eq:analytic_continuation_IF_to_fully_retarded}:
$
2F_{\sigma\sigma'}\sseta(\vec{\omega})
\!=\!
F_{\sigma\sigma'}(\mi\vec{\omega})
\big|_{\mi\vec{\omega} \to \vec{\omega}^{[\eta]}} 
$.
For \Eq{eq:vertex_SOPT}, this is easily verified using standard Keldysh diagrammatic techniques.
For \Eqs{eq:vertex_AL}, it has recently been shown explicitly \cite{Ge2020a}.
The last two lines of \Eqs{eq:vertex_AL} consist of anomalous terms (proportional to Kronecker $\delta$'s) and cannot be analytically continued to retarded components. However, the real-frequency anomalous contributions (proportional to Dirac $\delta$'s) are contained in other Keldysh components, as further discussed in App.~\ref{sec:app:HA_2ndOrder}.
The bare vertex $F^{k_1k_2k_3k_4}_{0;\sigma\sigma'}$
equals $\tfrac{1}{2} U \delta_{\sigma,\bar{\sigma}'}$
if $k_{1234}$ is odd and zero otherwise.
This can be seen starting from the contour basis \cite{Jakobs2009,Jakobs2010a},
where the bare interaction requires all operators to be on the same branch
and comes with a minus sign on the backward branch.
A fourfold Keldysh rotation then generates the prefactor
$[1 - (-1)^{k_{1\cdots 4}}]/4$.

\section{Numerical vertex results}
\label{sec:results}

In the following, we present results obtained by numerically computing local $4$p functions with NRG.
Generally, NRG allows one to construct a complete basis of approximate eigenstates of the Hamiltonian \cite{Anders2005,Anders2006}
and thus directly evaluate the spectral representations 
(see Refs.~\onlinecite{Weichselbaum2007,Peters2006} for computing retarded $2$p functions).
In the accompanying paper \cite{Lee2021}, we develop a new NRG scheme 
to treat $3$p and $4$p functions.
We refer interested readers to Secs.~IV--V of that paper for how PSFs are computed with NRG as sums of discrete $\delta$ peaks, and to Sec.~VI for how 
these are broadened to smooth curves and how connected correlators and the full vertex are obtained from them.
Here, we show the final results of the $4$p vertex, 
for the MIM in the \ZF/ and for the AIM, with both 
a boxed-shaped and a DMFT self-consistent hybridization, in the \MF/ and \KF/. 

\subsection{\ZF/ for MIM: Power laws}
\label{eq:ZF-results-powerlaw-vertex}

\begin{figure}
\includegraphics[width=\linewidth]{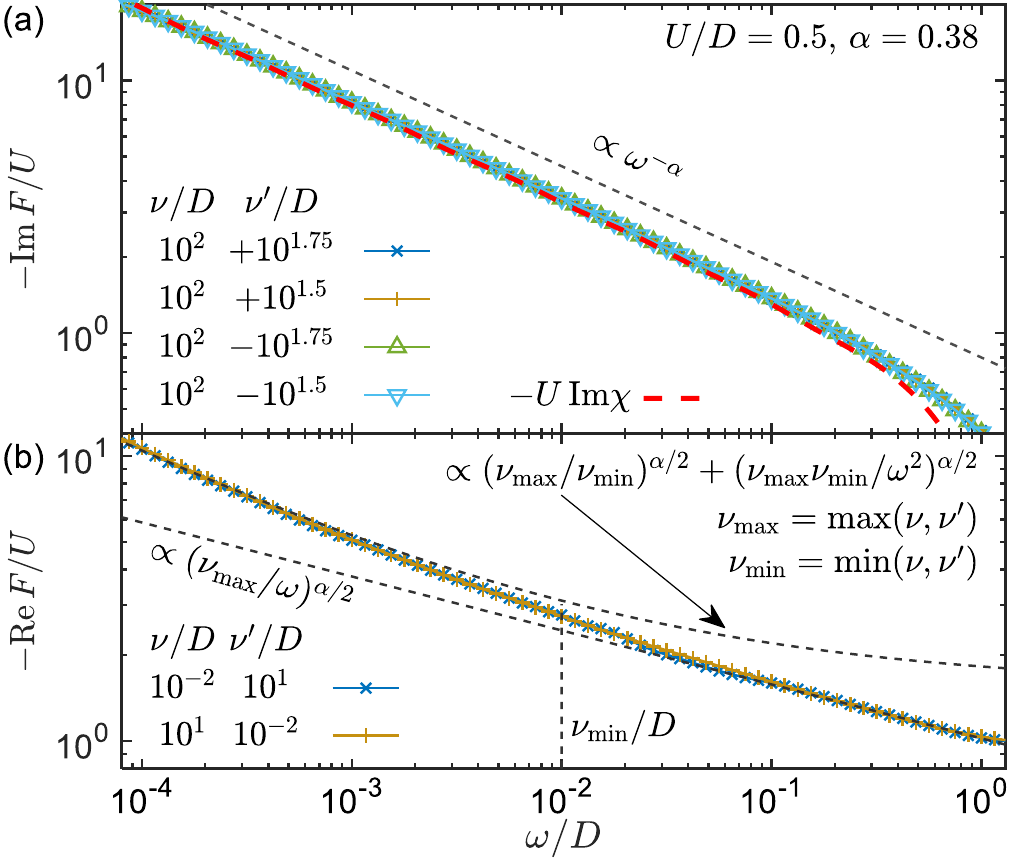}
\caption{%
The \ZF/ $4$p vertex $-F/U$ of the MIM, 
plotted as a function of the bosonic transfer frequency $\omega$ 
for fixed fermionic frequencies $\nu$ and $\nu'$ [cf.~\Eq{eq:freq_ph_representation-XES}].
(a) The imaginary part of $F$ at $\omega \!>\! 0$.
For $|\nu|, |\nu'|, |\nu \!\pm\! \nu'| \!\gg\! D$, 
all results collapse onto a single curve.
This curve follows a power law $\omega^{-\alpha}$ 
matching the ($2$p) susceptibility $\chi$ [\Eq{eq:XES_analytic}],
which can be independently computed by NRG (red dashed curve).
(b) The real part of $F$ at $\omega \!<\! 0$.
If only one of $\nu$ and $\nu'$ is large, $F$ follows two distinct scaling behaviors for $\nu_{\min} \!<\! -\omega \!<\! \nu_{\max}$ and $-\omega \!<\! \nu_{\min}$, 
given, respectively, by Eqs.~(36) and (40) of Ref.~\onlinecite{Roulet1969}.}
\label{fig:RF_XEM} 
\end{figure}

The MIM is a prototypical model for the \ZF/;
see e.g.\ Refs.~\onlinecite{Roulet1969,Nozieres1969a,Nozieres1969b}.
For this model, NRG proved very successful in computing power-law divergences of $2$p functions
\cite{Oliveira1981,Costi1997,Anders2005,Bulla2008,Weichselbaum2011,Muender2012},
but it was never used to investigate similar singularities in $4$p functions.
As explained in Sec.~\ref{sec:quantities}, we can extract the famous power law 
$\omega^{-\alpha}$ in terms of the (bosonic) transfer frequency
of the particle-hole susceptibility $\chi$ 
directly from the $4$p vertex $F$ by setting the fermionic frequencies $\nu$, $\nu'$ to very large values.
Figure~\ref{fig:RF_XEM}(a) shows various cross sections of $\mathrm{Im} \, F(\nu,\nu',\omega)$, where $|\nu|$, $|\nu'|$, and $|\nu \!\pm\! \nu'|$ are much larger than the half-bandwidth $D$. 
All of them collapse onto the same curve.

This curve meets two consistency checks. First, as expected from \Eq{eq:XES_vertex_limit}, it matches $\mathrm{Im} \, \chi(\omega)$ (red dashed line), computed separately as a $2$p correlator. 
The discrepancy at $\omega \gtrsim U$ is due to the broadening of discrete PSFs (described in Ref.~\onlinecite{Lee2021}, Sec.~VI.B); it can be removed by reducing broadening, but then wiggly discretization artifacts appear. 
Second, as expected from Eq.~\eqref{eq:XES_analytic}, 
\Fig{fig:RF_XEM}(a) shows a power-law divergence for $\omega$ close to the threshold.
The exponent $\alpha \!=\! 0.38$ in $\omega^{-\alpha}$ (black dashed line, guide to the eye)
was obtained from \Eq{eq:XES_analytic} using the phase shift $\delta \!=\! \arctan(\pi g) = 0.67$.
We obtained the same value,
$\delta \!=\! \pi \Delta_\mathrm{h} \!= \! 0.67$, 
when computing the $\Delta_\mathrm{h}$, the charge drawn in to the
core hole, with NRG following Ref.~\onlinecite{Weichselbaum2011}. 

Next, we probe further power laws in the vertex $F$ by setting
one fermionic frequency to a large value, $\nu_{\mathrm{max}}$,
and the other one to a small value, $\nu_{\mathrm{min}}$.
As $|\omega|$ is reduced from $|\omega| \!>\! \nu_{\mathrm{min}}$ 
to $|\omega| \!<\! \nu_{\mathrm{min}}$,
$F$ crosses over between two power laws, given by 
Eqs.~(36) and (40) of Ref.~\onlinecite{Roulet1969}, respectively.
Both of them are very well reproduced by our NRG results in \Fig{fig:RF_XEM}(b).
The analytic power laws are shown as dashed lines.
Their prefactors are 0.66 and 0.42, respectively,
in reasonable agreement with the predictions 1 and 0.5,
obtained in logarithmic accuracy in Ref.~\onlinecite{Roulet1969}.

These consistency checks, with matching results
for highly nontrivial $4$p and $2$p functions, on the one hand,
and agreement between numerical $4$p results and analytic predictions, on the other hand, 
provide confidence that NRG is well suited to compute local $4$p functions in the \ZF/. 
Moreover, it successfully meets the particularly tough challenge of the regimes 
$\omega \ll |\nu| \simeq |\nu'|$ and 
$\omega \ll |\nu| \ll |\nu'|$, 
namely resolving a small frequency with exponential accuracy 
while also keeping track of two larger ones.

\subsection{\texorpdfstring{\MF/}{MF} for AIM: Temperature evolution}

\begin{figure}
\includegraphics[width=\linewidth]{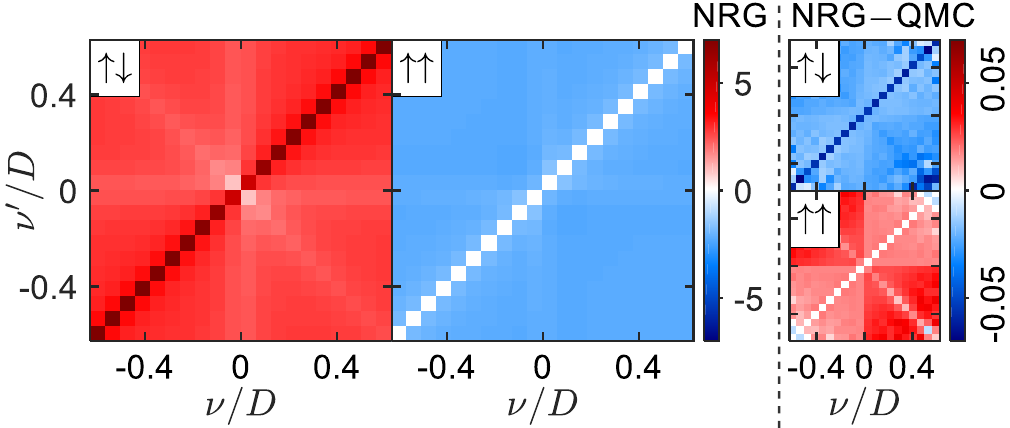}
\caption{%
\MF/ $4$p vertex $F_{\sigma \sigma'} (\mi\vec{\omega})/U$ of the AIM
as a function of $\nu$ and $\nu'$ at $\omega \!=\! 0$
and a moderately low temperature $T \!=\! 10^{-2} D$.
The labels $\uparrow\uparrow$ and $\uparrow\downarrow$ indicate the spin indices $\sigma \sigma'$.
The left panels show NRG results $F^{\mathrm{NRG}}_{\sigma\sigma'}(\mi\vec{\omega})/U$.
The right panel shows their difference to QMC data,
$(F_{\sigma\sigma'}^{\mathrm{NRG}} \!-\! F_{\sigma\sigma'}^{\mathrm{QMC}})/U$,
which is two orders of magnitude smaller than the original signal.}
\label{fig:IF_NRG_QMC} 
\end{figure}

Most numerical work involving nonperturbative $4$p functions is obtained in the \MF/, where,
thanks to the steady progress in quantum Monte Carlo (QMC) techniques,
local $4$p functions can nowadays be computed with high precision
(see \cite{Rohringer2018} for a recent list of references).
Using the spectral representation as given by \Eqs{eq:G_iw_KS_IF}, 
\eqref{eq:S_w}, \eqref{eq:K_Omega_IF},
they can be computed with NRG, too.
For the next parts, we consider the half-filled AIM with box-shaped hybridization 
and large interaction $U/\Delta \!=\! 5$, where $U/D \!=\! 1/5$,
focusing on the (full) $4$p vertex 
$F_{\sigma\sigma'}$ (cf.\ Sec.~\ref{sec:quantities}).

We start with a moderately low temperature $T \!=\! 10^{-2} D$.
Figure~\ref{fig:IF_NRG_QMC} shows our \MF/ NRG results for the 
two spin components of $F_{\sigma\sigma'}$
as a function of $\nu$ and $\nu'$ at $\omega \!=\! 0$
in the particle-hole representation~\eqref{eq:freq_ph_representation}.
One can still see the discrete nature of the Matsubara frequencies
$\mi\nu^{(\prime)} \!\in\! \mi\pi T (2\mathbb{Z} \!+\! 1)$.
Furthermore, one observes the typical structure in the frequency dependence of the full vertex \cite{Rohringer2012,Wentzell2020,Li2016,Rohringer2018}
composed of a background value (independent of $\nu^{(\prime)}$), 
a distinct signal on the diagonal and antidiagonal (though weaker),
and a plus-shaped feature ($\nu^{(\prime)} \!=\! \pm\pi T$)
(the latter is more pronounced at lower $T$).
Note that $F_{\uparrow\uparrow}$ vanishes identically for $\nu \!=\! \nu'$.
This is intuitive from the Pauli principle since, then, all quantum numbers of both fermions involved match.
It also follows from the symmetry relation mentioned below \Eq{eq:vertex_AL},
since exchanging either $\omega_1 \!\leftrightarrow\! \omega_3$ or $\omega_2 \!\leftrightarrow\! \omega_4$ at $\omega \!=\! 0$ leaves the diagonal ($\nu \!=\! \nu'$) invariant.
At a temperature $T \!=\! 10^{-2} D$, we can compare our results to highly accurate QMC data
\cite{Chalupa2021,Chalupa2018,Wallerberger2019}.
We find that the results differ on the level of 1\%,
which confirms the reliability of our new NRG scheme at moderately low temperatures.

\begin{figure}
\includegraphics[width=\linewidth]{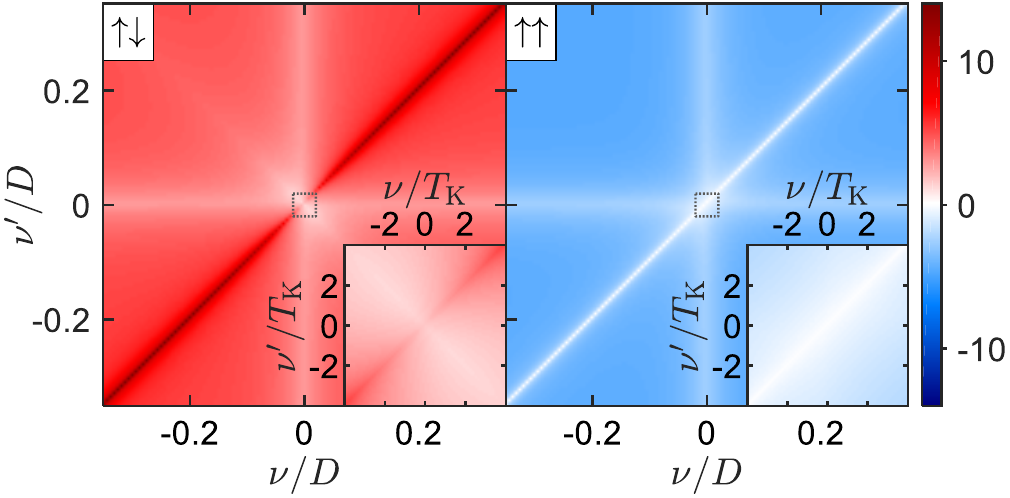}
\caption{%
$F_{\sigma\sigma'}(\mi\vec{\omega})/U$ computed by NRG,
analogous to \Fig{fig:IF_NRG_QMC},
but at a much lower temperature $T \!=\! 10^{-4} D$.
The inset enlarges the low-energy window marked by the small square.
For $\nu^{(\prime)} \!\lesssim\! \TK$, the vertex is strongly suppressed, giving rise to the Fermi-liquid regime of weakly interacting quasiparticles.}
\label{fig:IF_NRG_lowT} 
\end{figure}

Typical QMC algorithms scale unfavorably with inverse temperature. 
An important advantage of our \MF/ NRG scheme is hence that it extends to arbitrarily low $T$.
For the given parameters, the Kondo temperature is
$\TK \!\simeq\! 5 \!\times\! 10^{-3} D$
\footnote{We use the standard formula for the Kondo temperature,
$\TK = 0.4107 \, (U\Delta/2)^{1/2} \exp( - \tfrac{\pi U}{8\Delta} + \tfrac{\pi\Delta}{2U} )$;
see Eq.~(6.109) and the following text in Ref.~\onlinecite{Hewson1993}.}.
\nocite{Hewson1993}
Accordingly, $T \!=\! 10^{-2} D$, as used above, is not low enough to enter the strong-coupling regime,
but $T \!=\! 10^{-4} D$, used for \Fig{fig:IF_NRG_lowT}, is. 
There, we find that the features already observed in \Fig{fig:IF_NRG_QMC} become sharper and more pronounced.
Particularly interesting is the region $\nu^{(\prime)} \!\lesssim\! \TK$,
describing the Fermi liquid with an impurity screened by the Kondo cloud.
Indeed, the inset in \Fig{fig:IF_NRG_lowT} shows that the vertex is strongly reduced in this regime,
thus giving way to weakly interacting quasiparticles.

To explore this concept further,
renormalized perturbation theory (RPT) offers a way of extracting the quasiparticle interaction
(local Landau parameter) directly from the NRG low-energy spectrum \cite{Hewson1993a,Hewson2004,Pandis2015,Kugler2020}.
This proceeds by comparing the eigenenergies of states with two excited quasiparticles to those with only one;
it does not require knowledge of frequency-dependent correlation functions.
The resulting values $\tilde{U}_{\sigma\sigma'} \!=\! \tilde{U} \delta_{\sigma,\bar{\sigma}'}$ should match
the low-energy limit of the effective interaction 
(i.e.\ the $4$p vertex $F_{\sigma\sigma'}$ with all frequencies sent to zero),
multiplied by the quasiparticle weight $Z$:
$\tilde{U}_{\sigma\sigma'} \!=\! Z^2 F_{\sigma\sigma'}(\mi\vec{\omega} \to \vec{0})$
\cite{Hewson2004}.

\begin{figure}
\includegraphics[width=\linewidth]{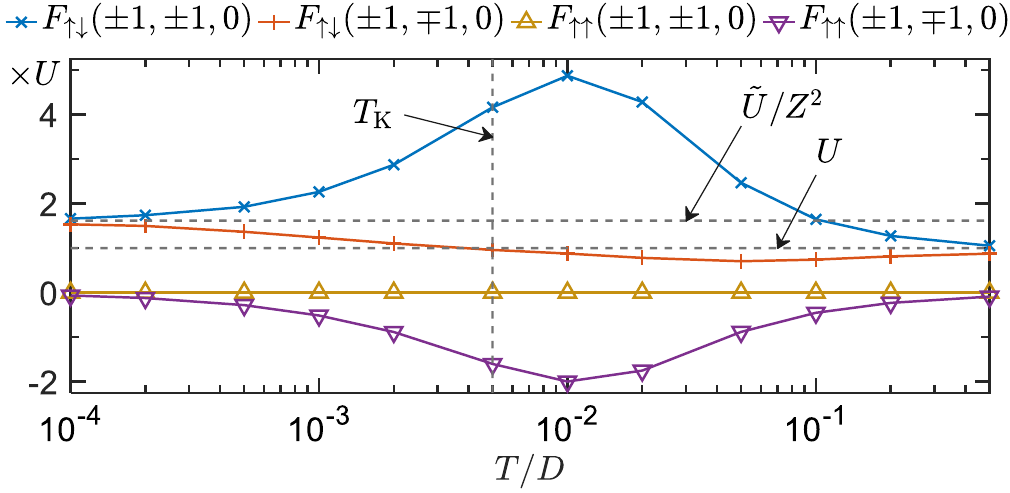}
\caption{%
The vertex evaluated at the lowest Matsuabra frequencies
$\nu^{(\prime)} \!=\! \pm \pi T$ and $\omega \!=\! 0$, denoted by
$F_{\sigma\sigma'}(\pm 1, \pm 1, 0)$,
as a function of decreasing temperature.
For $T \!\gg\! D$, the vertex reduces to the bare interaction
$F_{0;\sigma\sigma'} \!=\! U \delta_{\sigma,\bar{\sigma}'}$.
Upon lowering $T$, it exhibits strong renormalization effects and undergoes a crossover
from increasing to decreasing magnitude for $T \!>\! \TK$ and $T \!<\! \TK$, respectively.
For $T \!\ll\! \TK$, it converges to the
quasiparticle interaction 
$\tilde{U}_{\sigma\sigma'} \!=\! \tilde{U} \delta_{\sigma,\bar{\sigma}'}$, 
divided by twice the quasiparticle weight $Z$.
The parameters $\tilde{U} \!\simeq\! 0.20 U$ 
and $Z \!\simeq\! 0.36$ were found from RPT and NRG, 
independent of the $4$p computation,
and thus provide a strong consistency check.}
\label{fig:IF_NRG_RPT} 
\end{figure}

Figure \ref{fig:IF_NRG_RPT}
shows the vertex evaluated at the lowest Matsubara frequencies,
$\nu^{(\prime)} \!=\! \pm \pi T$, $\omega \!=\! 0$,
as a function of decreasing temperature.
At very large temperatures, $T \!\gg\! D$, correlation effects are suppressed,
and the vertex reduces to the bare interaction,
$F_{\sigma\sigma'} \!\to\! U \delta_{\sigma,\bar{\sigma}'}$.
As we lower temperature much below $D$,
there are strong renormalization effects, and particularly 
$F_{\uparrow\downarrow}$ for $\nu \!=\! \nu'$
and
$F_{\uparrow\uparrow}$ for $\nu \!=\! -\nu'$ grow in magnitude.
(Recall that $F_{\uparrow\uparrow}$ vanishes identically for $\nu \!=\! \nu'$.)
Now, for $T$ on the order of the Kondo temperature $\TK$,
this trend comes to a halt, and the low-energy components of the vertex start to decrease again.
This is the nonperturbative crossover from strongly interacting particles to weakly interacting quasiparticles, as one enters the Fermi-liquid regime.
Indeed, for temperatures sufficiently below $\TK$, we find that the low-energy vertex has precisely the same form as the bare vertex: 
it is nonzero only for different spins, with a value independent of the signs of the frequencies $\nu^{(\prime)} \!=\! \pm \pi T$.
This value precisely matches the RPT estimate,
$F_{\sigma\sigma'} \!\to\! \tilde{U}_{\sigma\sigma'}/Z^2$.
Vice versa, $\tilde{U}_{\sigma\sigma'}$ can be determined from $Z^2 F_{\sigma\sigma'}(\mi\vec{\omega} \to \vec{0}$).
Both the decrease of $F_{\sigma\sigma'}(\mi\vec{\omega})$ 
for $|\omega_i|, T \!<\! \TK$ 
and the small quasiparticle weight $Z \!\simeq\! 0.36$
lead to a small (but finite) quasiparticle interaction of 
$\tilde{U} \!\simeq\! 0.20 U$. %0.04 \, D.

To our best knowledge, comparing RPT to the low-energy limit of a nonperturbative vertex computation
has not been realized before.
It is a stringent consistency check for the underlying Fermi liquid 
and confirms the reliability of our NRG scheme down to very low temperatures.

\begin{figure}[b]
\includegraphics[width=\linewidth]{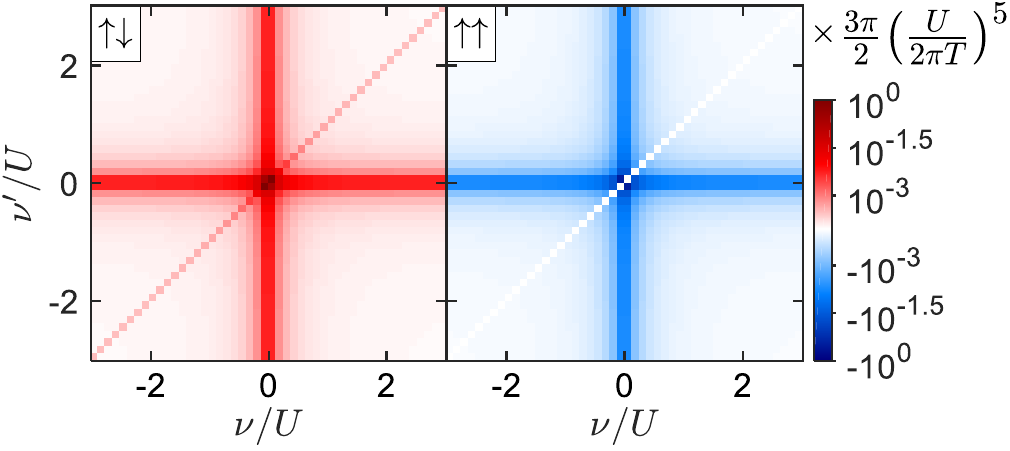}
\caption{%
\MF/ $4$p vertex $(F_{\sigma\sigma'} - F_{0;\sigma\sigma'})/U$ of the HA
(AIM at infinitely strong interaction, $\Delta/U \!=\! 0$),
at $\omega \!=\! 0$.
It is calculated from the analytic result \eqref{eq:vertex_AL} at a temperature $T/U \!=\! 1/50$;
the bare vertex $F_{0;\sigma\sigma'} \!=\! U \delta_{\sigma,\bar{\sigma}'}$ is subtracted for clarity.}
\label{fig:IF_HA} 
\end{figure}

\subsection{\KF/ for AIM: Benchmarks and strong coupling}

\begin{figure*}
\includegraphics[width=\linewidth]{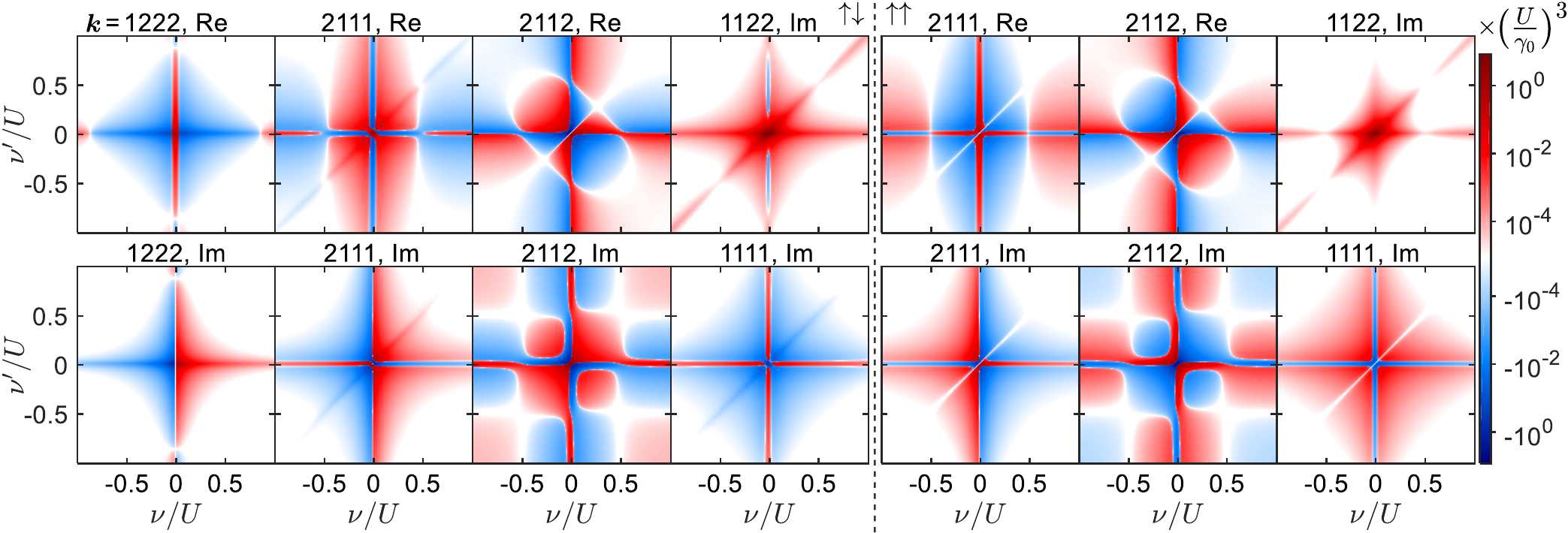}
\caption{%
\KF/ $4$p vertex 
$(F^{\vec{k}}_{\sigma\sigma'} \!-\! F^{\vec{k}}_{0;\sigma\sigma'})/U$ of the HA 
at $\omega \!=\! 0$ and $T/U \!=\! 1/50$ (using $\gamma_0 \!=\! T$).
The left (right) panels show $\sigma \! \neq \! \sigma'$ ($\sigma \! = \! \sigma'$),
the top (bottom) rows the real (imaginary) part, except for the purely imaginary 
$\vec{k} \!=\! 1122$ and $1111$ components.
Keldysh components not plotted look similar to those that are plotted:
components 
where $\vec{k}$ has only one $1$ are 
analogous to $1222$ 
(these vanish for $\sigma \! = \! \sigma'$);
those with only one $2$ are analogous to $2111$;
$1122$ corresponds to $2211$,
while other components with two $1$'s and two $2$'s are similar to $2112$.
The $2222$ vertex vanishes identically.}
\label{fig:KF_HA} 
\end{figure*}

To our best knowledge, nonperturbative results for \KF/ $4$p functions have not been obtained in the literature before.
Hence, we begin our analysis with two benchmark cases.
The first concerns the limit of infinitely strong interaction,
$\Delta/U \!=\! 0$, i.e., the atomic limit of the AIM or simply the Hubbard atom (HA).
In this case, the many-body basis consists of only four states,
and a NRG computation reduces to a simple, exact diagonalization.
The PSFs involve only a few $\delta$ singularities,
which, at half filling, are placed at multiples of $U/2$.
Any real-frequency correlation function directly inherits these singularities.
In numerical calculations, one sets a minimal imaginary part $\gamma_0$
in the spectral representation. 
As the denominators contain three factors of the type
$(\omega_i \!+\! \mi\gamma_0 \!-\! \omega_i')$,
the poles reach a magnitude of $\gamma_0^{-3}$.

The \MF/ vertex of the HA is well understood analytically \cite{Thunstroem2018},
see \Eq{eq:vertex_AL},
and reveals many features shared by general \MF/ vertices; see \Fig{fig:IF_HA}.
We now analyze the \KF/ vertex of the HA.
To this end, we compute the spectral representation 
involving 24 permutations, subtract the disconnected part, and amputate
the external legs (cf.\ Sec.~\ref{sec:quantities}).
As the calculation in this limit is \textit{exact}, these steps pose no further difficulties. 
The result are 16 Keldysh components, each having a real and imaginary part.
Figure~\ref{fig:KF_HA} shows the huge variety of features that can be observed
as a function of $\nu$ and $\nu'$ at $\omega \!=\! 0$ 
[in the particle-hole representation \eqref{eq:freq_ph_representation}].
We display five different Keldysh components;
all others, related by permutations of their Keldysh indices, 
show analogous features.
A very compact, analytic result for the fully retarded components $F\sseta$ can be deduced 
from the known Matsubara result \eqref{eq:vertex_AL} and 
the analytic continuation \eqref{eq:analytic_continuation_IF_to_fully_retarded}.
Our numerical results match those to floating-point precision.
As explained in Sec.~\ref{sec:benchmarks}, the typical diagonal features of the vertex
become $\delta$ functions in the atomic limit. They do not appear
in $F\sseta$ but in other Keldysh components; see \Fig{fig:KF_HA}.
For these other components, too, analytic results have recently been obtained \cite{Ge2020a},
and they perfectly match our numerical ones.

Next, we turn on the coupling to the bath.
As we now have a continuous spectrum, 
discrete PSFs, obtained from a finite number of terms in 
the sum of \Eq{eq:S_w}, must be \textit{broadened}.
The subtraction of the disconnected part $G^\mathrm{dis}_{\sigma\sigma'}$, 
which requires exact cancellations of large terms, 
is then numerically difficult. 
Note that, even if the retarded components $G\sseta$ 
do not have a disconnected part from an analytical perspective, 
this again relies on exact cancellations that are easily violated numerically.
To obtain the vertex via $G^\mathrm{con}_{\sigma\sigma'}$ most accurately, 
we employ a twofold strategy \cite{Lee2021}.
First, we compute $G^\mathrm{con}_{\sigma\sigma'}$ directly from 
$S^\mathrm{con} \!=\! S - S^\mathrm{dis}$
(as discussed at the end of Sec.~\ref{sec:RF}),
where $S^\mathrm{dis}$ is deduced
from the full $S$ through appropriate sum rules
and subtracted prior to broadening.
This eliminates the disconnected part to a large extent 
but, due to imperfect numerical accuracy, not completely.
Second, we use a Keldysh analog of the equation-of-motion method 
as presented in Ref.~\onlinecite{Hafermann2012}.
It is based on expressing $G^\mathrm{con}_{\sigma\sigma'}$ through
auxiliary correlators, obtained by differentiating $G_{\sigma\sigma'}$ w.r.t., say, the first time argument.
Clearly, the choice of this time argument introduces a bias.
While this does not appear important in \MF/ applications,
we found it to be highly relevant in the \KF/.
As a consequence, the connected part of only those Keldysh components that similarly single out
one time argument, i.e.\ $G\sseta$, could be reliably determined.
For other components, we expect a symmetric version of the equation of motion to be beneficial,
similar to the one presented in Ref.~\onlinecite{Kaufmann2019} for the \MF/.
This is, however, left for future work.

\begin{figure}[b]
\includegraphics[width=\linewidth]{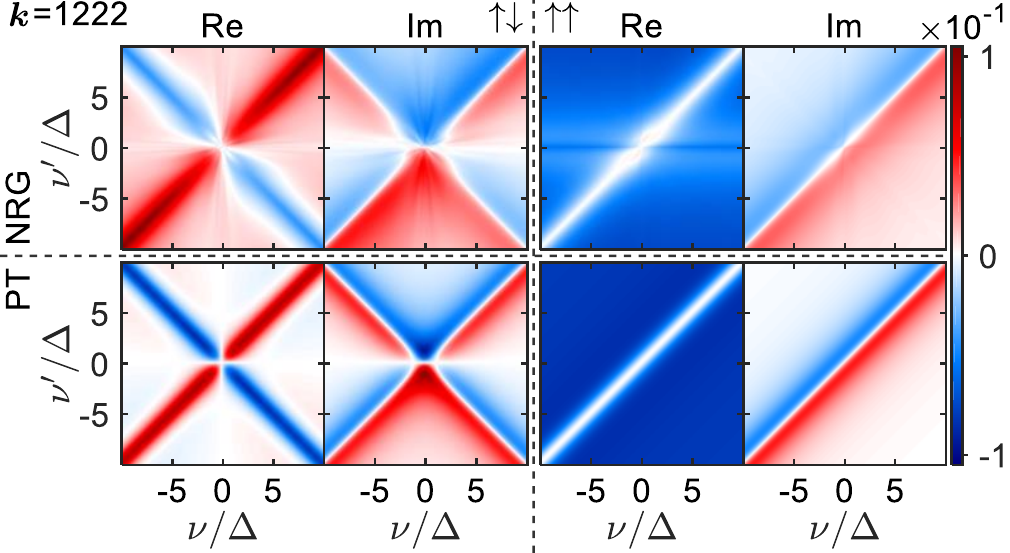}
\caption{%
\KF/ $4$p point vertex 
$(F^{\vec{k}}_{\sigma\sigma'} \!-\! F^{\vec{k}}_{0;\sigma\sigma'})/U$
at $\omega \!=\! 0$ and $\vec{k} \!=\! 1222$
in the AIM at weak interaction, 
$U/\Delta \!=\! 1/2$.
The left (right) panels show the real and imaginary parts for $\sigma \! \neq \! \sigma'$ ($\sigma \! = \! \sigma'$).
The top row shows NRG data,
with $U/D \!=\! 1/20$ and $T \!=\! 10^{-3}D$;
the bottom row shows results of 
second-order perturbation theory (PT)
in the wideband and zero-temperature limit.
We find very good qualitative agreement
(the overall magnitude depends on the 
broadening prescription),
even though the limit of weak interaction is highly nontrivial for the diagonalization-based NRG.}
\label{fig:KF_SOPT} 
\end{figure}

\begin{figure}
\includegraphics[width=\linewidth]{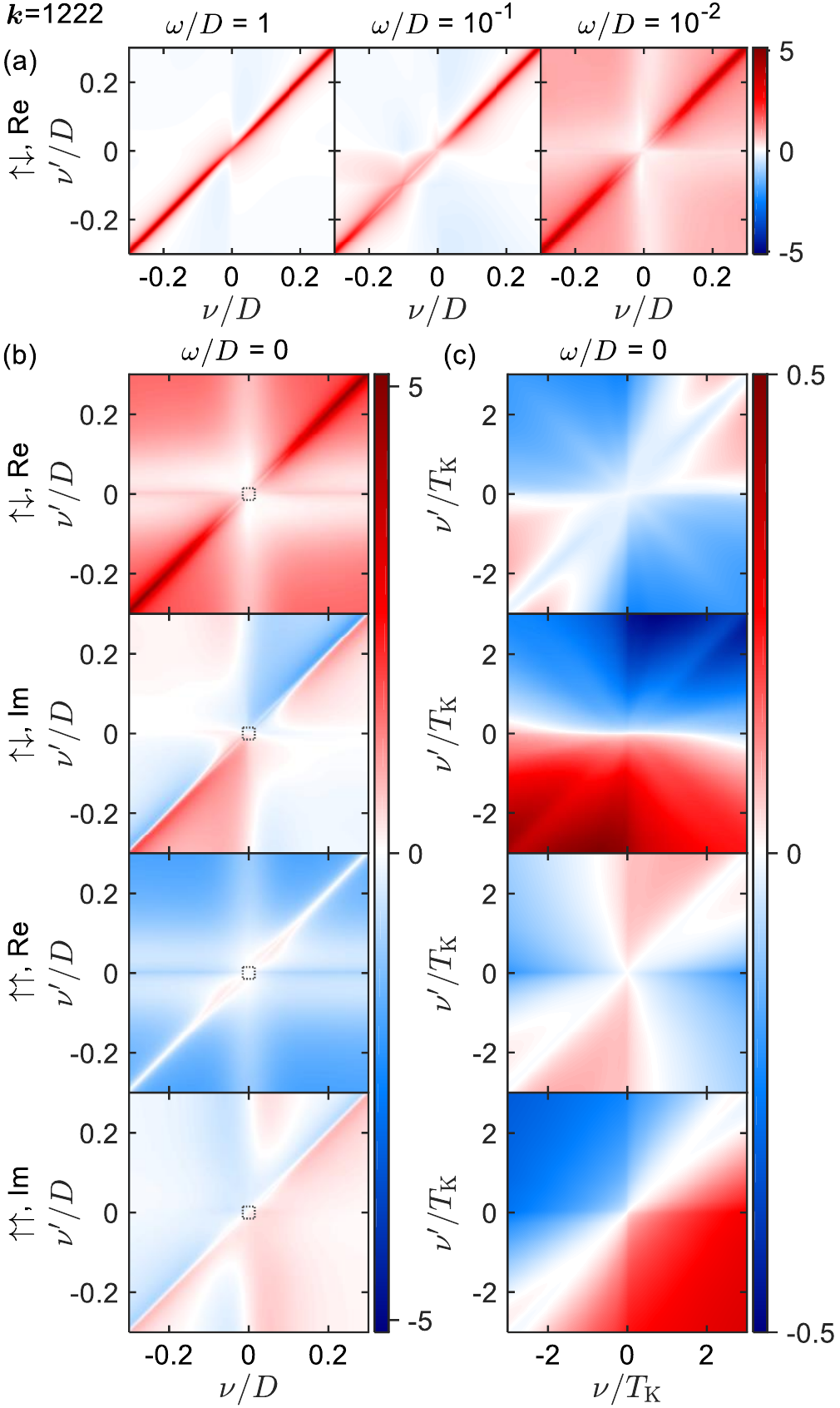}
\caption{%
\KF/ $4$p vertex $F_{\sigma\sigma'}^{[1]}$ 
in the strongly interacting AIM, 
$U/\Delta\!=\!5$, $U/D\!=\!1/5$, $T\!=\!10^{-4}D$.
(a) $\mathrm{Re} \, (F_{\uparrow\downarrow}^{[1]} \!-\! F_{0;\uparrow\downarrow}^{[1]})/U$ 
for three choices of $\omega \!>\! \TK$.
Lowering $\omega$, more features build up.
(b) 
All components of
$(F_{\sigma\sigma'}^{[1]} \!-\! F_{0;\sigma\sigma'}^{[1]})/U$ at $\omega \!=\! 0$.
(c) Enlargement on the window 
$|\nu|, |\nu'| \!\lesssim\! \TK$
[dashed squares in (b)],
showing $[F_{\sigma\sigma'}^{[1]} \!-\! \tilde{U}_{\sigma\sigma'}/(2Z^2)]/U$.
The values in (c) are an order of magnitude smaller than in (b)
and vanish for $\nu^{(\prime)} \!\to\! 0$,
showing that the low-energy retarded vertex reproduces RPT.}
\label{fig:KF_SIAM} 
\end{figure}

Our second \KF/ benchmark involves the limit of weak interaction, $U/\Delta \!=\! 1/2$.
There, one has an analytic result from second-order perturbation theory (PT),
conveniently obtained in the wide-band and zero-temperature limit,
which can be found by analytic continuation of \Eq{eq:vertex_SOPT}.
One should keep in mind that weak interaction, and even the noninteracting limit,
is highly nontrivial for a diagonalization-based algorithm like NRG.
In \Fig{fig:KF_SOPT}, we compare our NRG results of the weakly interacting AIM to PT,
considering the real and imaginary parts of $F_{\sigma\sigma'}^{[1]}$,
as a function of $\nu$ and $\nu'$ at $\omega \!=\! 0$ for both spin components.
While the precise values of the NRG vertex depend on the broadening prescription,
we overall find very good qualitative agreement between NRG and PT.
There are dominant diagonal structures that arise whenever 
one of the (bosonic) frequency arguments in (the analytic continuation of) \Eq{eq:vertex_SOPT} vanishes.
According to \Eq{eq:bubble_IF}, their real part describes a
peak of width $\sim \! \Delta$, as reproduced in \Fig{fig:KF_SOPT}.
Upon subtraction of the bare vertex,
the $\uparrow\downarrow$ component has no background contribution,
since $\chi_0(\omega_{12} \!=\! \omega)$ in \Eq{eq:vertex_SOPT} comes with a factor $\delta_{\sigma,\bar{\sigma}'}$.
The $\uparrow\uparrow$ component has a background value of $\chi_0(0)$;
its diagonal vanishes upon cancellation of
$\chi_0(\omega_{12} \!=\! \omega)$
and
$\chi_0(\omega_{14} \!=\! \nu \!-\! \nu')$.
Whereas these features are easily explained from a perturbative perspective on the vertex,
obtaining them in a numerical approach that starts from the spectral representation of the correlator
is a stringent test for the whole machinery.

After passing both of these benchmarks, we can confidently present our results
for the retarded vertex in the strongly interacting regime.
We use identical parameters as above:
$U/\Delta \!=\! 5$, $U/D \!=\! 1/5$, and $T \!=\! 10^{-4}D$ below the Kondo temperature
$\TK \!\simeq\! 5 \!\times\! 10^{-3} D$.
In \Fig{fig:KF_SIAM}(a), we show
$\mathrm{Re} \, F_{\uparrow\downarrow}^{[1]} \!-\! F_{0;\uparrow\downarrow}^{[1]}$
as a function of $\nu$ and $\nu'$, for three choices of $\omega$.
At large $\omega$, the vertex mostly involves only one diagonal.
This structure can again be understood from a diagrammatic perspective \cite{Wentzell2020}
related to \Eq{eq:vertex_SOPT}:
contributions from both the particle-hole channel
corresponding to $\chi_0(\omega_{12} \!=\! \omega)$
and the particle-particle channel,
corresponding to $\chi_0(\omega_{13} \!=\! \omega \!+\! \nu \!+\! \nu')$,
are suppressed at large $\omega$,
while
those of the other particle-hole channel
persist independent of it,
similar to $\chi_0(\omega_{14} \!=\! \nu \!-\! \nu')$.
As we decrease $\omega$, more features build up:
$\mathrm{Re} \, F_{\uparrow\downarrow}^{[1]}$
develops an increasing background value,
independent of $\nu$ and $\nu'$,
except for distinct values forming a plus shape.

Figure~\ref{fig:KF_SIAM}(b) shows the real and imaginary parts of
$F_{\sigma\sigma'}^{[1]} \!-\! F_{0;\sigma\sigma'}^{[1]}$,
for both spin components, at $\omega \!=\! 0$.
We again enlarge the low-energy window
$\nu^{(\prime)} \!\lesssim\! \TK$ (dashed square).
Similar to the \MF/ results (cf.\ \Fig{fig:IF_NRG_RPT}),
the low-energy limit of $F\sseta$ should reproduce the RPT prediction.
Figure~\ref{fig:KF_SIAM}(c) shows the difference
$F_{\sigma\sigma'}^{[1]} \!-\! \tilde{U}_{\sigma\sigma'}/(2Z^2)$
(the factor of 2 comes from the Keldysh rotation).
Indeed, this difference is an order of magnitude smaller than $F_{\sigma\sigma'}^{[1]} \!-\! F_{0;\sigma\sigma'}^{[1]}$,
and, for $\nu^{(\prime)} \!\to\! 0$, it goes to zero.
While the real parts of both spin components show a rather extended 
regime of small values, the imaginary parts, particularly regarding $\uparrow\downarrow$,
exhibit only a thin line of zero values. 

\begin{figure}
\includegraphics[width=\linewidth]{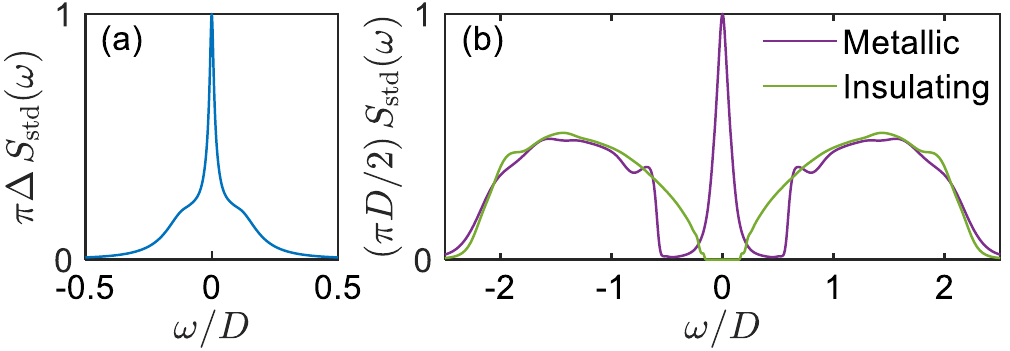}
\caption{%
Standard ($2$p) spectral functions $S_\mathrm{std}$ [cf.~\Eq{eq:Sstd}] 
for (a) the AIM chosen for Figs.~\ref{fig:IF_NRG_lowT} and \ref{fig:KF_SIAM} and 
(b) the DMFT solutions of the HM chosen for \Fig{fig:IF_KF_1HM}.
Here, $S_\mathrm{std} (\omega)$ is obtained by 
the adaptive broadening scheme of Ref.~\onlinecite{Lee2016}.}
\label{fig:Sstd} 
\end{figure}

\begin{figure*}
\includegraphics[width=\linewidth]{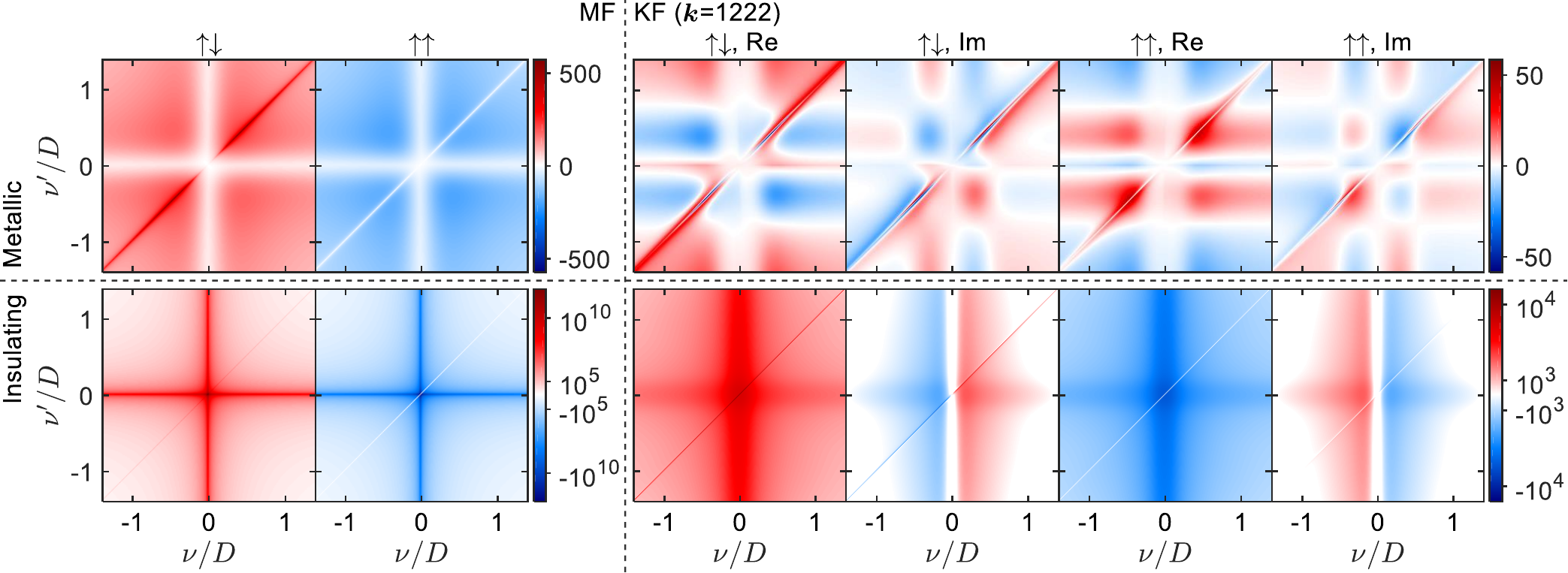}
\caption{%
\MF/ and retarded \KF/ $4$p vertices, $(F_{\sigma\sigma'} \!-\! F_{0;\sigma\sigma'})/U$ and 
$(F_{\sigma\sigma'}^{[1]} \!-\! F_{0;\sigma\sigma'}^{[1]})/U$, respectively, 
at $\omega \!=\! 0$ in the DMFT solutions of the HM 
(Bethe lattice, half-bandwidth $D$).
The first (second) rows show the metallic (insulating) solutions at 
$U/D \!=\! 2.6$, $T/D \!=\! 10^{-4}$,
using a linear (logarithmic) color scale.
The \KF/ vertex for the metallic solution exhibits distinct features at intermediate frequencies $0.1 D \lesssim |\nu|, |\nu'| \lesssim 0.5D$, similar to where the standard ($2$p) spectral function $S_{\mathrm{std}}(\omega)$ has dips separating quasiparticle peak and Hubbard bands (Fig.~\ref{fig:Sstd}).
The insulating solution has very sharp diagonal lines ($\nu \!=\! \nu'$).}
\label{fig:IF_KF_1HM} 
\end{figure*}

\subsection{\texorpdfstring{\MF/}{MF} and \KF/ for HM: DMFT solution}

For the strongly interacting AIM, the \MF/ vertex in Fig.~\ref{fig:IF_NRG_lowT} and the real part of the retarded \KF/ vertex in Fig.~\ref{fig:KF_SIAM} look somewhat similar. 
[Recall that \KF/ $4$p functions inherit a factor $1/2$ from the Keldysh rotation; see e.g.\ \Eq{eq:analytic_continuation_IF_to_fully_retarded}.]
This is drastically different for the results we present in the following.
There, we analyze the (half filled) one-band HM within 
DMFT, which amounts to solving a self-consistently determined AIM \cite{Georges1996}.
We take $U/D \! = \! 2.6$ and $T/D \!=\! 10^{-4}$ in the coexistence region of a metallic and insulating solution.
Figure~\ref{fig:Sstd} shows the standard ($2$p) spectral function $S_\mathrm{std}$ of 
the strongly interacting AIM [Fig.~\ref{fig:Sstd}(a)], considered previously, 
and the metallic and insulating solution of the self-consistent AIM describing the HM [Fig.~\ref{fig:Sstd}(b)].
The self-consistent metallic solution has much more pronounced features, where the spectral weight between the quasiparticle peak and the Hubbard bands almost goes to zero.
The insulating solution has a gap at $\omega \!=\! 0$ and broad Hubbard bands around the positions of the atomic peaks, $\pm U/2$.

Figure~\ref{fig:IF_KF_1HM} shows the local $4$p vertex for the DMFT solution of the HM, displaying the \MF/ vertex on the left, and a retarded Keldysh component, $\vec{k} \!=\! 1222$, on the right. 
In the top row, depicting the metallic solution, the \MF/ vertex is reminiscent of the AIM results in Fig.~\ref{fig:IF_NRG_lowT}, albeit with significantly larger values and a more extended plus-shaped structure.
By contrast, the \KF/ vertex reveals entirely new features not found in the AIM results of Fig.~\ref{fig:KF_SIAM}.
Next to the typical structure consisting of a background value, diagonal line, and plus-shaped structure, it exhibits sign changes at intermediate frequencies $0.1 \!\lesssim\! |\nu^{(\prime)}|/D \!\lesssim\! 0.5$.
They are thus at similar frequencies as the aforementioned dips in $S_\mathrm{std}$ (Fig.~\ref{fig:Sstd}) and likely related to these.
In the bottom row, depicting the insulating solution, the \MF/ vertex is very similar to the 
HA solution (Fig.~\ref{fig:IF_HA}), with almost divergent values.
The \KF/ vertex also looks similar to that, but different from the \KF/ HA results in Fig.~\ref{fig:KF_HA}.
The reason is that, for the HA, the diagonal structure expected in $F^{[1]}$ has a width that vanishes as $\Delta \!\to\! 0$ and enters as a $\delta$ function in other Keldysh components. 
However, for the insulating solution of the HM, the hybridization, though gapped, remains finite.
It thus allows for a diagonal signal in $F^{[1]}$, even if it is only a very sharp line.

The overall magnitudes of the \KF/ results in Fig.~\ref{fig:KF_HA} are rather different from the \MF/ results. 
On the one hand, this may result from the fact that the real-frequency results do have a richer structure in this case.
On the other hand, the precise values of the \KF/ results still depend on the broadening prescription, particularly for the insulating solution.
Further improving the broadening of $4$p real-frequency data is planned for follow-up work.
This will enable a thorough analysis of the DMFT real-frequency vertex and promises valuable insight into strong-correlation effects on the two-particle level.

\section{Summary and outlook}
\label{sec:conclusion}

\subsection{Summary}
\label{sec:concl:summary}

The many-body problem is typically addressed by either 
deriving $\ell$-point ($\ell$p) correlation and vertex functions from an action
or by working with operators and states in relation to a Hamiltonian.
The connection between both approaches through spectral or Lehmann representations
is well known for various $2$p functions and for imaginary-time functions with $\ell \!\leq\! 4$. 
Here, we completed this picture in a generalized framework, 
providing spectral representations for arbitrary $\ell$p functions
in three commonly used many-body frameworks:
the real-frequency zero-temperature formalism (\ZF/),
imaginary-frequency Matsubara formalism (\MF/),
and real-frequency Keldysh formalism (\KF/).

Through the spectral representations, we elucidated how $\ell$p correlators $G$ 
in the \ZF/, \MF/, and \KF/ are related to one another.
We expressed $G$ as a convolution of partial spectral functions (PSFs) $S$ 
and kernel functions $K$.
The PSFs are formalism independent and contain the dynamical information of a given system.
By contrast, the kernels are system independent and encode the time-ordering prescriptions
of the formalism at hand.
We first derived the spectral representation in the \ZF/ where it is most compact.
We proceeded with the \MF/, using analogous arguments and the same PSFs, 
and discussed anomalous terms that arise when vanishing (bosonic) Matsubara frequencies 
and degeneracies in the spectrum occur together.
For the \KF/ in both the contour and Keldysh bases, 
we identified a (fully) retarded kernel $K\sseta$ through which all other \KF/ kernels can be expressed.
Among the \KF/ correlators, those with a solitary Keldysh index equal to $2$, $G\sseta$, have the simplest spectral representation.
It precisely matches the one of the \MF/ (without anomalous parts) up to a replacement of imaginary frequencies by real frequencies with infinitesimal imaginary parts,
$\mi\vec{\omega} \!\to\! \vec{\omega}\sseta$,
making the analytic continuation between \MF/ and retarded \KF/ $\ell$p functions manifest.

We used a novel NRG method, described in the accompanying paper \cite{Lee2021}, 
to evaluate the spectral representations of $4$p functions for selected quantum impurity models. Starting with a simple model for x-ray absorption treated in the \ZF/, we analyzed multiple power laws in the $4$p vertex.
Proceeding with the Anderson impurity model (AIM) 
in the \MF/, we benchmarked our NRG results against Monte Carlo data at intermediate temperatures.
The NRG technique is beneficial at very 
low temperatures, where we found the vertex to exhibit strongly reduced values for Matsubara frequencies below the Kondo temperature $\TK$.
Indeed, by decreasing the temperature $T$ from above half-bandwidth $D$ to below $\TK$,
we studied the crossover from strongly interacting particles to weakly interacting quasiparticles 
on the level of the $4$p vertex $F$ (effective interaction).
At $T \!\ll\! \TK$, we deduced from $F$ the quasiparticle interaction, 
which has a form identical to the bare electronic interaction, 
with values predictable from renormalized perturbation theory (RPT) 
and the NRG low-energy spectrum.

For the AIM in the \KF/, 
we first tested our NRG results in the solvable limits of weak and infinitely strong interaction.
Finding qualitative agreement with perturbation theory for weak interaction 
was a demanding test for the diagonalization-based NRG.
For infinitely strong interaction, i.e.\ the Hubbard atom (HA), 
we found rich features in the different Keldysh components of the vertex.
In the intermediate regime of the strongly interacting AIM,
we observed the formation of renormalization effects in the retarded vertex $F^{[1]}$
with decreasing the transfer frequency $\omega$
and showed that $F^{[1]}$ reproduces RPT for temperature and frequencies below $\TK$.
Overall, the retarded vertex showed features that might be expected from the \MF/ result 
and perturbation theory.
This was drastically different in our final results
where we analyzed the one-band Hubbard model 
within DMFT in the regime of coexisting metallic and insulating solutions.
For the strongly correlated metal, 
the \KF/ vertex showed distinct features at intermediate frequencies, 
absent in the \MF/ counterpart.
We speculate that these go hand in hand with the dips 
separating the quasiparticle peak and the Hubbard bands in the standard ($2$p) spectral function.
Regarding the insulating solution, the \MF/ vertex is similar to the \MF/ HA result. 
The \KF/ vertex also looks similar to that, but the finite hybridization, albeit gapped,
leads to notable deviations from the \KF/ HA.

\subsection{Outlook}
\label{sec:concl:outlook}

In this work, we have focused on
(i) the formal properties of spectral representations and
(ii) benchmark tests of our NRG scheme \cite{Lee2021}
to compute the $4$p vertex of quantum impurity models.
This sets the stage for a variety of intriguing applications.

Let us start with more formal aspects.
In the Introduction, 
we mentioned the 
divergences of two-particle irreducible vertices in the \MF/.
Mathematically, these divergences arise through the matrix inversion of a generalized susceptibility $\chi$, 
a function of discrete Matsubara frequencies,
whose eigenvalues become negative and thus cross zero at some point \cite{Chalupa2018}.
Expressed through real frequencies, the former sums over matrix elements
become integrals over complex functions.
At finite temperature, there is an additional matrix structure of Keldysh indices.
The tools presented here allow one to compute $\chi$ in the \ZF/ or \KF/,
and to investigate if divergences of the 2PI vertex persist for real frequencies.

Moreover, real frequencies provide the natural language for Fermi-liquid theory.
Indeed, the original Fermi-liquid works used the zero-temperature approach, 
defining the Landau parameters in terms
of the \ZF/ (full) $4$p vertex \cite{Landau1980}.
In the mean time, the $T \!=\! 0$ Landau parameters were expressed through Matsubara vertices, too \cite{Hewson2001a,Krien2019,Melnick2020}.
Such \MF/ vertices can also be computed at $T \!>\! 0$. 
However, a stringent extension of Fermi-liquid theory to finite temperature should ideally use real-frequency Keldysh objects.
Their nonperturbative evaluation becomes accessible through this work.

On the practical side, an important application of local $4$p functions is given by diagrammatic extensions of DMFT.
We briefly recall that DMFT describes local correlations, assuming a purely local self-energy. 
Diagrammatic extensions of DMFT employ diagrammatic relations to further incorporate 
correlations of arbitrary wavelength \cite{Rohringer2018}. 
For instance, in the ladder dynamical vertex approximation \cite{Toschi2007,Held2008,Galler2017}, 
dual fermion formalism \cite{Rubtsov2008, Brener2008, Hafermann2009a}, 
or a functional renormalization group (fRG) flow \cite{Metzner2012} starting from DMFT \cite{Taranto2014},
a momentum-dependent self-energy and vertex are constructed from the local 
(full) $4$p vertex. 
Many successful results along these lines have been obtained in the \MF/ \cite{Rohringer2018}.
Yet, the commonly used Monte Carlo-based DMFT impurity solvers are not able to reach very low temperatures.
Our \MF/ NRG results, which extend to arbitrarily low temperature, can remedy this limitation.
Furthermore, to properly interpret the intriguing phases of strongly correlated electron systems,
real-frequency diagrammatic extensions of DMFT would be invaluable.
The momentum dependence could be generated by real-frequency diagrammatic techniques,
such as the rather well-developed Keldysh fRG
\cite{Jakobs2009,Jakobs2010,Schimmel2017}.
The necessary building blocks are real-frequency local $4$p functions, such as the \KF/ NRG results shown here.

Another interesting
topic requiring the computation of real-frequency $4$p functions is 
the theory of resonant inelastic x-ray scattering (RIXS) of strongly correlated materials~\cite{Ament2011}. We present some proof-of-principle RIXS results 
in the accompanying paper \cite{Lee2021}.

\section*{Acknowledgments}
We thank Anxiang Ge, Gabriel Kotliar, and Andreas Weichselbaum for insightful discussions.
We also thank A.\ Ge for making his analytical \KF/ results for the 
Hubbard atom available to us as benchmark data.
Furthermore, we thank Alessandro Toschi and Patrick Chalupa for fruitful discussions and for providing the QMC benchmark data.
Discussions with Daniel Springer and Clemens Watzenb\"ock in early stages of this work are appreciated. 
For our numerical computations,
we exploited symmetries using the QSpace library developed 
by A.\ Weichselbaum \cite{Weichselbaum2012a,Weichselbaum2020}.
We were supported by the Deutsche Forschungsgemeinschaft (DFG, German Research Foundation) under Germany's Excellence Strategy EXC-2111 (Project No.\ 390814868) and through Project No.\ 409562408. S.-S.B.L.\ acknowledges the DFG grant LE3883/2-1 (Project No.\ 403832751).
F.B.K.\ acknowledges support by the Alexander von Humboldt Foundation through the Feodor Lynen Fellowship.

\appendix

\section{Explication of \ZF/ formula for \texorpdfstring{$\ell = 2$}{l=2}}
\label{app:RF_G2}

Here, we show how the general $\ell$p formula \eqref{eq:G_w_RF}
reproduces the familiar $2$p result \eqref{eq:G2_RF} for $G(\omega)$
in the \ZF/.
For $\ell=2$ and $\vec{\Oc} = (\Ac,\Bc)$, \Eq{eq:G_w_RF} reads
\begin{align}
G(\omega)
& =
\nint \md \omega_1' \,
K(\omega_1 - \omega_1')
S[\Ac,\Bc](\omega_1')
\nonumber
\\
& \ 
+ \zeta
\nint \md \omega_2' \,
K(\omega_2 - \omega_2')
S[\Bc,\Ac](\omega_2')
. 
\label{eq:GL_RF_app}
\end{align}
For brevity, we wrote $G(\vec{\omega}) \!=\! G(\omega)$, 
$K(\vec{\omega}_p) \!=\! K(\omega_{\ovb{1}})$,
and $S(\vec{\omega}_p') \!=\! S(\omega'_{\ovb{1}})$, 
hiding the second frequency argument, 
with $\omega_1 \!=\! -\omega_2 \!= \!\omega$
and $\omega_2' \!=\! - \omega_1'$ implicit.
The ingredients needed above are $K(\omega_i) =
1/\omega^+_i$ and, using \Eq{eq:Sc_w_explicit},
\begin{subequations}
\label{subeq:psf_21_app}
\begin{align}
S[\Ac,\Bc](\omega_1')
& =
\sum_{\ub{1}\,\ub{2}}
\rho_{\ub{1}} A_{\ub{1}\ub{2}} B_{\ub{2}\ub{1}} 
\delta(\omega'_1 - E_{\ub{2}\ub{1}})
, 
\label{eq:psf_12_app}
\\
S[\Bc,\Ac](\omega_2')
& =
\sum_{\ub{1}\,\ub{2}}
\rho_{\ub{2}} B_{\ub{2}\ub{1}} A_{\ub{1}\ub{2}} 
\delta(\omega'_2 - E_{\ub{1}\ub{2}})
. 
\label{eq:psf_21_app}
\end{align}
\end{subequations}
Inserting \Eqs{subeq:psf_21_app} into \Eq{eq:GL_RF_app},
we obtain \Eq{eq:G2_RF}:
\begin{align}
\nonumber
G(\omega) & = 
\nint \md \omega_1'
\frac{S[\Ac,\Bc](\omega_1')}{\omega^+_1 - \omega_1'}
+ \zeta
\nint \md \omega_2'
\frac{S[\Bc,\Ac](\omega_2')}{\omega^+_2 - \omega_2'}
\\
& = \sum_{\ub{1}\,\ub{2}}
A_{\ub{1}\ub{2}} B_{\ub{2}\ub{1}}
\Big[
\frac{\rho_{\ub{1}} }{\omega^+ - E_{\ub{2}\ub{1}}}
- \zeta
\frac{\rho_{\ub{2}}}{\omega^- - E_{\ub{2}\ub{1}}}
\Big] 
. 
\label{eq:exampleGell=2explicit}
\end{align}

\section{Direct \texorpdfstring{\MF/}{MF} calculation}
\label{app:IF_direct}

In the main text, we derived the \MF/ formulas somewhat indirectly, 
arguing that all contributions 
from the upper integration boundaries in \Eq{eq:Kc_Omega_integral} cancel. 
For completeness, 
we here give a direct derivation of the \MF/ Lehmann representations for $\ell \!=\! 2, 3, 4$. 
We also discuss anomalous terms arising when denominators vanish. 
For this purpose, we focus on correlators for which at most one frequency 
$\omega_{\ovb{1} \cdots \ovb{i}}$, with $i \!<\! \ell$, is bosonic,
and derive the expressions \eqref{eq:K_Omega_IF_compact}
and \eqref{eq:K_Omega_IF} for the full \MF/
kernel $K(\vec{\Omega}_p)$ given in the main text.
The reasons for this focus are stated before \Eq{eq:K_Omega_IF_compact}; other cases can be treated analogously.

We begin by discussing the computation for general $\ell$.
We exploit time-translational invariance, 
define $G(\vec{\tau})=\Gc(\vec{\tau})|_{\tau_\ell=0}$,
use $\omega_{1\cdots \ell}=0$ in $G(\mi \vec{\omega})$,
and directly compute 
\begin{align}
\label{eq:G-IF-l=3-direct}
G(\mi \vec{\omega}) = 
\nbint{0}{\beta} \md \tau_1 \, \ndots \, \md \tau_{\ell-1}
\, e^{\sum_{i=1}^{\ell-1} \mi \omega_i \tau_i} 
G(\vec{\tau})
.
\end{align}
The $(\ell \!-\! 1)$-fold integral involves $(\ell \!-\! 1)!$ different time orderings.
Hence, the sum over $p$ in the representation \eqref{eq:Gc_tau_KS_IF_total} for $\Gc(\vec{\tau})$ 
reduces to $(\ell \!-\! 1)!$ permutations, say
$q = (\ovb{1} \ovb{2} \, \ndots \ovb{\ell \!-\! 1} \ell)$
(last index fixed), 
with a new kernel $k(\vec{\tau}_q) = (-1)^{\ell-1} \prod_{i=1}^{\ell-2}
\theta(\tau_{\ovb{i}} - \theta_{\ovb{i+1}})$,
as $\tau_\ell \!=\! 0$ always is the smallest time.
We obtain
\begin{flalign}
G(\mi\vec{\omega})
& =
\sum_q \vec{\zeta}^q \! 
\nint \md^{\ell-1} \omega'_q \,
k(\mi \vec{\omega}_q \!-\! \vec{\omega}'_q) \, 
S[\vec{\Oc}_q](\vec{\omega}'_q)
, \hspace{-1cm} & 
\label{eq:MatsubaraGellConvolutation-explicit-l=3}
\\
k(\vec{\Omega}_q)
& =
(-1)^{\ell-1} \! 
\nbint{0}{\beta} 
\! \md \tau_{\ovb{1}} 
\nbint{0}{\tau_{\ovb{1}}} 
\! \md \tau_{\ovb{2}} 
\: \ndots 
\nbint{0}{\tau_{\ovb{\ell-2}}} 
\! \md \tau_{\ovb{\ell-1}} \, 
e^{\sum_{i=1}^{\ell-1} \Omega_{\ovb{i}} \tau_{\ovb{i}}} ,
\label{eq:MatsubaraGellConvolution-explicit-kernel}
\end{flalign}
where $\vec{\Omega}_q = \mi \vec{\omega}_q - 
\vec{\omega}'_q$, with $\omega'_{1 \cdots \ell} = 0$
and thus $\Omega_{1 \cdots \ell} = 0$. 
We will often use the shorthand $S_p = 
S[\vec{\Oc}_p](\vec{\omega}'_p)$ for permuted versions of $S$. 
The $\delta$ functions in $S$, as given 
by \Eq{eq:Sc_w_explicit}, ensure 
$\omega'_{\ovb{i}} = E_{\ub{\ovb{i+1}} \, \ub{\ovb{i}}}$
and $\omega'_{\ovb{1} \cdots \ovb{i}} = E_{\ub{\ovb{i+1}} \, \ub{\ovb{1}}}$; 
hence, $\omega'$ variables serve as shorthands for energy differences.

In the following, we consider the cases $\ell = 2$, $3$, and $4$ in turn.
For each, we start by computing the regular contributions, 
signified by a tilde on $\ovwt{k}$ and $\ovwt{G}$, 
for which all denominators arising from
the $\tau_{\ovb{i}}$ integrals are assumed to be nonzero. 
We subsequently discuss anomalous
cases, signified by a hat on $\ovwh{k}$, for which this assumption does not hold. For these, we recompute the corresponding integrals more carefully. Alternatively and more elegantly, the
anomalous terms can also be found from the regular ones 
by a limiting procedure, treating nominally 
vanishing denominators as infinitesimal rather than zero.

\subsection{\texorpdfstring{\MF/}{MF}, \texorpdfstring{$\ell = 2$}{l=2}}

We begin with the case $\ell = 2$, involving
two bosonic or fermionic operators. 
Although it was already covered in Sec.~\ref{sec:motivation}, 
we discuss it again, to set the stage for the
subsequent analogous treatments of the cases $\ell = 3$, $4$.

For $\ell = 2$, there are only two permutations, $p=(\ovb{1} \ovb{2})$, namely $(12)$ and $(21)$, with $\vec{\zeta}^{(12)}\!=\! 1$, $\vec{\zeta}^{(21)}\!=\! \zeta$, and $S_{(12)} = S[\Oc^1,\Oc^2](\omega'_1,\omega'_2)$, $S_{(21)} = S[\Oc^2,\Oc^1](\omega'_2,\omega'_1)$.
The kernel $k(\vec{\Omega}_p)$ has arguments
$(\Omega_{\ovb{1}}, \Omega_{\ovb{2}})$,
with $\Omega_\ovb{i} = \mi \omega_\ovb{i} - \omega'_\ovb{i}$, 
and it is understood \textit{a priori} that 
$\omega_{12} = 0$ and $\omega'_{12} = 0$. 
The sum over $q$ in \Eq{eq:MatsubaraGellConvolutation-explicit-l=3} has only a single
term, $q=(12)$, with $k (\vec{\Omega}_{(12)}) =
k (\Omega_1, \Omega_2)$ defined by 
\begin{align}
\label{eq:l=2IFK-integral-explicit}
k (\vec{\Omega}_{(12)}) & = -
\nbint{0}{\beta} 
\md \tau_1 \, e^{\Omega_1 \tau_1}
. 
\end{align}

\textit{Regular part:} For $\Omega_{1} \neq 0$,
\Eq{eq:l=2IFK-integral-explicit} evaluates to
\begin{align}
\label{eq:l=2IFK-integral-explicit-result}
\ovwt{k} (\vec{\Omega}_{(12)})
& =
\frac{1 - e^{\beta \Omega_{1}}}{\Omega_{1}}
=
\frac{1}{\Omega_{1}}
+
\frac{\zeta e^{-\beta \omega'_1}}{\Omega_{2}}
, 
\end{align}
where we used $\Omega_{12}=0$ for the second step. 
Now, consider the product
$k (\vec{\Omega}_{(12)}) S_{(12)}$
in \Eq{eq:MatsubaraGellConvolutation-explicit-l=3}
for $G(\mi \vec{\omega})$. The cyclicity relation \eqref{eq:Sc_w_cyclicity}
implies 
$S_{(21)} = e^{-\beta \omega'_1} S_{(12)}$;
hence, the regular part of this product takes the form 
\begin{flalign}
\label{eq:K2S3-explicit-b1b23}
\ovwt{k} (\vec{\Omega}_{(12)}) S_{(12)}
= 
\frac{S_{(12)}}{\Omega_{1}}
+
\frac{\zeta S_{(21)}}{\Omega_{2}}
. 
\end{flalign}
The regular part of $G$ can thus be expressed as 
\begin{flalign}
\label{eq:K2S3-explicit-p}
& \ovwt{G}(\mi\vec{\omega})
 = \nint \md \omega'_1 
\ovwt{k} (\vec{\Omega}_{(12)}) S_{(12)}
= 
\sum_p \vec{\zeta}^p
\nint \md \omega'_{\ovb{1}} 
\frac{S_{(\ovb{1}\ovb{2})}}
{\Omega_{\ovb{1}}} 
, 
\hspace{-1cm} & 
\end{flalign}
where the sum now runs over the two permutations
$p= (\ovb{1}\ovb{2}) \in \{(12), (21) \}$. This reproduces
the general result stated by 
\Eqs{eq:G_iw_KS_IF} and
\eqref{eq:Ktilde_Omega_IF} from the main text.

\textit{Anomalous part:}
The integral \eqref{eq:l=2IFK-integral-explicit} 
is anomalous 
if $\Omega_1 = - \Omega_2 = 0$; indeed,
both terms in \Eq{eq:l=2IFK-integral-explicit-result}
for $\ovwt{k}$ would then involve vanishing denominators.
This situation can arise if $\omega_1 = - \omega_2 = 0$, possible only for bosonic frequencies ($\zeta=1$), 
and if simultaneously $\omega'_1 = - \omega'_2 = 0$ due to 
a degeneracy, $E_{\ub{2}} = E_{\ub{1}}$, in the spectrum
[cf.\ \Eq{eq:S_w}].
In this case, the integral in \Eq{eq:l=2IFK-integral-explicit} yields 
\begin{align}
\label{eq:l=2IFK-integral-anomalous}
\ovwh{k} (\vec{0}) = 
- \nbint{0}{\beta} 
\md \tau_1 = - \beta 
. 
\end{align}
Alternatively, this result can be directly obtained
from the regular part $\ovwt{k}$ 
by taking $\Omega_1 \!= \! - \Omega_2$ not zero
but infi\-ni\-tesimal. We set 
$\omega_1 \! = \! - \omega_2 \! = \! 0$ 
and 
$- \omega'_1 \! = \! \omega'_2 \! = \! \epsilon$,
so that $\Omega_1 \! = \! - \Omega_2 \! = \! \epsilon$, 
and compute 
$\ovwh{k}$ as $\ovwt{k}_{\epsilon \to 0}$, using
\Eq{eq:l=2IFK-integral-explicit-result}:
\begin{align}
\ovwh{k} (\vec{0}) 
=
\bigg[
\frac{1}{\epsilon} 
+ \frac{e^{\beta \epsilon}}{-\epsilon} 
\bigg]_{\epsilon \to 0} 
= - \beta .
\label{eq:K-IF-l=2-divercences-cancel}
\end{align}

It will be convenient to split the anomalous term equally
among both terms in the sum $\sum_p$ for $G$.
Since $S_{(21)} = S_{(12)}$ when $\vec{\omega}'_p=\vec{0}$,
this can be done by symmetrizing the anomalous contribution 
to $k (\vec{\Omega}_{(12)}) S_{(12)}$ as
\begin{align}
\label{eq:K-IF-l=2-anomalous-symm}
\ovwh{k}(\vec{0}) S_{(12)} = 
- (\beta/2) 
\big[ S_{(12)} + S_{(21)} \big] 
.
\end{align} 
Equations~\eqref{eq:K-IF-l=2-anomalous-symm} and \eqref{eq:K2S3-explicit-p}
imply that $G(\mi\vec{\omega})$ of \Eq{eq:MatsubaraGellConvolutation-explicit-l=3} 
can be expressed in the form \eqref{eq:G_iw_KS_IF}, with a kernel defined as
\begin{align}
K(\vec{\Omega}_q) = 
\begin{cases}
1/\Omega_{\ovb{1}}
& \text{if} \quad \Omega_{\ovb{1}} 
\neq 0
, 
\\
- \beta/2
& \text{if} \quad \Omega_{\ovb{1}} = 0
, 
\end{cases}
\end{align}
consistent with the general \Eq{eq:K_Omega_IF}
from the main text.

Since the divergences from the denominators of
the regular $\ovwt{k}$ cancel after summing over
permutations [cf.\ \Eq{eq:K-IF-l=2-divercences-cancel}], 
we may also represent the full kernel as 
\begin{align}
K(\vec{\Omega}_p)
= 
\frac{1}{\Omega_{\ovb{1}}} - \delta_{\Omega_{\ovb{1}},0} \, \frac{\beta}{2} 
,
\label{eq:K-IF-l=2-compact-App}
\end{align}
corresponding to \Eq{eq:K_Omega_IF_compact}. 
Here,
$\delta_{\Omega_{\ovb{1}},0}$ is symbolic notation 
indicating that the anomalous second term is nonzero 
only if $\Omega_{\ovb{1}} = 0$. 
No restriction is placed on the $\Omega_{\ovb{1}}$
values for the first term, with the understanding
that vanishing denominators should be treated using infinitesimals.

\subsection{\texorpdfstring{\MF/}{MF}, \texorpdfstring{$\ell = 3$}{l=3}}

Consider $\ell = 3$. Without loss of generality, we choose 
$\Oc^3$ bosonic, with $\Oc^1$ and $\Oc^2$ either 
both bosonic ($\zeta=1$) or both fermionic ($\zeta=-1$). 
Hence, $\vec{\zeta}^p$ takes the values
\begin{align*}
\vec{\zeta}^{(123)}\!=
\vec{\zeta}^{(132)}\!=
\vec{\zeta}^{(312)}\!=1, \quad 
\vec{\zeta}^{(231)}\!=
\vec{\zeta}^{(213)}\!=
\vec{\zeta}^{(321)}\!= \zeta
.
\end{align*}
The twofold time integral in \Eq{eq:G-IF-l=3-direct}
now involves two time orderings, $\tau_1 > \tau_2 $ and $\tau_2 > \tau_1$;
hence, the sum $\sum_q$ in 
\Eq{eq:MatsubaraGellConvolutation-explicit-l=3}
involves two permutations
$q=(\ovb{1} \ovb{2} 3)$, namely $(123)$ and $(213)$. 
The kernel $k$ of \Eq{eq:MatsubaraGellConvolution-explicit-kernel} reads
\begin{align}
\label{eq:G3_direct_K3-revised-integrated-once-full}
k (\vec{\Omega}_{(\ovb{1}\ovb{2}3)}) & =
\nbint{0}{\beta} 
\md \tau_{\ovb{1}} \, e^{\Omega_{\ovb{1}} \tau_{\ovb{1}}} 
\nbint{0}{\tau_{\ovb{1}}} 
\md \tau_{\ovb{2}} \, e^{\Omega_{\ovb{2}} \tau_{\ovb{2}}}
.
\end{align}

\textit{Regular part:} 
Performing the integral \eqref{eq:G3_direct_K3-revised-integrated-once-full} 
and assuming all resulting denominators to be nonzero,
we obtain
\begin{subequations}
\begin{align}
\label{eq:G3_direct_K3-revised-integrated-once}
\ovwt{k} (\vec{\Omega}_{(\ovb{1}\ovb{2}3)})
& =
\frac{1}{\Omega_{\ovb{2}}}
\nbint{0}{\beta} 
\md \tau_{\ovb{1}} \,
\Big(
e^{\Omega_{\ovb{1}\ovb{2}} \tau_{\ovb{1}}} 
-
e^{\Omega_{\ovb{1}} \tau_{\ovb{1}}} 
\Big) 
\\
& =
\frac{e^{\beta \Omega_{\ovb{1}\ovb{2}}} -1}{\Omega_{\ovb{2}}
\Omega_{\ovb{12}}}
-
\frac{e^{\beta \Omega_{\ovb{1}}} -1}{\Omega_{\ovb{2}}
\Omega_{\ovb{1}}}
\nonumber 
\\
& =
\frac{e^{-\beta \omega'_{\ovb{1}\ovb{2}}}}{\Omega_{3} \Omega_{3\ovb{1}}}
+ 
\frac{\zeta e^{-\beta \omega'_{\ovb{1}}}}{\Omega_{\ovb{2}}
\Omega_{\ovb{2}3}}
+
\frac{1}{\Omega_{\ovb{1}}
\Omega_{\ovb{12}}} 
. 
\label{eq:G3_direct_K3-revised}
\end{align}
\end{subequations}
For the last line, we combined the $\beta$-independent
terms using $(1/\Omega_{\ovb{1}} - 1/\Omega_{\ovb{1}\ovb{2}})/
\Omega_{\ovb{2}} = 1/(\Omega_{\ovb{1}}\Omega_{\ovb{1}\ovb{2}})$.
We also exploited $\Omega_{\ovb{1}\ovb{2}3}=0$ to
rewrite the denominators of the 
$\beta$-dependent terms in a way that reveals 
the denominators of all three terms to be cyclically related by 
$(3\ovb{1}\ovb{2}) \to (\ovb{2}3\ovb{1}) \to (\ovb{1}\ovb{2}3)$. 
Since the cyclicity relation~\eqref{eq:Sc_w_cyclicity}
implies 
$e^{-\beta \omega'_{\ovb{1}\ovb{2}}} S_{(3 \ovb{1}\ovb{2})} = 
e^{-\beta \omega'_{\ovb{1}}} S_{(\ovb{2}3 \ovb{1})} 
 = S_{(\ovb{1}\ovb{2}3)}$,
we conclude that
\begin{flalign}
\label{eq:K3S3-explicit-b1b23}
\ovwt{k} (\vec{\Omega}_{(\ovb{1}\ovb{2}3)}) S_{(\ovb{1}\ovb{2}3)}
= 
\frac{S_{(3 \ovb{1}\ovb{2})}}{\Omega_{3} \Omega_{3\ovb{1}}}
+ 
\frac{\zeta S_{(\ovb{2}3 \ovb{1})}}{\Omega_{\ovb{2}}
\Omega_{\ovb{2}3}}
+
\frac{S_{(\ovb{1}\ovb{2}3)}}{\Omega_{\ovb{1}}
\Omega_{\ovb{12}}} 
, 
& 
\end{flalign}
which again has cyclically related index structures 
on the right. The sum over the two choices for $q=(\ovb{1} \ovb{2}3)$ in \Eq{eq:MatsubaraGellConvolutation-explicit-l=3}
for $G(\mi \vec{\omega})$, namely $(123)$ and $(213)$, thus yields 
\begin{flalign}
\ovwt{G}(\mi\vec{\omega})
& =
\sum_p \vec{\zeta}^p \! 
\nint \md \omega'_{\ovb{1}} \md \omega'_{\ovb{2}}\,
\frac{
S_{(\ovb{1}\ovb{2}\ovb{3})}}
{\Omega_{\ovb{1}} \Omega_{\ovb{12}}} 
, 
\end{flalign}
where the sum now runs over all six permutations
$p= (\ovb{1}\ovb{2}\ovb{3})$. This is consistent
with \Eqs{eq:G_iw_KS_IF} and
\eqref{eq:Ktilde_Omega_IF}. 

\textit{Anomalous part:} 
If $\omega_{\ovb{1}}$ and 
$\omega_{\ovb{2}}$ are fermionic frequencies, 
$\Omega_{\ovb{1}}$ and 
$\Omega_{\ovb{2}}$ are always nonzero. However, since
$\omega_3$ is bosonic, 
$\Omega_{\ovb{1}\ovb{2}} = -\Omega_3$ can vanish,
yielding anomalous contributions. Indeed,
the denominators in the first and third terms of \Eq{eq:G3_direct_K3-revised} for $\ovwt{k}$
vanish if $\Omega_{\ovb{1}\ovb{2}} \! = \! -\Omega_3 \! = \! 0$.
This happens if 
$\omega_{\ovb{1}\ovb{2}} \! =\! - \omega_3 \! = \! 0$
and also $\omega'_{\ovb{1}\ovb{2}} 
= - \omega'_3 = 0$.
Recomputing the integral 
\eqref{eq:G3_direct_K3-revised-integrated-once} for this case, we obtain 
\begin{flalign}
\label{eq:l=3-Khat-unsymmetrized-start}
& \ovwh{k} (\vec{\Omega}_{(\ovb{1}\ovb{2}3)}) =
\frac{\beta}{\Omega_{\ovb{2}}}
-
\frac{e^{\beta \Omega_{\ovb{1}}} -1}{\Omega_{\ovb{2}}
\Omega_{\ovb{1}}}
= - \frac{1}{\Omega_{\ovb{1}}} 
\! \left[\beta + \frac{1}{\Omega_{\ovb{1}}} \right]\!
+ \frac{\zeta e^{-\beta \omega'_{\ovb{1}}}}{\Omega_{\ovb{2}}
\Omega_{\ovb{2}3}} 
. 
\end{flalign}
The last term matches the second term of the regular $\ovwh{k}$
of \Eq{eq:G3_direct_K3-revised}; the first two are anomalous contributions. 
These can also be deduced directly from the first and third terms of $\ovwt{k}$ by treating 
$\Omega_{\ovb{1}\ovb{2}} = - \Omega_{3}$ there as infinitesimal. 
To this end, we set $\omega_{\ovb{1}\ovb{2}} = - \omega_{3} = 0$ and
$- \omega'_{\ovb{1}\ovb{2}} = \omega_{3}' = 
\epsilon$, so that $\Omega_{\ovb{1}\ovb{2}} = - \Omega_3= \epsilon$, 
and compute $\ovwh{k}$ as 
$\ovwt{k}|_{\epsilon \to 0}$.
The first and third terms of \Eq{eq:G3_direct_K3-revised} then give
\begin{align}
\frac{e^{\beta \epsilon}}{
-\epsilon (\Omega_{\ovb{1}} - \epsilon) }
+
\frac{1}{\Omega_{\ovb{1}} \epsilon} 
\, \to \, - \frac{1}{\Omega_{\ovb{1}}} 
\!\! \
\bigg[
\beta \! + \! \frac{1}{\Omega_{\ovb{1}}} 
\bigg] 
,
\end{align}
reproducing the anomalous terms in \Eq{eq:l=3-Khat-unsymmetrized-start}.

Since $\omega'_{\ovb{1}\ovb{2}}=0$ implies 
$S_{(3 \ovb{1}\ovb{2})}= S_{(\ovb{1}\ovb{2}3)}$, 
we can express the product $\ovwh{k} (\vec{\Omega}_{(\ovb{1}\ovb{2}3)}) S_{(\ovb{1}\ovb{2}3)}$ as 
\begin{flalign}
\label{eq:l=3-Khat-symmetrized}
& 
-\frac{1}{2}\!\! \left[\beta \!+\! \frac{1}{\Omega_{3\ovb{1}}} 
\right]\!\! \frac{S_{(3 \ovb{1}\ovb{2})}}{\Omega_{3\ovb{1}}}
+ 
\frac{\zeta S_{(\ovb{2}3 \ovb{1})}}{\Omega_{\ovb{2}}
\Omega_{\ovb{2}3}}
-\frac{1}{2}\!\! \left[\beta \!+\! \frac{1}{\Omega_{\ovb{1}}} 
\right]\!\! \frac{S_{(\ovb{1}\ovb{2}3)}}{\Omega_{\ovb{1}}} .
\hspace{-1cm} & 
\end{flalign}
Here, we split the anomalous contribution
equally between the first and third terms, 
using $\Omega_3=0$ to rewrite
the denominators such that their $\Omega$'s 
match the nonzero 
$\Omega$'s in the first and third terms of 
\Eq{eq:G3_direct_K3-revised} for $\ovwt{k}$. 

Jointly, \Eqs{eq:l=3-Khat-symmetrized} and \eqref{eq:G3_direct_K3-revised} imply that 
$G(\mi\vec{\omega})$ of \Eq{eq:MatsubaraGellConvolutation-explicit-l=3} 
has the form \eqref{eq:G_iw_KS_IF}, with a kernel defined as
\begin{align}
K(\vec{\Omega}_p) = 
\begin{cases}
{\displaystyle 
\frac{1}{\Omega_{\ovb{1}} \Omega_{\ovb{1}\ovb{2}}}} 
& \text{if} \quad \Omega_{\ovb{1}} \Omega_{\ovb{1}\ovb{2}} 
\neq 0 
, 
\\[3mm]
{\displaystyle 
- \frac{1}{2} \! 
\left[\beta + \frac{1}{\Omega_{\ovb{1}}}
\right]\! \frac{1}{\Omega_{\ovb{1}}}}
& \text{if} \quad \Omega_{\ovb{1}\ovb{2}} = 0 
, 
\\[3mm]
{\displaystyle 
- \frac{1}{2} \! 
\left[\beta + \frac{1}{\Omega_{\ovb{1}\ovb{2}}} 
\right]\! \frac{1}{\Omega_{\ovb{1}\ovb{2}}}}
& \text{if} \quad \Omega_{\ovb{1}} = 0 
, 
\end{cases}
\label{eq:K-IF-l=3-anomalous-explicit}
\end{align}
for any of the six permutations $p=(\ovb{1}\ovb{2}\ovb{3})$.
This is consistent with the general \Eq{eq:K_Omega_IF}
from the main text.

Since the anomalous terms
not proportional to $\beta$ stem from the numerators
of the regular part $\ovwt{k}$, they need not be displayed separately---they are generated automatically when treating vanishing
denominators as infinitesimal and summing over permutations.
The full kernel can thus be expressed 
in the following form, equivalent to \Eq{eq:K-IF-l=3-anomalous-explicit}, but written in the notation of
\Eq{eq:K_Omega_IF_compact}: 
\begin{align}
K(\vec{\Omega}_p) =
\frac{1}{\Omega_{\ovb{1}} \Omega_{\ovb{1}\ovb{2}}} 
 - 
\frac{\beta}{2}
\bigg[
\frac{\delta_{\Omega_{\ovb{1}\ovb{2}},0}}{\Omega_{\ovb{1}}}
+ \frac{\delta_{\Omega_{\ovb{1}},0}}{\Omega_{\ovb{12}}}
\bigg] 
. 
\label{eq:K-IF-l=3-compact-App}
\end{align}

\subsection{\texorpdfstring{\MF/}{MF}, \texorpdfstring{$\ell = 4$}{l=4}}

Finally, we consider the case $\ell=4$,
with four bosonic or four fermionic operators. 
Now, \Eq{eq:MatsubaraGellConvolutation-explicit-l=3} for $G(\mi\vec{\omega})$
involves a sum over six permutations, $q=(\ovb{1}\ovb{2}\ovb{3}4)$, with 
\begin{align}
\label{eq:G4_direct_K4-revised-integrated-once-full}
k (\vec{\Omega}_{(\ovb{1}\ovb{2}\ovb{3}4)}) & =
-
\nbint{0}{\beta} \!
\md \tau_{\ovb{1}} \, e^{\Omega_{\ovb{1}} \tau_{\ovb{1}}} 
\nbint{0}{\tau_{\ovb{1}}} \! 
\md \tau_{\ovb{2}} \, e^{\Omega_{\ovb{2}} \tau_{\ovb{2}}} 
\nbint{0}{\tau_{\ovb{2}}} \! 
\md \tau_{\ovb{3}} \, e^{\Omega_{\ovb{3}} \tau_{\ovb{3}}} 
.
\end{align}

\textit{Regular part:} 
Ignoring anomalous cases, this yields 
\begin{flalign}
& \ovwt{k} (\vec{\Omega}_{(\ovb{1}\ovb{2}\ovb{3}4)}) =
-
\nbint{0}{\beta} \!
\md \tau_{\ovb{1}} \, e^{\Omega_{\ovb{1}} \tau_{\ovb{1}}} 
\nbint{0}{\tau_{\ovb{1}}} \!\!
\md \tau_{\ovb{2}} \; 
\frac{ 1 }{\Omega_{\ovb{3}} } \Big[
e^{\Omega_{\ovb{23}} \tau_{\ovb{2}}} 
-
e^{\Omega_{\ovb{2}} \tau_{\ovb{2}}} 
\Big] \hspace{-1cm} 
& 
\nonumber
\\ 
& =
\nbint{0}{\beta} 
\md \tau_{\ovb{1}} \;
\frac{ 1 }{\Omega_{\ovb{3}} }
\! \left[
\frac{
-e^{\Omega_{\ovb{123}} \tau_{\ovb{1}}} 
\!+\! e^{\Omega_{\ovb{1}} \tau_{\ovb{1}}}
}{
\Omega_{\ovb{23}}
}
+
\frac{
e^{\Omega_{\ovb{12}} \tau_{\ovb{1}}}
\!-\! e^{\Omega_{\ovb{1}} \tau_{\ovb{1}}} 
}{
\Omega_{\ovb{2}}
}
\right] \hspace{-1cm} & 
\nonumber
\\ 
& =
\frac{-e^{\beta \Omega_{\ovb{1}\ovb{2}\ovb{3}}} 
+ 1}{\Omega_{\ovb{3}}\Omega_{\ovb{23}}\Omega_{\ovb{123}}}
+ 
\frac{e^{\beta \Omega_{\ovb{1}\ovb{2}}} - 1}
{\Omega_{\ovb{3}}\Omega_{\ovb{2}}\Omega_{\ovb{12}}}
+
\frac{e^{\beta \Omega_{\ovb{1}}}- 1}
{\Omega_{\ovb{3}}\Omega_{\ovb{1}}}
\! \left[
\frac{ 1 }{\Omega_{\ovb{23}}}
\!-\! 
\frac{ 1 }{\Omega_{\ovb{2}}}
\right] \hspace{-1cm} & 
\nonumber
\\
& =
\frac{\zeta 
e^{-\beta \omega'_{\ovb{1}\ovb{2}\ovb{3}}} }
{\Omega_{4} \Omega_{4\ovb{1}} \Omega_{4\ovb{12}}}
\!+\!
\frac{e^{-\beta \omega'_{\ovb{1}\ovb{2}}}}{\Omega_{\ovb{3}}\Omega_{\ovb{3}4}\Omega_{\ovb{3}4\ovb{1}}}
\!+\!
\frac{\zeta 
e^{-\beta \omega'_{\ovb{1}}}}{\Omega_{\ovb{2}}\Omega_{\ovb{23}}\Omega_{\ovb{2}\ovb{3}4}}
\!+\!
\frac{1}{\Omega_{\ovb{1}}\Omega_{\ovb{12}}\Omega_{\ovb{123}}} . &
\label{eq:FinalResultTimeIntegralGIFl=4-revised}
\end{flalign}
The $\beta$-independent term was obtained using
\begin{align}
\nonumber
\frac{1}{\Omega_{\ovb{3}}}
\Biggl[\, 
\underbrace{
\frac{ 1 }{\Omega_{\ovb{23}}\Omega_{\ovb{123}}}
\!-\!
\frac{ 1 }{\Omega_{\ovb{1}}\Omega_{\ovb{23}}}
}_{
-1/(\Omega_{\ovb{1}}\Omega_{\ovb{123}})
}
+
\underbrace{
\frac{ 1 }{\Omega_{\ovb{1}}\Omega_{\ovb{2}}}
\!-\!
\frac{ 1 }{\Omega_{\ovb{2}}\Omega_{\ovb{12}}}
}_{
1/(\Omega_{\ovb{1}}\Omega_{\ovb{12}})
}\, \Biggr]
= \frac{1}
{\Omega_{\ovb{1}}\Omega_{\ovb{2}}\Omega_{\ovb{123}}} , 
\end{align}
while, for the $\beta$-dependent ones, we exploited 
$\Omega_{\ovb{1}\ovb{2}\ovb{3}4}=0$ to 
obtain four cyclically related denominators, 
with the first three obtainable from the fourth via the permutations 
$(4\ovb{1}\ovb{2}\ovb{3})$, $(\ovb{3}4\ovb{1}\ovb{2})$, and $(\ovb{2}\ovb{3}4\ovb{1})$.
Using the cyclicity relation~\eqref{eq:Sc_w_cyclicity},
we find that 
$\ovwt{k} (\vec{\Omega}_{(\ovb{1}\ovb{2}\ovb{3}4)}) 
S_{(\ovb{1}\ovb{2}\ovb{3}4)}$ has the form 
\begin{align}
\frac{\zeta 
S_{(4\ovb{1}\ovb{2}\ovb{3})} }
{\Omega_{4} \Omega_{4\ovb{1}} \Omega_{4\ovb{12}}}
\!+\!
\frac{S_{(\ovb{3}4\ovb{1}\ovb{2})}}{\Omega_{\ovb{3}}\Omega_{\ovb{3}4}\Omega_{\ovb{3}4\ovb{1}}}
\!+\!
\frac{\zeta S_{(\ovb{2}\ovb{3}4\ovb{1})}}{\Omega_{\ovb{2}}\Omega_{\ovb{23}}\Omega_{\ovb{2}\ovb{3}4}}
\!+\!
\frac{S_{(\ovb{1}\ovb{2}\ovb{3}4)}}{\Omega_{\ovb{1}}\Omega_{\ovb{12}}\Omega_{\ovb{123}}} 
. 
\label{eq:ell=4KS-regular}
\end{align}
The sum over all six permutations $q = ( \ovb{1}\ovb{2}\ovb{3}4)$ yields 
\begin{flalign}
\ovwt{G}(\mi\vec{\omega})
& =
\sum_p \vec{\zeta}^p \! 
\nint 
\md \omega'_{\ovb{1}} 
\md \omega'_{\ovb{2}}
\md \omega'_{\ovb{3}} \,
\frac{
S_{(\ovb{1}\ovb{2}\ovb{3}\ovb{4})}}
{\Omega_{\ovb{1}} \Omega_{\ovb{12}} \Omega_{\ovb{123}}}
, 
\label{eq:GIF-final-l=4-appendix}
\end{flalign}
where the sum now runs over all $24$ permutations
$p= (\ovb{1}\ovb{2}\ovb{3}\ovb{4})$. This is again consistent
with \Eqs{eq:G_iw_KS_IF} and
\eqref{eq:Ktilde_Omega_IF}.

\textit{Anomalous part:} 
If all four operators are fermionic, 
$\omega_{\ovb{i}}$ and $\Omega_{\ovb{i}}$ are always nonzero, 
but $\Omega_{\ovb{ij}}$ can vanish. 
Indeed, in \Eq{eq:FinalResultTimeIntegralGIFl=4-revised},
denominators vanish 
(i) in the second and fourth terms if 
$\Omega_{\ovb{1}\ovb{2}} = -\Omega_{\ovb{3}4} = 0$, 
and (ii) in the first and third terms if $\Omega_{\ovb{2}\ovb{3}} = 
- \Omega_{4 \ovb{1}}= 0$. 
We discuss these cases by taking the vanishing frequencies to be infinitesimal; 
recomputing the integral \eqref{eq:G4_direct_K4-revised-integrated-once-full} yields the same results. 

For case (i), we set $\omega_{\ovb{1}\ovb{2}} = - \omega_{\ovb{3}4} = 0$ and
$- \omega'_{\ovb{1}\ovb{2}} = \omega_{\ovb{3}4}' = \epsilon$, 
so that $\Omega_{\ovb{1}\ovb{2}} = - \Omega_{\ovb{3}4} = \epsilon$,
and compute $\ovwh{k}$ as $\ovwt{k}|_{\epsilon \to 0}$.
The second and fourth terms of \Eq{eq:FinalResultTimeIntegralGIFl=4-revised} then yield 
\begin{align}
\nonumber
& \frac{e^{\beta \epsilon}}{
\Omega_{\ovb{3}} (-\epsilon)(\Omega_{\ovb{1}} \! - \! \epsilon)}
\! + \! 
\frac{1}{\Omega_{\ovb{1}} \epsilon 
(\Omega_{\ovb{3}} \! + \! \epsilon)}
\to - \bigg[
\beta \! + \! 
\frac{1}{\Omega_{\ovb{1}}} \! + \! 
\frac{1}{\Omega_{\ovb{3}}}
\bigg] \frac{1}{\Omega_{\ovb{1}}\Omega_{\ovb{3}}} , 
\end{align}
while the first and third terms remain unchanged.
Since $\omega'_{\ovb{1}\ovb{2}}=0$ implies 
$S_{(\ovb{3}4\ovb{1}\ovb{2})}= S_{(\ovb{1}\ovb{2}\ovb{3}4)}$, 
we can thus obtain $\ovwh{k} 
(\vec{\Omega}_{(\ovb{1}\ovb{2}\ovb{3}4)}) 
S_{(\ovb{1}\ovb{2}\ovb{3}4)}$ from $\ovwt{k} 
(\vec{\Omega}_{(\ovb{1}\ovb{2}\ovb{3}4)}) 
S_{(\ovb{1}\ovb{2}\ovb{3}4)}$ of \Eq{eq:ell=4KS-regular} 
by replacing the
second and fourth terms of the latter by
\begin{align}
& 
- \! \frac{1}{2} 
\! \! \left[ 
\beta 
\! + \! \frac{1}{\Omega_{\ovb{3}4\ovb{1}}}
\! + \! \frac{1}{\Omega_{\ovb{3}}}
\right] \! 
\frac{S_{(\ovb{3}4\ovb{1}\ovb{2})}}
{\Omega_{\ovb{3}4\ovb{1}} \Omega_{\ovb{3}}} 
- \! \frac{1}{2} 
\! \! \left[ 
\beta \! + \! \frac{1}{\Omega_{\ovb{1}}}
\! + \frac{1}
{\Omega_{\ovb{1}\ovb{2}\ovb{3}}} 
\right] \! 
\frac{S_{(\ovb{1}\ovb{2}\ovb{3}4)}}
{\Omega_{\ovb{1}}\Omega_{\ovb{1}\ovb{2}\ovb{3}}}
. \rule[-4.5mm]{0mm}{0mm}
\label{eq:l=4-Khat-symmetrized}
\end{align}
Here, we split the anomalous contribution
equally between these two terms, 
using $\Omega_{\ovb{1}} = \Omega_{\ovb{3}4\ovb{1}}$
and $\Omega_{\ovb{3}} = \Omega_{\ovb{123}}$
to rewrite
the denominators such that their $\Omega$'s 
match the nonzero 
$\Omega$'s in the second and fourth terms of 
\Eq{eq:ell=4KS-regular}.

For case (ii), we similarly 
set $\Omega_{\ovb{2}\ovb{3}} = 
- \Omega_{4 \ovb{1}}= \epsilon$. Then, 
the first and third terms of $\ovwt{k}$ in \Eq{eq:FinalResultTimeIntegralGIFl=4-revised} yield 
\begin{align}
\nonumber
& \frac{\zeta e^{\beta \epsilon} e^{- \beta \omega'_{\ovb{1}}}}
{\Omega_{4} (-\epsilon)(\Omega_{\ovb{2}} \! - \! \epsilon)}
\! + \! 
\frac{\zeta e^{- \beta \omega'_{\ovb{1}}}}{\Omega_{\ovb{2}} \epsilon 
(\Omega_{4} \! + \! \epsilon)}
\to - \bigg[
\beta \! + \! \frac{1}{\Omega_{\ovb{2}}} \! + \! 
\frac{1}{\Omega_{4}}
\bigg] \frac{\zeta e^{- \beta \omega'_{\ovb{1}}}}
{\Omega_{\ovb{2}}\Omega_{4}} ,
\end{align}
while the second and fourth terms remain unchanged.
Using $\Omega_{4} = \Omega_{\ovb{23}4}$
and $\Omega_{2} = \Omega_{4\ovb{12}}$,
we can thus obtain $\ovwh{k} 
(\vec{\Omega}_{(\ovb{1}\ovb{2}\ovb{3}4)}) 
S_{(\ovb{1}\ovb{2}\ovb{3}4)}$ from $\ovwt{k} 
(\vec{\Omega}_{(\ovb{1}\ovb{2}\ovb{3}4)}) 
S_{(\ovb{1}\ovb{2}\ovb{3}4)}$ of \Eq{eq:ell=4KS-regular} 
by replacing the first and third terms of the latter by
\begin{align}
& 
- \! \frac{1}{2} 
\! \! \left[ 
\beta 
\! + \! \frac{1}{\Omega_{4\ovb{1}\ovb{2}}}
\! + \! \frac{1}{\Omega_{4}}
\right] \! \! 
\frac{\zeta 
S_{(4\ovb{1}\ovb{2}\ovb{3})} }
{\Omega_{4} \Omega_{4\ovb{1}\ovb{2}}} 
- \! \frac{1}{2} 
\! \left[ 
\beta 
\! + \! \frac{1}{\Omega_{\ovb{2}\ovb{3}4}}
\! + \! \frac{1}{\Omega_{\ovb{2}}}
\right] \! \! 
\frac{\zeta S_{(\ovb{2}\ovb{3}4\ovb{1})}}
{\Omega_{\ovb{2}}\Omega_{\ovb{2}\ovb{3}4}} , 
\label{eq:l=4-Khat-symmetrized-Omega23=0}
\rule[-4.5mm]{0mm}{0mm}
\end{align}
with denominators matching the nonzero 
$\Omega$'s in \Eq{eq:ell=4KS-regular}.

If both (i) and (ii) hold simultaneously, 
the product $\ovwh{k} 
(\vec{\Omega}_{(\ovb{1}\ovb{2}\ovb{3}4)}) 
S_{(\ovb{1}\ovb{2}\ovb{3}4)}$ is given by the sum
of the anomalous terms in \Eqs{eq:l=4-Khat-symmetrized} and
\eqref{eq:l=4-Khat-symmetrized-Omega23=0}.
Note that the four parts of these two equations
are cyclically related, in that the second part of 
\eqref{eq:l=4-Khat-symmetrized} yields the
other three via the permutations
$(\ovb{3}4\ovb{1}\ovb{2})$, $(4\ovb{1}\ovb{2}\ovb{3})$
or $(\ovb{2}\ovb{3}4\ovb{1})$ (with $\vec{\zeta}^{(4\ovb{1}\ovb{2}\ovb{3})}
= \vec{\zeta}^{(\ovb{2}\ovb{3}4\ovb{1})} = \zeta$).

Jointly, the expressions \eqref{eq:ell=4KS-regular},
\eqref{eq:l=4-Khat-symmetrized},
and \eqref{eq:l=4-Khat-symmetrized-Omega23=0}
obtained above for the regular 
$\ovwt{k}(\vec{\Omega}_{(\ovb{1}\ovb{2}\ovb{3}4)})$
or the anomalous $\ovwh{k}(\vec{\Omega}_{(\ovb{1}\ovb{2}\ovb{3}4)}) $ times $ S_{(\ovb{1}\ovb{2}\ovb{3}4)}$ 
imply a kernel of the form 
\begin{align}
K(\vec{\Omega}_p) = 
\begin{cases}
{\displaystyle 
\frac{1}{\Omega_{\ovb{1}} \Omega_{\ovb{1}\ovb{2}}
\Omega_{\ovb{1}\ovb{2}\ovb{3}}}} 
& \text{if} \quad \Omega_{\ovb{1}\ovb{2}} 
\neq 0
, 
\\[3mm]
{\displaystyle 
- \frac{1}{2} \! 
\left[\beta + \frac{1}{\Omega_{\ovb{1}}} + 
\frac{1}{\Omega_{\ovb{1}\ovb{2}\ovb{3}}}
\right]\! \frac{1}{
\Omega_{\ovb{1}}\Omega_{\ovb{1}\ovb{2}\ovb{3}}}}
 & \text{if} \quad \Omega_{\ovb{1}\ovb{2}} = 0 
 , 
\end{cases}
\label{eq:K-IF-l=4-anomalous-explicit}
\end{align}
for any of the 24 permutations $p=(\ovb{1}\ovb{2}\ovb{3}\ovb{4})$, consistent with the general \Eq{eq:K_Omega_IF}
from the main text.

Since the anomalous terms not proportional to $\beta$ follow
from expanding denominators of $\ovwt{k}$, they need not
be displayed explicitly.
The full kernel can thus be expressed 
in the following form, 
equivalent to \Eq{eq:K-IF-l=4-anomalous-explicit}, but written in the notation of \Eq{eq:K_Omega_IF_compact}:
\begin{align}
K(\vec{\Omega}_p) =
\frac{1}{\Omega_{\ovb{1}} \Omega_{\ovb{1}\ovb{2}}
\Omega_{\ovb{1}\ovb{2}\ovb{3}}} 
-
\delta_{\Omega_{\ovb{1}\ovb{2}},0}
\frac{\beta}{2}
\frac{1}{
\Omega_{\ovb{1}}\Omega_{\ovb{1}\ovb{2}\ovb{3}}} 
. 
\label{eq:K-IF-l=4-compact-App}
\end{align}
This concludes our derivation of \Eqs{eq:K_Omega_IF_compact}
and \eqref{eq:K_Omega_IF}.

\section{Explication of \KF/ formula for \texorpdfstring{$\ell = 2$}{l=2}}
\label{app:KF_G2}
\label{app:KF}

Here, we show how the general $\ell$p formulas \eqref{eq:G_w_KF_eta_j_total}
reproduce the well-known $2$p correlators
in the Keldysh basis of the \KF/.
For $\ell=2$, \Eqs{eq:G_w_KF_eta_j_total} read
\begin{flalign}
\nonumber
G^{[\eta_1 \nidots\eta_\alpha]} (\omega)
& =
\sum_p \vec{\zeta}^p 
\nint \md \omega_{\ovb{1}}' \,
K^{[\ovh{\eta}_1 \nidots \ovh{\eta}_\alpha]}
(\omega_{\ovb{1}} - \omega_{\ovb{1}}' )
S[\vec{\Oc}_p](\omega_{\ovb{1}}') , 
\\
K\sseta(\omega_{\ovb{1}}) & 
= \frac{1}{\omega\sseta_{\ovb{1}}},
\quad 
K^{[12]}(\omega_{\ovb{1}}) 
= \frac{1}{\omega^{[\ovb{1}]}_{\ovb{1}}} -
\frac{1}{\omega^{[\ovb{2}]}_{\ovb{1}}} 
. & 
\label{eq:GL_KF_keldysh-eta-l=2-Appendix}
\end{flalign}
We hid the second frequency argument
of $G$, $K$, and $S$, as done in \Eq{eq:GL_RF_app}, 
with $\omega_1 \!=\! -\omega_2 \!= \!\omega$ and
$\omega_1' \!=\! - \omega_2'$ being understood. 
Summing over $p \!\in\! \{(12), (21)\}$, 
inserting \Eqs{subeq:psf_21_app} for the PSFs, 
and recalling $\omega\sseta_{12} \!=\! 0$
to replace $\omega\sseta_2$ by $- \omega\sseta_1$, 
we obtain
\begin{subequations}
\label{subeq:KF-l=2-G[1][2][12]-general-gamma0}
\begin{flalign} 
\nonumber
G^{[\eta]}(\omega)
& = \sum_{\ub{1}\,\ub{2}}
A_{\ub{1}\ub{2}} B_{\ub{2}\ub{1}} \! 
\pigg[
\frac{ \rho_{\ub{1}}}
{\omega_1^{[\eta]} \! - \! E_{\ub{2}\ub{1}}} 
+ 
\frac{ \zeta \rho_{\ub{2}}}
{\omega_2^{[\eta]} \! - \! E_{\ub{1}\ub{2}}} 
\pigg] 
, 
\\
& = \sum_{\ub{1}\,\ub{2}}
A_{\ub{1}\ub{2}} B_{\ub{2}\ub{1}} \, \frac{ \rho_{\ub{1}} - \zeta \rho_{\ub{2}}}
{\omega_1^{[\eta]} \! - \! E_{\ub{2}\ub{1}}} 
,
\label{eq:GL_KF0_app} 
\\ 
\nonumber
G^{[12]}(\omega)
& = \sum_{\ub{1}\,\ub{2}}
 A_{\ub{1}\ub{2}} B_{\ub{2}\ub{1}}
\pigg[ 
\rho_{\ub{1}}
\pigg( 
\frac{1}{\omega^{[1]}_{1} - E_{\ub{2}\ub{1}}} 
-
\frac{1}{\omega^{[2]}_{1} - E_{\ub{2}\ub{1}}}
\pigg)
\\
& \hspace{2cm} 
+ 
\zeta \rho_{\ub{2}}
\pigg(
\frac{1}{\omega^{[2]}_{2} \! - \! E_{\ub{1}\ub{2}}} 
-
\frac{1}{\omega^{[1]}_{2} \! - \! E_{\ub{1}\ub{2}}}
\pigg) 
\pigg]
\nonumber
\\
& = \sum_{\ub{1}\,\ub{2}}
A_{\ub{1}\ub{2}} B_{\ub{2}\ub{1}} 
\pigg[
\frac{\rho_{\ub{1}} + \zeta \rho_{\ub{2}}}{\omega^{[1]}_{1} \! - \! E_{\ub{2}\ub{1}}} 
-
\frac{\rho_{\ub{1}} + \zeta \rho_{\ub{2}}}{\omega^{[2]}_{1} \! - \! E_{\ub{2}\ub{1}}}
\pigg] 
. 
\hspace{-0.5cm} & 
\label{eq:G[12]App} 
\end{flalign}
\end{subequations}

To make contact with the expressions from Sec.~\ref{sec:motivation},
we now set $\omega\sseta_{\ovb{1}=\eta}
= \omega_{\ovb{1}} + \mi \gamma_0$ and 
$\omega\sseta_{\ovb{1} \neq \eta}
= \omega_{\ovb{1}} - \mi \gamma_0$ [cf.~\Eq{eq:choice_imaginary_parts}] and
use $\omega_1 = - \omega_2 = \omega$ for their real parts. 
Choosing $\gamma_0$ infinitesimal, in which case $K^{[12]}(\omega) = - 2 \pi \mi 
\delta (\omega)$, we obtain the standard 
expressions for $G^{21}= G^{[1]}$ [cf.~\Eq{eq:G2_w_KF_retarded}], 
$G^{12}= G^{[2]}$, and $G^{22}= G^{[12]}$:
\begin{subequations}
\label{eq:G2-KF-app-gamma_0-infinitesimal}
\begin{flalign}
\begin{rcases}
G^{[1]}(\omega)\\
G^{[2]}(\omega)
\end{rcases} 
 & = 
\sum_{\ub{1}\,\ub{2}}
A_{\ub{1}\ub{2}} B_{\ub{2}\ub{1}}
\frac{ \rho_{\ub{1}}- \zeta \rho_{\ub{2}}}
{\omega^\pm - E_{\ub{2}\ub{1}}} 
,
&
\label{eq:G[1,2]App} 
\\
G^{[12]}(\omega) 
& =
-2\pi\mi \!
\sum_{\ub{1}\,\ub{2}} \!
A_{\ub{1}\ub{2}} B_{\ub{2}\ub{1}}
\big(
\rho_{\ub{1}} \!+\! \zeta \rho_{\ub{2}}
\big) 
\delta(\omega \!-\! E_{\ub{2}\ub{1}})
. 
\hspace{-1cm} & 
\label{eq:G[12]App_inf}
\end{flalign}
\end{subequations}
These fulfill the fluctuation-dissipation theorem (FDT)
\begin{align}
G^{[12]}(\omega) = 
\big[
\coth(\beta \omega/2)
\big]^{\zeta} 
\big[
G^{[1]}(\omega) - G^{[2]}(\omega)
\big] 
,
\label{eq:G22_FDT}
\end{align}
since $\rho_{\ub{2}} = e^{-\beta \omega} \rho_{\ub{1}}$ 
if $\omega = E_{\ub{2}\ub{1}}$. Note that if 
$\gamma_0$ is finite, the latter condition does not hold,
and neither does \Eq{eq:G22_FDT}.

\section{Imaginary shifts for external legs}
\label{app:explicit_amputation}

In this Appendix, we explain how the imaginary frequency shifts 
needed for \ZF/ and \KF/ correlators should
be chosen in the external legs of a connected $4$p correlator when amputating
these to extract the $4$p vertex. As in Sec.~\ref{sec:quantities}, we
display only the first $\ell \!-\! 1$ frequency arguments of $\ell$p functions,
with $\omega_\ell = - \omega_{1 \cdots \ell-1}$. 
We begin by discussing the \KF/ case; the \ZF/ case follows by analogy.

In using \Eq{eq:Keldysh_vertex_part} to numerically extract the \KF/ vertex 
$F(\omega_1, \omega_2, \omega_3)$ from the connected 
correlator $G^{\mathrm{con}}(\omega_1, \omega_2, \omega_3)$ 
by dividing out the external legs, the \textit{same} 
imaginary frequency shifts must be used for the external-leg $2$p correlators on the right as for the $4$p correlator $G^{\mathrm{con}}$ on the left. 
This may seem daunting, since 
the spectral representation \eqref{eq:G_w_KF_Keldysh} of a $4$p correlator
involves a permutation sum $\sum_p$, 
and the imaginary shifts accompanying the arguments $\vec{\omega}_p$ of the \KF/ kernels 
$K^{ \vec{k}_p }( \vec{\omega}_p \!-\! \vec{\omega}_p' )$ depend on $p$. 
However, $K^{ \vec{k}_p}$, given by \Eqs{eq:Kc_KF_Keldysh_via_Keta} and \eqref{eq:Keta_finite_gamma}, 
depends on its frequency arguments $\vec{\omega}_p$
only via the complex 4-tuples 
$\vec{\omega}_p^{[1]} , \ndots, \vec{\omega}_p^{[4]}$ of \Eq{eq:choice_imaginary_parts}, 
whose components $\omega\sseta_\ovb{i}$ have only two
possible imaginary parts, $+ 3 \mi \gamma_0$
or $- \mi \gamma_0$. Hence, each argument
$\omega_i$ of $G^{\mathrm{con}}(\omega_1, \omega_2, \omega_3)$
enters as either $\omega_i \!+\! 3\mi\gamma_0$ or $\omega_i \!-\! \mi\gamma_0$. 
Thus, we just have to ensure that the frequency arguments of the external-leg $2$p correlators enter in the same manner. 

For $2$p correlators $G(\omega)$, the argument $\omega$ enters via the complex 2-tuples 
$\vec{\omega}^{[1]}$ and $\vec{\omega}^{[2]}$.
We choose these as 
\begin{align}
\label{eq:2p-imaginary-shifts-external-legs}
\vec{\omega}^{[1]} \!=\! (\omega \!+\! a\mi\gamma_0, -\omega \!-\! a\mi\gamma_0), 
\quad 
\vec{\omega}^{[2]} \!=\! (\omega \!-\! b\mi\gamma_0, -\omega \!+\! b\mi\gamma_0),
\end{align} 
consistent with
\Eq{eq:choice_imaginary_parts}, but involving
two prefactors, $a, b \!>\! 0$, to be specified below.
The $2$p correlators on the right of \Eq{eq:Keldysh_vertex_part}
occur in two ways, 
(1) $G({\omega}_i)$ ($i \!=\! 1,3$) and (2) $G(-{\omega}_i)$ ($i \!=\! 2,4$), with 
${\omega}_i$ one of the $4$p arguments. For case (1), we choose $a \!=\! 3$, $b \!=\! 1$, 
such that $\vec{\omega}^{[1]}$ depends on ${\omega}_i + 3 \mi \gamma_0$
and $\vec{\omega}^{[2]}$ on ${\omega}_i - \mi \gamma_0$. 
For case (2), we choose $a \!=\! 1$, $b \!=\! 3$, such that $\vec{\omega}^{[1]}$ depends on 
$-{\omega}_i \!+\! \mi\gamma_0 \!=\! -({\omega}_i \!-\! \mi\gamma_0)$ and $\vec{\omega}^{[2]}$ on
$-{\omega}_i \!-\! 3\mi\gamma_0 \!=\! -({\omega}_i \!+\! 3\mi\gamma_0)$.
By \Eqs{subeq:KF-l=2-G[1][2][12]-general-gamma0}, 
the external-leg $2$p correlators for cases (1) and (2) thus read:
\begin{align}
G^{[1]}(\omega_i)
& \!\stackrel{{(\rm 1)}}{=}\! 
A_{\ub{1}\ub{2}} B_{\ub{2}\ub{1}} 
\frac{
\rho_{\ub{1}} - \zeta \rho_{\ub{2}} 
}
{
\omega_i + 3\mi\gamma_0 - E_{\ub{2}\ub{1}}
}
,
\\
G^{[2]}(\omega_i)
& \!\stackrel{{(\rm 1)}}{=}\! 
A_{\ub{1}\ub{2}} B_{\ub{2}\ub{1}} 
\frac{
\rho_{\ub{1}} - \zeta \rho_{\ub{2}} 
}
{
\omega_i - \mi\gamma_0 - E_{\ub{2}\ub{1}}
}
,
\nonumber
\\
G^{[12]}(\omega_i)
& \!\stackrel{{(\rm 1)}}{=}\! 
A_{\ub{1}\ub{2}} B_{\ub{2}\ub{1}} 
\bigg[
\frac{
\rho_{\ub{1}} + \zeta \rho_{\ub{2}}
}
{
\omega_i \!+\! 3\mi\gamma_0 \!-\! E_{\ub{2}\ub{1}}
}
\!-\!
\frac{
\rho_{\ub{1}} + \zeta \rho_{\ub{2}}
}
{
\omega_i \!-\! \mi\gamma_0 \!-\! E_{\ub{2}\ub{1}}
}
\bigg]
\! ,
\nonumber
\\
G^{[1]}(-\omega_i)
& \!\stackrel{{(\rm 2)}}{=}\! 
A_{\ub{1}\ub{2}} B_{\ub{2}\ub{1}} 
\frac{
\rho_{\ub{1}} - \zeta \rho_{\ub{2}} 
}
{
-\omega_i + \mi\gamma_0 - E_{\ub{2}\ub{1}}
}
,
\nonumber
\\
G^{[2]}(-\omega_i)
& \!\stackrel{{(\rm 2)}}{=}\! 
A_{\ub{1}\ub{2}} B_{\ub{2}\ub{1}} 
\frac{
\rho_{\ub{1}} - \zeta \rho_{\ub{2}} 
}
{
-\omega_i - 3\mi\gamma_0 - E_{\ub{2}\ub{1}}
}
,
\nonumber
\\
G^{[12]}(-\omega_i)
& \!\stackrel{{(\rm 2)}}{=}\! 
A_{\ub{1}\ub{2}} B_{\ub{2}\ub{1}} 
\bigg[
\frac{
\rho_{\ub{1}} + \zeta \rho_{\ub{2}}
}
{
-\omega_i \!+\! \mi\gamma_0 \!-\! E_{\ub{2}\ub{1}}
}
\!-\!
\frac{
\rho_{\ub{1}} + \zeta \rho_{\ub{2}}
}
{
-\omega_i \!-\! 3\mi\gamma_0 \!-\! E_{\ub{2}\ub{1}}
}
\bigg]
\! .
\nonumber
\end{align}
Here, a summation over underlined indices is implicit.

Next, we discuss the \ZF/ case. 
As pointed out at the end of Sec.~\ref{sec:KF:retarded_kernel}, 
the \ZF/ imaginary frequency shifts can be defined in two ways: 
for $4$p correlators, we can replace $\vec{\omega}_p$ either 
(i) by $\vec{\omega}^{[\ovb{1}]}_p$ or 
(ii) by $\big( \vec{\omega}^{[\ovb{4}]}_p \big)^*$. 
For choice (i), 
$\omega_\ovb{i}$ enters as either $\omega_\ovb{i} + \mi 3 \gamma_0$ or $\omega_\ovb{i} - \mi \gamma_0$, 
for choice (ii) as $\omega_\ovb{i} + \mi \gamma_0$ or $\omega_\ovb{i} - 3\mi \gamma_0$. 
Choice (i) matches the situation encountered above for the \KF/;
choice (ii) requires an additional interchange $3 \leftrightarrow 1$ for the size of the positive or negative shifts.
Thus, the amputation of $2$p correlators with first argument $\omega$ may again proceed via \Eq{eq:2p-imaginary-shifts-external-legs}. 
For choice (i), we take $a$, $b$ as specified above, 
and for choice (ii) we modify that specification by $3 \leftrightarrow 1$.

\section{Hubbard atom at second order}
\label{sec:app:HA_2ndOrder}

To understand how anomalous contributions in the \MF/ are translated into the \KF/, 
we consider the (half-filled) Hubbard atom at second order in $U$ as an instructive example.
The vertex at second order follows from
the bare susceptibility $\chi_0$,
cf.\ \Eq{eq:vertex_SOPT}. 
We can also deduce the second-order self-energy $\Sigma$ from $\chi_0$.
In the \MF/,
\begin{flalign}
\chi_0(\mi\omega) 
& \!=\! 
-(G_0 
\!\ast\!
G_0)(\mi\omega)
,
\ \
\Sigma(\mi\nu) 
\!=\! 
U^2
(G_0 
\!\ast\! 
\chi_0)(\mi\nu)
,
\hspace{-0.5cm}
&
\label{eq:HA_PT_IF}
\end{flalign}
where
$(A \!\ast\! B)(\mi\omega) \!=\! \frac{1}{\beta} \sum_{\nu} A(\mi\nu \!+\! \mi\omega) B(\mi\nu)$.
Using $G_0(\mi\nu) \!=\! 1/(\mi\nu)$,
we find
$\chi_0(\mi\omega) \!=\! \tfrac{1}{4}\beta \delta_{\omega,0}$.
Indeed, this result for $\chi_0$ is also seen 
by expanding the vertex \eqref{eq:vertex_AL} to order $U^2$,
$
F_{\uparrow\downarrow}
\!=\!
U
\!+\!
\tfrac{1}{4} \beta U^2
(
\delta_{\omega_{14}} 
\!-\!
\delta_{\omega_{13}}
)
$
and
$
F_{\uparrow\uparrow}
\!=\!
\tfrac{1}{4} \beta U^2
(
\delta_{\omega_{14}}
\!-\!
\delta_{\omega_{12}}
)
$,
and comparing it to \Eq{eq:vertex_SOPT}.
Computing the self-energy yields
$\Sigma(\mi\nu) 
\!=\!
U^2/(4\mi\nu)$.
The second-order result for $\Sigma$ turns out to be exact,
as can be checked from $\Sigma \!=\! G_0^{-1} - G^{-1}$
with the exact
$G(\mi\nu) = \tfrac{1}{2} \sum_{\pm} (\mi\nu \pm U/2)^{-1}$.

Now, we compute $\chi_0$ in the \KF/.
As a bosonic $2$p function, it has three nonzero Keldysh components
$\chi_0^R$, $\chi_0^A$, $\chi_0^K$.
The formula for the retarded component is
\begin{align}
\chi_0^R(\omega) 
& = 
-(G^R_0 \ast G^K_0 + G^K_0 \ast G^A_0)(\omega)
,
\end{align}
where
$(A \ast B)(\omega) \!=\! \frac{1}{2\pi\mi} \nint \md \nu A(\nu \!+\! \omega) B(\nu)$.
From 
$G_0^R(\nu) \!=\! 1/(\nu^+)$,
where $\nu^+ = \nu + \mi0^+$,
and the fermionic FDT [\Eq{eq:G22_FDT}], we find
$G_0^K(\nu) \!\propto\! \tanh (\beta\nu/2) \delta(\nu) \!=\! 0$.
Hence, $\chi_0^R(\omega) \!=\! 0$.

The \KF/ result $\chi_0^R(\omega) \!=\! 0$ does not reflect the 
nonzero part of $\chi_0(\mi\omega) \!=\! \tfrac{1}{4}\beta \delta_{\omega,0}$.
[One may attribute this to the fact that the analytic continuation 
$\chi_0^R(\omega) \!=\! \chi_0(\mi\omega \!\to\! \omega^+)$ 
smoothly approaches $\omega \!=\! 0$ from above the real axis, where
$\chi_0(\mi\omega)=0$.]
Still, the two expressions are consistent in that both
arise from a time-independent operator, 
like $\Oc = d^\dag_\sigma d_\sigma$ with $[\Hc,\Oc]=0$ for the Hubbard atom.
On the one hand, in the \MF/ time domain,
$\Oc(\tau) \!=\! \Oc$ yields 
$\chi(\tau) \!\propto\! \langle \TO \Oc(\tau) \Oc \rangle \!\propto\! \textrm{const}$;
hence, $\chi(\mi \omega) \!\propto\! \delta_{\omega,0}$. 
On the other hand, in the \KF/ time domain, $\Oc(t) \!=\! \Oc$ yields
$\chi^R(t) \!\propto\! \langle [\Oc(t), \Oc] \rangle = 0$, 
since the commutator vanishes.
Importantly, the information contained in 
$\chi_0(\tau) \!\propto\! \textrm{const}$ 
and $\chi(\mi\omega) \!\propto\! \delta_{\omega,0}$ is not lost;
the \KF/ encodes it via the anticommutator of the Keldysh component,
$\chi^K(t) \!=\! -\mi\langle \{\Oc(t), \Oc\} \rangle \!\propto\! \textrm{const}$
yielding 
$\chi^K(\omega) \!\propto\! \delta(\omega)$.
Indeed, the formula for the Keldysh component of $\chi_0$,
\begin{flalign}
\chi_0^K(\omega) 
& = 
-(G^R_0 \ast G^A_0 + G^A_0 \ast G^R_0 + G^K_0 \ast G^K_0)(\omega)
,
\hspace{-0.5cm}
&
\end{flalign}
gives $\chi_0^K(\omega) \!=\! 2\pi\mi\delta(\omega)$.
Note that these results for $\chi_0$ in the \KF/
are consistent with the bosonic FDT [\Eq{eq:G22_FDT}],
rearranged as follows to avoid the singularity of $\coth (\beta\omega/2)$ at $\omega \!=\! 0$:
$\tanh (\beta\omega/2) \chi^K(\omega) \!=\! \chi^R(\omega) \!-\! \chi^A(\omega) \!=\! 0$.

Finally, we remark that there are attempts in the literature to incorporate
anomalous contributions known from the \MF/, as the one in $\chi_0(\mi\omega)$,
into retarded functions like $\chi_0^R(\omega)$
(see Ref.~\onlinecite{Shvaika2006}, and references therein).
This is not necessary when working within the full-fledged \KF/, 
as the information of anomalous \MF/
contributions is encoded in other Keldysh components,
such as $\chi_0^K(\omega)$.
Indeed, the second-order retarded self-energy,
\begin{align}
\Sigma^R(\nu) 
& = 
( U/2 )^2 \,
( G^K_0 \ast \chi_0^A
+
G^R_0 \ast \chi_0^K )
(\nu)
,
\end{align}
yields the correct result
$\Sigma^R(\nu) 
\!=\!
U^2/(4\nu^+)$ upon using
$\chi_0^A(\omega) \!=\! [\chi_0^R(\omega)]^* \!=\! 0$ and
$\chi_0^K(\omega) \!=\! 2\pi\mi\delta(\omega)$.
Any artificial modification of $\chi_0^R$ would give an incorrect result.

\bibliography{references}

\end{document}